\documentclass[11pt]{article}
\pdfoutput=1 % if your are submitting a pdflatex (i.e. if you have
\usepackage{jheppub} % for details on the use of the package, please
\usepackage{braket,slashed,bm}
\usepackage{array,multirow}
\usepackage[normalem]{ulem}
\usepackage{xcolor,cancel}
\usepackage{amsmath}
\usepackage{bbold}
\usepackage{dsfont}
\usepackage{cleveref}
\usepackage{xspace}
\usepackage{verbatim}
\usepackage{soul}

\crefname{table}{Table}{Tables}
\crefname{equation}{Eq.}{Eqs.}
\crefname{appendix}{App.}{Apps.}
\crefname{section}{Sec.}{Secs.}
\crefname{figure}{Fig.}{Figs.}

\usepackage{physics}
\usepackage{stackengine,xcolor,scalerel}

\usepackage[T1]{fontenc} % if needed
\usepackage{mathrsfs}
\usepackage{tikz}
\usetikzlibrary{arrows,decorations.pathmorphing,backgrounds,positioning,fit,petri,automata,shadows,calendar,mindmap,
decorations.markings,calc}

\def\Tom{\mathscr{T}}
\def\Lag{\mathcal{L}}
\def\nn{\nonumber\\}

\def\hyp{\mathsf{y}}

\def\sc{{\overline s}}

\def\12{\frac{1}{2}}

\def\dag{\dagger}

\def\tX{{\widetilde X}}
\def\tG{{\widetilde G}}
\def\tW{{\widetilde W}}
\def\tB{{\widetilde B}}
\def\tH{{\widetilde H}}
\def\lrD{\overleftrightarrow{D}}
\def\hc{\text{h.c.}}

\def\Op{O}
\def\WlC{a}

\def\lR{l_R}
\def\lRbar{\bar{l}_R}

\def\qR{q_R}
\def\qRbar{\bar{q}_R}

\definecolor{nucolor}{rgb}{0.0, 0.0, 1.0}
\definecolor{Hvcolor}{rgb}{1.0, 0.0, 0.0}
\definecolor{qvcolor}{rgb}{1.0, 0.5, 0.0}
\definecolor{lvcolor}{rgb}{0.0, 0.75, 0.0}

\newcommand{\shownu}[1]{{\color{nucolor}#1}}
\newcommand{\showHv}[1]{{\color{Hvcolor}#1}}
\newcommand{\showqv}[1]{{\color{qvcolor}#1}}
\newcommand{\showlv}[1]{{\color{lvcolor}#1}}

\newcommand{\Blcolor}{{\bf{black}}}
\newcommand{\nucolor}{{\bf\shownu{blue}}}
\newcommand{\Hvcolor}{{\bf\showHv{red}}}
\newcommand{\qvcolor}{{\bf\showqv{orange}}}
\newcommand{\lvcolor}{{\bf\showlv{green}}}

\def\cpmH{{\mathbin{\ThisStyle{\ensurestackMath{\abovebaseline[-\dimexpr1pt+2.0\LMpt]{%
	\stackunder[-\dimexpr1pt+2.0\LMpt]{\SavedStyle+}{\showHv{\SavedStyle-}}}}}}}}

\def\cpmq{{\mathbin{\ThisStyle{\ensurestackMath{\abovebaseline[-\dimexpr1pt+2.0\LMpt]{%
	\stackunder[-\dimexpr1pt+2.0\LMpt]{\SavedStyle+}{\showqv{\SavedStyle-}}}}}}}}
\def\cmpq{{\mathbin{\ThisStyle{\ensurestackMath{\abovebaseline[\dimexpr1pt-2.0\LMpt]{%
	\stackunder[-\dimexpr1pt+2.0\LMpt]{\showqv{\SavedStyle-}}{\SavedStyle+}}}}}}}
\def\cpml{{\mathbin{\ThisStyle{\ensurestackMath{\abovebaseline[-\dimexpr1pt+2.0\LMpt]{%
	\stackunder[-\dimexpr1pt+2.0\LMpt]{\SavedStyle+}{\showlv{\SavedStyle-}}}}}}}}

\def\cpmdouble{{\mathbin{\ThisStyle{\ensurestackMath{\abovebaseline[-\dimexpr1pt +2.0\LMpt]{%
	\stackunder[-\dimexpr1pt+1.5\LMpt]{+}{\textbf{\hspace{0.7pt}\showlv{-}\showqv{-}}}}}}}}}
\def\cmpqtwo{{\mathbin{\ThisStyle{\ensurestackMath{\abovebaseline[\dimexpr1pt-2.0\LMpt]{%
	\stackunder[-\dimexpr1pt+2.0\LMpt]{\SavedStyle-}{\showqv{\SavedStyle+}}}}}}}}
\def\cmpql{{\mathbin{\ThisStyle{\ensurestackMath{\abovebaseline[\dimexpr1pt-2.0\LMpt]{%
	\stackunder[-\dimexpr1pt+2.0\LMpt]{\showqv{\SavedStyle-}}{\showlv{\SavedStyle+}}}}}}}}

\title{Custodial symmetry (violation) in SMEFT}

\author[a]{Graham D. Kribs,}
\author[a]{Xiaochuan Lu,}
\author[b]{Adam Martin,}
\author[a,c]{and Tom Tong}

\affiliation[a]{Institute for Fundamental Science and Department of Physics, \\
  University of Oregon, Eugene, OR, 97403 USA}
\affiliation[b]{Department of Physics, University of Notre Dame, South Bend, IN
  46556 USA}
\affiliation[c]{Amherst Center for Fundamental Interactions, Department of Physics,
 University of Massachusetts Amherst, Amherst, MA 01003, USA}

\abstract{We investigate precision observables sensitive to custodial symmetric/violating UV physics beyond the Standard Model. We use the SMEFT framework which in general includes non-oblique corrections that requires a generalization of the Peskin-Takeuchi $T$ parameter to unambiguously detect custodial symmetry/violation. We take a first step towards constructing a SMEFT reparameterization-invariant replacement, that we call $\Tom$, valid at least for tree-level custodial violating contributions. We utilize a new custodial basis of $\nu$SMEFT (SMEFT augmented by right-handed neutrinos) which explicitly identifies the global $SU(2)_R$ symmetries of the Higgs and fermion sectors, that in turn permits easy identification of higher-dimensional operators that are custodial preserving or violating. We carefully consider equation-of-motion redundancies that cause custodial symmetric operators in one basis to be equivalent to a set of custodial symmetric and/or violating operators in another basis. Utilizing known results about tree/loop operator generation, we demonstrate that the basis-dependent appearance of custodial-violating operators does not invalidate our $\Tom$ parameter at tree-level. We illustrate our results with several UV theory examples, demonstrating that $\Tom$ faithfully identifies custodial symmetry violation, while $T$ can fail.}

\begin{document}

\maketitle

%  The following is needed *after* \maketitle so that JHEP page numbers
%  align with PDF page numbers, i.e., the title page is page 1.
\setcounter{page}{2}
\flushbottom

\section{Introduction}
\label{sec:Introduction}

It is widely anticipated that there is new physics beyond the Standard Model (SM). In the absence of directly producing the new particles of the Beyond the SM (BSM) sector, we would like to maximize the information we can glean about the UV physics from indirect probes. In this approach, the LEP era established the importance of electroweak precision data \cite{Z-Pole,LEP-2}, which could test the SM to an accuracy of $\sim 0.1\%$. Constraints on the scale of new physics can be $\Lambda \gtrsim 10$~TeV for those contributing to electroweak precision observables at the order $v^2/\Lambda^2$.

Directly calculating the contributions to electroweak precision observables from a given UV theory is in principle straightforward. However, it must be done on a case-by-case basis and consequently, does not (necessarily) provide general insights about the new physics. Peskin and Takeuchi demonstrated that the new physics effects can be efficiently categorized by utilizing three \emph{precision parameters} $S$, $T$, and $U$ \cite{Peskin:1991sw}. These parameters provide a simple, stunningly ubiquitous bridge between the effects of a new UV sector and electroweak precision data, and have become popular tests in determining the phenomenologically viable parameter space for BSM theories. In particular, the $T$ parameter is identified as the manifestation of ``custodial'' symmetry breaking effects from the UV sector. Theories beyond the SM are often constructed to respect custodial symmetry in order to avoid the strong bounds on the $T$ parameter, including originally technicolor \cite{Sikivie:1980hm} (for a review \cite{Hill:2002ap}), as well as composite Higgs, e.g., \cite{Georgi:1984af, Dugan:1984hq, Agashe:2004rs, Agashe:2006at, Giudice:2007fh}, little Higgs theories \cite{Csaki:2002qg, Hewett:2002px, Chang:2003zn, Cheng:2003ju, Chen:2003fm}, dark matter theories \cite{Delgado:2015aha, Kribs:2018oad, Choi:2019zeb}, etc.

The Peskin-Takeuchi $T$ parameter can be constrained from a variety of electroweak data. There are two observables that are often associated with directly constraining $T$: $\hat\rho_*(0)$, the ratio of charged current to neutral current (CC/NC) in the limit of zero momentum \cite{Peskin:1991sw}; and, the Veltman $\hat\rho\equiv \frac{m_W^2}{m_Z^2 \cos^2\theta}$ \cite{Veltman:1977kh}. We emphasize that these two are quite different observables\footnote{We thank S.~Chang for emphasizing this point to us.} despite often being confused with each other (see e.g.\ PDG \cite{PDG}). In particular, $\hat\rho_*(0)$ only depends on $T$, where a nonzero value can be directly associated with custodial violation. Veltman $\hat\rho$, on the other hand, depends on all of $S$, $T$, and $U$, and can deviate from 1 due to custodial \emph{symmetric} UV effects (see \cite{Peskin:1991sw}):
\begin{subequations}
\begin{align}
\hat\rho_*(0) -1 &= \alpha T \,, \label{eq:T} \\[8pt]
\hat\rho - 1 &= \frac{\alpha}{\cos2\theta} \left( -\frac12\, S +\cos^2\theta\, T + \frac{\cos2\theta}{4\sin^2\theta}\, U \right) \,. \label{eq:rhoSTU}
\end{align}
\end{subequations}
In determining the strongest experimental constraints on $T$ (and $S$, $U$), the simplicity of \cref{eq:T} may be outweighed by the precision on the Veltman $\hat\rho$ observable and associated $Z$-pole observables that can simultaneously constrain $S, T$, and $U$.

The $S$, $T$, $U$ parameters, however, have limitations. In particular, a key assumption, clearly stated at the time, is that the UV physics contributes only ``obliquely'', i.e., via the 2-point functions of the SM electroweak gauge bosons. Another assumption is that the analysis only accounts for up to $p^2$ terms in gauge boson two-point functions. As precision increases, the framework has been generalized to $p^4$ order, by introducing the new parameters $V$, $W$, $X$, and $Y$ \cite{Maksymyk:1993zm, Burgess:1993mg, Kundu:1996ah, Barbieri:2004qk}, though the oblique assumption remains in place.

Following the discovery of the Higgs boson \cite{Aad:2012tfa, Chatrchyan:2012ufa}, the Standard Model Effective Field Theory (SMEFT) \cite{Weinberg:1979sa, Weinberg:1980bf, Buchmuller:1985jz} has become a new popular framework for model-independent analyses of BSM physics, especially given the null results for the various direct BSM searches at the LHC\@. In this framework, new physics is considered as sufficiently heavy, such that it can be integrated out, resulting in higher dimensional operators, supplementing the SM Lagrangian.\footnote{Classifying the general form of these operators has had a long history \cite{Weinberg:1980bf,Buchmuller:1985jz}. The `Warsaw' basis \cite{Grzadkowski:2010es}, for instance, provides a non-redundant parameterization of the set of all dimension-six (dim-6) operators. Other operator bases, e.g. SILH basis \cite{Giudice:2007fh} can be related through integration-by-parts (IBP) and equations-of-motion (EOM) redundancies \cite{Elias-Miro:2013eta}. A systematic classification and counting of SMEFT operators has been recently achieved using the Hilbert series technique \cite{Jenkins:2009dy, Hanany:2010vu, Lehman:2015via, Henning:2015daa, Henning:2017fpj} up to dim-8 and beyond \cite{Lehman:2014jma, Lehman:2015coa, Henning:2015alf, Marinissen:2020jmb}. The number of operators grows rapidly with the dimension \cite{Henning:2015alf, Marinissen:2020jmb}. At dim-6, SMEFT contains 3045 operators \cite{Alonso:2013hga, Henning:2015alf}, assuming all of the global symmetries of the SM are broken.}

From the SMEFT point of view, only a very restricted set of UV theories (the so-called \emph{universal} theories \cite{Barbieri:2004qk, Wells:2015uba, Wells:2015cre}) contribute only obliquely; fully general UV sectors captured by SMEFT also have non-oblique corrections \cite{Trott:2014dma}. In addition, even for universal theories, oblique corrections do not remain oblique --- non-oblique corrections are generated as soon as (renormalization group) RG effects are included \cite{Trott:2017yhn, Brivio:2017bnu}. Therefore, a generalization of the $T$ parameter that does not rely on restricting to oblique-only corrections would be of significant importance to determine custodial symmetry or its violation of a generic UV theory.

In this paper we present a first step in resolving this issue. Note that once we generalize beyond the oblique assumption, exactly what one means by custodial symmetry becomes more subtle and needs to be revisited. The CC/NC ratio, universal for purely oblique corrections, now depends on what fermions are considered. Given this ambiguity in defining custodial symmetry, we make a choice that resembles the definition from $\hat\rho_*(0)$. Specifically, we define UV physics to be custodial symmetric when an $SU(2)_R$ global symmetry is preserved (in the limit of zero hypercharge coupling) by all UV interactions with the Higgs sector of the SM\@.

In the SMEFT framework, one works with effective operators whose constituents have manifest transformation properties under the global $SU(2)_R$ symmetries in the SM Higgs and/or fermion sectors. In this sense, the Wilson coefficients are superior to the $S$, $T$, $U$ parameters, as they can directly indicate $SU(2)_R$ symmetries or their violation. This is a simplification compared to the oblique framework, where one has to \emph{infer} the $SU(2)_R$ symmetry from the CC/NC ratio. To better utilize this feature of SMEFT, we take linear combinations of operators in dimension-six (dim-6) Warsaw basis---extended to include right-handed neutrinos (thus $\nu$SMEFT, rather than SMEFT), and map them into a new \emph{custodial basis} where all operators have manifest transformation properties under the global $SU(2)_R$ symmetries in the SM. This allows us to directly identify UV custodial symmetry/violation from the Wilson coefficients generated by matching.

Assisted by the custodial ($\nu$)SMEFT basis, we construct a linear combination of Wilson coefficients that we call $\Tom_l$, a new precision parameter that serves as a generalization of the $T$ parameter to indicate custodial symmetry/violation in non-oblique theories. We show that $\Tom_l$ can be constructed from $Z$-pole observables and $m_W$, faithfully determining at tree-level if the UV sector contains ``hard'' custodial violation (that persists even when the hypercharge gauge coupling vanishes), \emph{independent} of whether the UV sector contributes only obliquely. Importantly, as our new $\Tom_l$ parameter involves multiple electroweak observables.  As a consequence, the constraint on custodial violation that it sets is only as strong as the weakest link, namely that the least precise component observable determines the true bound on custodial violation of UV physics.

Our new parameter $\Tom_l$ is a first step only, as it does not capture loop corrections from the SM nor from SMEFT\@. In particular, modifications to the parameter are required to account for the known SM violation of custodial symmetry at one-loop level (arising mainly from top loop contributions). Furthermore, incorporating loop-level effects in SMEFT requires a substantial amount of additional effort due to an ambiguity that arises from equation of motion (EOM) redundancies. When custodial symmetric UV physics is integrated out, it generates custodial symmetric operators, but not necessarily in the Warsaw or custodial basis of ($\nu$)SMEFT\@. Ordinarily one simply utilizes integration-by-parts (IBP) and EOM redundancies to rewrite the UV generated operators in terms of whatever basis one prefers, in our case, our custodial basis of ($\nu$)SMEFT\@. However, the EOM redundancy can trade custodial symmetric operators for custodial violating operators proportional to the SM violation of custodial symmetry. This is simply because the EFT as a whole does not respect custodial symmetry, even if the integrated-out UV physics does. This could have sunk any chance to isolate observables only sensitive to \emph{UV sector} violations of custodial symmetry. Fortunately, from known results about tree/loop operator generation \cite{Arzt:1994gp, Einhorn:2013kja, Craig:2019wmo}, we find that restricting to tree level generated operators, our set of observables, and hence our $\Tom_l$ parameter, remain faithful in identifying hard custodial violation of UV physics. This is the main result of this paper.

The layout of the rest of this paper is as follows: In \cref{sec:CustodialSymmetries} we establish notation and review the global $SU(2)_R$ symmetries of the Higgs and fermion sectors of the SM, as well as how they are broken by various interactions. This will help us lay out our working definition of custodial symmetry. Next, in \cref{sec:CustodialBasis}, we introduce the custodial basis for ($\nu$)SMEFT and classify operators in that basis according to their properties under our custodial symmetry. We also provide mapping between this basis and conventional SMEFT bases which will be useful for quickly importing past results. In \cref{sec:Observables}, we select a set of electroweak precision observables and determine a particular combination of them that is sensitive to hard UV custodial violation at tree level. Our new electroweak precision parameter $\Tom_l$ arises from this combination, and serves as a generalization of the $T$ parameter to UV theories with non-oblique corrections. This section contains our main results. The impact (or lack thereof) of EOM redundancies is the subject of \cref{sec:EOM}. In \cref{sec:UVTheories} we investigate several example UV theories to demonstrate the validity of our new $\Tom_l$ parameter constructed in \cref{sec:Observables}. Finally, we conclude in \cref{sec:Discussion}.

\section{$SU(2)_R$ Symmetries in ($\nu$)SM and Custodial Symmetry}
\label{sec:CustodialSymmetries}

In this section, we discuss the (approximate) global $SU(2)_R$ symmetries in ($\nu$)SM [\cref{subsec:SU2RH,subsec:SU2Rf}]; identify their breaking sources [\cref{subsec:SU2RBreaking}]; and then introduce our definition of the custodial symmetry [\cref{subsec:CustodialDef}].

Let us first establish our notation for the group theory and field content. We use $\tau^a=\tau_R^a=\sigma^a$ with $a=1,2,3$ to denote Pauli matrices. The $SU(2)_L$ and $SU(2)_R$ generators in the fundamental representation are hence $t^a=\frac{1}{2}\tau^a$ and $t_R^a = \frac{1}{2} \tau_R^a$ respectively. The $SU(3)_c$ generators in the fundamental representation are denoted by $T^A$ with $A=1,\cdots,8$. The SM covariant derivative is
\begin{equation}
D_\mu=\partial_\mu -ig_3 G_\mu^A T^A - ig_2 W_\mu^a t^a - ig_1 B_\mu \hyp \,,
\end{equation}
with $\hyp$ denoting the hypercharge, $G_\mu^A, W_\mu^a, B_\mu$ denoting the gauge fields, and $g_3, g_2, g_1$ denoting the gauge couplings.
A general field strength is denoted as $X_{\mu\nu} \in \left\{G_{\mu\nu}^A, W_{\mu\nu}^a, B_{\mu\nu}\right\}$. For the dual, we adopt the convention $\tX_{\mu\nu} \equiv \frac{1}{2} \epsilon_{\mu\nu\alpha\beta} X^{\alpha\beta}$, with $\epsilon_{0123}=+1$. We use the usual Dirac matrices $\gamma^\mu$, and $\sigma^{\mu\nu} \equiv \frac{i}{2} \comm{\gamma^\mu}{\gamma^\nu}$.

Our notation for the SM Lagrangian is
\begin{align}
\Lag_\text{SM} &= - \frac14 G_{\mu\nu}^A G^{A\mu\nu}  - \frac14 W_{\mu\nu}^a W^{a\mu\nu} - \frac14 B_{\mu\nu} B^{\mu\nu}
+ \left|DH\right|^2 - \lambda \left( \left|H\right|^2 - \frac12 v^2 \right)^2 \notag\\[5pt]
&\quad + \sum_\psi{\bar\psi i\slashed{D} \psi}\, - \left( Y_u\, {\bar q}{\tilde H}u + Y_d\, {\bar q}Hd + Y_e\, {\bar l}He + \hc \right) \,,
\label{eqn:LagSM}
\end{align}
where for the SM fermions $\psi$, we follow Ref.~\cite{Grzadkowski:2010es} to use $\{q, l\}$ for left-handed $SU(2)_L$-doublets, and $\{u, d, e\}$ for right-handed $SU(2)_L$-singlets. In the above, the Yukawa couplings $Y_u, Y_d, Y_e$ are $3\times3$ matrices in the generation space, but we have suppressed the generation indices for compactness. We can also extend the SM to include right-handed neutrinos $\nu$---what we refer to as $\nu$SM\@. In this case, the Lagrangian is augmented as
\begin{equation}
\Lag_{\nu\text{SM}} = \Lag_\text{SM} + \bar\nu i\slashed{D} \nu\, - \left( Y_\nu\, {\bar l}{\tilde H}\nu + \hc \right) \,.
\label{eqn:LagnuSM}
\end{equation}

\subsection{Higgs sector: $SU(2)_{RH}$}
\label{subsec:SU2RH}

We begin our discussion of global $SU(2)_R$ symmetries with the Higgs doublet
\begin{equation}
H \;=\; \mqty( G^+ \\[3pt] \left( v + h + i G^0 \right)/\sqrt{2} ) \,.
\end{equation}
The Higgs potential is invariant under an $SO(4)$ symmetry
\begin{equation}
SO(4) \sim SU(2)_L \times SU(2)_{RH} \,,
\end{equation}
where the $SU(2)_L$ and the $t^3_R$ generator of $SU(2)_{RH}$ are gauged in ($\nu$)SM\@. This symmetry is spontaneously broken to $SO(3) \sim SU(2)_{V}$ when the Higgs develops a vacuum expectation value (vev).

We can re-express the Higgs field in terms of a $(\mathbf{2},\mathbf{2})$ bifundamental scalar field that transforms under the $(U_L, U_R) \in SU(2)_L \times SU(2)_{RH}$ as\footnote{Here $\tH\equiv i\sigma^2H^*=\epsilon H^*$, which transforms in the same way as $H$ itself under the $SU(2)_L$ symmetry; $\epsilon_{ij}=-\epsilon_{ji}$ is an $SU(2)$ invariant tensor, and we take $\epsilon_{12}=+1$.}
\begin{equation}
\Sigma \equiv \left( {\begin{array}{*{20}{c}}
{\tilde H}&H
                      \end{array}} \right) =
\left( \begin{array}{cc}
   (v + h - i G^0)/\sqrt{2} & G^+ \\
   - G^-                    & (v + h + i G^0)/\sqrt{2}
   \end{array} \right)  \quad\longrightarrow\quad  U_L\, \Sigma\, U_R^\dagger \,\,.
\label{eqn:sig-def}
\end{equation}
In principle, all interactions are built out of $\Sigma$, but it is sometimes helpful to make use of the identity
\begin{equation}
\Sigma^\dagger_{i_R i_L} \, \equiv \, \epsilon_{i_R j_R} \epsilon_{i_L j_L} \Sigma_{j_L j_R} \,,
\end{equation}
and write operators with $\Sigma^\dagger$, where the $SU(2)$ transformation properties are easier to recognize. For example, the SM Higgs potential can be written as
\begin{equation}
 V = \lambda\left( \left|H\right|^2 - \frac{v^2}{2}\right)^2 = \frac{\lambda}{4} \left[ \tr\left(\Sigma^\dagger \Sigma\right) - v^2 \right]^2 \,,
\end{equation}
where the $SU(2)_L\times SU(2)_{RH}$ symmetry is manifest. Similarly, one can rewrite the Higgs $SU(2)_L$ and $U(1)_Y$ currents into
\begin{subequations}
\begin{align}
H^\dagger i\lrD_\mu^a H &\;\equiv\;  H^\dagger \tau^a iD_\mu H + \hc \;=\; \tr \left( \Sigma^\dagger \tau^a iD_\mu \Sigma \right) \,, \\[5pt]
H^\dagger i\lrD_\mu H   &\;\equiv\;  H^\dagger iD_\mu H + \hc        \;=\; - \tr \left( \Sigma^\dagger iD_\mu \Sigma \tau_R^3 \right) \,,
\end{align}
\end{subequations}
where the $SU(2)$ preserving/violating structures are more explicit.

\subsection{Fermion sector: $SU(2)_{R\qR}$, $SU(2)_{R\lR}$}
\label{subsec:SU2Rf}

Turning to the fermion sector of the SM, there are several approximate $SU(2)_R$ symmetries that become exact in the limit of neglecting hypercharge coupling $g_1$ \emph{and} the Yukawa couplings. Focusing on one generation of fermions for the moment, the right-handed up-type quark $u$ and down-type quark $d$ can be grouped together to form a doublet
\begin{equation}
\qR \equiv \mqty(u \\ d) \,,
\end{equation}
which has an $U(2)_{\qR}$ global symmetry. We can break this symmetry up to the baryon number $U(1)_B$ and a global $SU(2)_R$ quark isospin symmetry that we will call $SU(2)_{R\qR}$:
\begin{equation}
U(2)_{\qR} = U(1)_B \times SU(2)_{R\qR} \,.
\end{equation}
Similarly, when the SM is extended to $\nu$SM, we can build a right-handed lepton doublet
\begin{equation}
\lR \equiv \mqty(\nu \\ e) \,,
\end{equation}
which has a global $U(2)_{\lR}$ symmetry that we identify as consisting of lepton number and isospin:
\begin{equation}
U(2)_{\lR} = U(1)_L \times SU(2)_{R\lR} \,.
\end{equation}
In case of three generations, we will get the quark isospin $SU(2)_{R\qR}$ and the lepton isospin $SU(2)_{R\lR}$ for each generation.

\subsection{$SU(2)_R$ violation in ($\nu$)SM}
\label{subsec:SU2RBreaking}

With the global $SU(2)_R$ symmetries in ($\nu$)SM identified, we can now classify the symmetry breaking sources. For simplicity, we will focus on the one generation case:\footnote{Similar arguments hold for the case of three generations.}
\begin{itemize}
  \item Yukawa couplings play two roles: (1) they tie the Higgs $SU(2)_{RH}$ symmetry to the isospin symmetries $SU(2)_{R\qR}, SU(2)_{R\lR}$; and (2) they break these symmetries. To disentangle these two effects, we can first write the Yukawa interactions in terms of the bi-doublet Higgs $\Sigma$, e.g. for quarks:
      \begin{equation}
      Y_u\, \bar{q} {\tilde H} u + Y_d\, \bar{q} H d  = \bar{q}\, \Sigma\, \mqty( Y_u & 0 \\ 0 & Y_d )\, \qR \,,
      \label{eqn:YukawaCombine}
      \end{equation}
      and then split the above Yukawa matrix as
      \begin{equation}
      \mqty( Y_u & 0 \\ 0 & Y_d ) = \frac{Y_u + Y_d}{2}\, \mathbb{1}_{2\times 2} + \frac{Y_u - Y_d}{2}\, \tau^3_R \,\,.
      \label{eqn:YukawaPattern}
      \end{equation}
      This way, the symmetry breaking pattern becomes clear. The term proportional to $\mathbb{1}_{2\times 2}$ leads to $SU(2)_{RH} \times SU(2)_{R\qR} \rightarrow SU(2)$, while the $\tau^3_R$ term breaks $SU(2)_{RH} \times SU(2)_{R\qR} \rightarrow U(1)$. By the same logic, the Yukawa interactions in the lepton sector of $\nu$SM can be grouped into a combination that ties $SU(2)_{RH}$ to $SU(2)_{R\lR}$ and a combination that breaks $SU(2)_{RH} \times SU(2)_{R\lR} \to U(1)$. The matrices $\mathbb{1}_{2\times 2}$ and $\tau^3_R$ will appear often in this work, so we will adopt the convenient shorthand
      \begin{equation}
      P_+ \equiv\mathbb{1}_{2\times2}\,,\quad  P_- \equiv\tau_R^3 \,\,.
      \end{equation}
      We will use $P_{\pm}$ to apply to both  $SU(2)_{R\qR}$ and $SU(2)_{R\lR}$ spaces---exactly which space we are working with should be clear from the context.
  \item Gauging hypercharge corresponds to gauging the $\tau_R^3$ generators of $SU(2)_{RH}$, $SU(2)_{R\qR}$, and $SU(2)_{R\lR}$. This breaks all of them simultaneously down to $U(1)_Y$---exactly the $U(1)$ left intact by $P_-$ from Yukawa breaking.
\end{itemize}

\subsection{Custodial $SU(2)_R$}
\label{subsec:CustodialDef}

Now that we have identified the $SU(2)_R$ symmetries and violation in ($\nu$)SM, we are ready to precisely define custodial symmetry in this paper:
\begin{center}
  \fbox{\parbox{0.92\textwidth}{\centering
  UV physics is \textbf{custodial symmetric} when there is a global $SU(2)_R$ symmetry preserved, in the limit $g_1 \to 0$, by all UV interactions with the Higgs sector of the SM\@.}}
\end{center}
Here, the preserved $SU(2)_R$ symmetry could be either $SU(2)_{RH}$ itself, or a diagonal subgroup of $SU(2)_{RH} \times SU(2)_{R\qR}$, $SU(2)_{RH} \times SU(2)_{R\lR}$, or $SU(2)_{RH} \times SU(2)_{R\qR} \times SU(2)_{R\lR}$. That is, the $SU(2)_R$ group must involve $SU(2)_{RH}$ in some way.\\\\
A few important comments are in order about this definition:
\begin{itemize}
\item Our definition is exclusively about the UV sector. Therefore, even in the case the UV sector respects custodial symmetry, the identified $SU(2)_R$ symmetry is still not an exact symmetry of the whole Lagrangian (the UV sector plus ($\nu$)SM\@). In particular, the hypercharge coupling $g_1$ and the mismatch in Yukawa couplings [$Y_u-Y_d$ and $Y_\nu-Y_e$; see e.g. \cref{eqn:YukawaPattern}] in $\nu$SM break it. Only in the limits $g_1\to0$, $Y_u-Y_d\to0$, $Y_\nu-Y_e\to0$, will the custodial $SU(2)_R$ become an exact symmetry of the entire UV + SM theory.
\item Also, because our definition is exclusive to the UV interactions, whether or not a UV sector is adjudicated to be  custodial symmetric does not depend on the presence or absence of ($\nu$)SM Yukawa couplings. By contrast, the hypercharge coupling $g_1$ could play a role, as it could participate in the UV interactions when some UV particles have nonzero hypercharge.
\item The breaking of custodial $SU(2)_R$ by the UV interactions can thus be categorized as:
    \begin{enumerate}
    \item ``Soft'' breakings that vanish in the limit $g_1\to 0$.
    \item ``Hard'' breakings that persist in the limit $g_1\to 0$.
    \end{enumerate}
    In our definition above, a UV sector with ``soft'' custodial $SU(2)_R$ breaking is defined as custodial symmetric. This is because our interest in this paper is ``hard'' custodial violation. In the rest of this paper, we will utilize this terminology strictly unless explicitly stated otherwise, namely that our ``custodial violating UV physics'' contains ``hard'' custodial breakings, and our ``custodial symmetric UV physics'' allows for soft breakings.
\end{itemize}

In the above, we have established a definition of the custodial symmetry for \emph{UV physics}. However, as explained in the introduction, we are not interested in analyzing UV theories case-by-case. Instead, we would like to follow the spirit of the electroweak precision parameters $S, T, U$, and use a general framework to analyze UV physics independent of the UV model. In this paper, the framework we use is dim-6 ($\nu$)SMEFT\@. This motivates us to divide the dim-6 ($\nu$)SMEFT operators into two categories: ``custodial preserving/violating \emph{operators}'' that can/cannot be possibly generated by custodial symmetric \emph{UV physics}.

Usually, a symmetry possessed by the UV theory gets inherited by the EFT (as long as the heavy states integrated out do not break it). In case of our custodial $SU(2)_R$, however, the situation is less straightforward, precisely because it is not an exact symmetry of the whole UV Lagrangian. Nevertheless, if we restrict ourselves to the leading matching order (which could be tree level, one-loop level, or even higher, depending on the UV theory), there are only heavy particle propagators in the contributing diagrams, and hence only UV interactions beyond ($\nu$)SM will participate. In this case, all the resulting ($\nu$)SMEFT operators will preserve the custodial $SU(2)_R$ symmetry. For the rest of the paper, we only consider the leading matching order unless explicitly stated otherwise. This allows us to make the above desired ($\nu$)SMEFT operator division simply based on their $SU(2)_R$ transformation properties, a task we will tackle in the next section.

\section{Custodial Basis of ($\nu$)SMEFT}
\label{sec:CustodialBasis}

In this section, we introduce a new basis for dim-6 ($\nu$)SMEFT---the \emph{custodial basis}, to facilitate the identification of operators that preserve/violate the custodial symmetry. Using this basis, we then identify the operators that can/cannot be possibly generated by integrating out custodial symmetric UV sectors.

Our presentation of the operator basis largely follows~\cite{Grzadkowski:2010es, Jenkins:2013zja, Jenkins:2013wua, Alonso:2013hga}, extended to include right-handed neutrinos~\cite{Liao:2016qyd}. As preparation, we first present in \cref{tbl:nuSMEFTWarsaw} all of the independent baryon-preserving operators in the Warsaw basis for $\nu$SMEFT (suppressing flavor indices).\footnote{For easy reading/contrasting, we have gathered all of the tables of operator bases and the relevant translation dictionaries in \cref{appsec:Tables}.} In addition to the $76=42+(17+\hc)$ SMEFT operators that we show in \Blcolor~color, there are $25=7+(9+\hc)$ new operators involving right-handed neutrinos $\nu$ that we show in \nucolor~color to allow for an easy recognition. Reducing $\nu$SMEFT back to SMEFT is straightforward by restricting appropriate Wilson coefficients to zero, which we show in \cref{tbl:SMEFTfromnuSMEFT}.

Now we build the custodial basis. Our basic approach is to recombine the Warsaw basis operators $Q_i$ such that their transformation properties under the global $SU(2)_{RH}$ and isospin symmetries $SU(2)_{R\qR}$, $SU(2)_{R\lR}$ become manifest, close to what we did for $\nu$SM in  \cref{eqn:YukawaCombine,eqn:YukawaPattern}. It is also worth mentioning that, similar rewriting of dim-6 SMEFT in other basis exists, see e.g.~\cite{Elias-Miro:2013mua}. Performing this recombination for all of the operators in \cref{tbl:nuSMEFTWarsaw}, we arrive at our custodial basis operators $\Op_i$ summarized in \cref{tbl:nuSMEFTCBasis}. An explicit translation dictionary between the two operator bases is further given in \cref{tbl:BasisTranslation}. Many operators do not change in going from the Warsaw basis to the custodial basis. In particular, operators built purely out of $SU(2)_R$-singlets translate trivially. All operators that involve exclusively the left-handed fermion fields of the SM fall into this category. On the other hand, significant differences from the Warsaw basis can be observed in the operators involving the right-handed fermion fields.

From the translation dictionary in \cref{tbl:BasisTranslation}, we can also easily determine the corresponding relations between the Wilson coefficients in these two bases:
\begin{equation}
\Lag_\text{EFT} - \Lag_\text{SM} = \sum\limits_i {\WlC_i \Op_i} = \sum\limits_i {C_i Q_i} \,\,.
\end{equation}
We provide explicit translation dictionaries between the Wilson coefficients in \cref{tbl:aFromC,tbl:CFroma}. Note that we have absorbed the scale suppressing the $Q_i$ and $O_i$ into the Wilson coefficients, making them dimensionful $[C_i ] = [a_i] = -2$. This is a bit unconventional, but it compactifies the notation. One can express our results in terms of dimensionless Wilson coefficients and a new physics scale $\Lambda$ by replacing $C_i \to \tilde C_i/\Lambda^2, a_i \to \tilde a_i/\Lambda^2$ everywhere.

We now wish to identify operators in \cref{tbl:nuSMEFTCBasis} that can/cannot be possibly generated by custodial symmetric UV physics. Recall that in the limit $g_1\to0$, custodial symmetric UV physics preserves an $SU(2)_R$ symmetry. Consequently, in this limit, only operators that preserve the same $SU(2)_R$ symmetry could be generated by matching (at the leading order). However, there are four possibilities for this $SU(2)_R$:
\begin{enumerate}
\item $SU(2)_{RH}$.
\item The diagonal subgroup of $SU(2)_{RH}\times SU(2)_{R\qR}$.
\item The diagonal subgroup of $SU(2)_{RH}\times SU(2)_{R\lR}$.
\item The diagonal subgroup of $SU(2)_{RH}\times SU(2)_{R\qR}\times SU(2)_{R\lR}$.
\end{enumerate}
Therefore, if an operator in \cref{tbl:nuSMEFTCBasis} preserves any of the four $SU(2)_R$'s above in the limit $g_1\to0$, then it \emph{can} potentially be generated, and should be categorized as a ``custodial preserving operator''. For example, in the limit $g_1 \to0$, the operator
\begin{equation}
\Op_{H\lR}^{(3)+} \equiv \tr\left(\Sigma^\dagger iD_\mu\Sigma\tau_R^a\right)\left(\lRbar \gamma^\mu \tau_R^a P_+ \lR\right) \,,
\end{equation}
preserves the diagonal subgroup of $SU(2)_{RH}\times SU(2)_{R\lR}$, and hence is a custodial preserving operator. Note in particular that any operator with an explicit $B_{\mu\nu}$ should be understood as accompanied by a power of $g_1$. Therefore, in the limit $g_1 \to0$, these operators vanish, and so they are classified as custodial preserving operators as well.

In \cref{tbl:nuSMEFTCBasis}, we use colors to distinguish custodial preserving and violating operators. Our convention is:
\begin{itemize}
\item We use \Hvcolor~color to denote custodial violating operators. They do not preserve any of the four $SU(2)_R$'s listed above, even in the limit $g_1 \to0$. Generating these operators at the leading matching order is a sign for hard custodial violation in the UV physics.
\item For custodial preserving operators, we largely use \Blcolor~color. However, some custodial preserving four-quark operators preserves $SU(2)_{RH}$ trivially (due to not involving the $H$ field), but violate the isospin $SU(2)_{R\qR}$ or $SU(2)_{R\lR}$ or both. In this case, we use \qvcolor~or \lvcolor~color or both to denote them to highlight the isospin violation.
\end{itemize}

\subsection{Flavor indices of the Wilson coefficients}
\label{subsec:a12}

In \cref{tbl:nuSMEFTWarsaw}-\cref{tbl:SMEFTfromnuSMEFT}, we have suppressed all the flavor indices, but it should be understood that each fermion field actually comes with a generation index, so are the corresponding Wilson coefficients. For example, the two-fermion operator $Q_{Hl}^{(3)}$ and four-fermion operator $Q_{ll}$ should actually read
\begin{subequations}
\begin{align}
Q_{\substack{Hl\\ pr}}^{(3)} &= \left(H^\dagger i\lrD_\mu^a H\right) \left({\bar l}_p \gamma^\mu \tau^a l_r\right) \,, \\[3pt]
Q_{\substack{ll\\ prst}} &= \left({\bar l}_p \gamma_\mu l_r\right) \left({\bar l}_s \gamma^\mu l_t\right) \,.
\end{align}
\end{subequations}
The EFT Lagrangian therefore has a sum over these generation indices:
\begin{equation}
\Lag_\text{EFT} \supset \sum\limits_{p, r=1}^3 C_{\substack{Hl\\ pr}}^{(3)} Q_{\substack{Hl\\ pr}}^{(3)} + \sum\limits_{p, r, s, t=1}^3 C_{\substack{ll\\ prst}} Q_{\substack{ll\\ prst}} = \sum\limits_{p, r=1}^3 \WlC_{\substack{Hl\\ pr}}^{(3)} \Op_{\substack{Hl\\ pr}}^{(3)} + \sum\limits_{p, r, s, t=1}^3 \WlC_{\substack{ll\\ prst}} \Op_{\substack{ll\\ prst}} \,.
\end{equation}
However, we often suppress the flavor indices when it is clear from the context.

As we will see, most four-fermion operators do not contribute to the observables to be discussed in \cref{sec:Observables}. However, one exception is the mixed first and second generation four-lepton operator, which contributes to $\hat G_F$. We give this operator a special name for future convenience:
\begin{equation}
Q_{12} \equiv \left({\bar l}_1 \gamma_\mu l_2\right) \left({\bar l}_2 \gamma^\mu l_1\right) \equiv \Op_{12} \,.
\label{eqn:Q12Op12Def}
\end{equation}
Clearly, the corresponding Wilson coefficients are related to our general notation as
\begin{equation}
C_{12} = C_{\substack{ll \\ 1221}} + C_{\substack{ll \\ 2112}} = \WlC_{\substack{ll \\ 1221}} + \WlC_{\substack{ll \\ 2112}} = \WlC_{12} \,.
\label{eqn:C12WlC12}
\end{equation}

\section{Observables Sensitive to Custodial Symmetry/Violation in ($\nu$)SMEFT}
\label{sec:Observables}

In this section, we study an \emph{example set} of precision observables that will allow us to identify whether the UV physics contain hard custodial violation:
\begin{align}
\left\{ \hat\alpha,\, \hat{G}_F,\, \hat{m}_Z^2,\, \hat{m}_W^2,\, \hat\Gamma_{Z\nu_L\bar\nu_L},\, \hat\Gamma_{Ze_L\bar{e}_L},\, \hat\Gamma_{Ze\bar{e}}  \right\} \,.
\label{eqn:ExampleObservables}
\end{align}
In order, these are the (electromagnetic) fine structure constant, the Fermi constant, the pole masses of $Z$ and $W$ bosons, the partial decay widths of the $Z$ boson to left-handed neutrinos, left-handed electrons and right-handed electrons.

In what follows, we compute corrections from dim-6 ($\nu$)SMEFT operators to these observables in \cref{subsec:ObvSM,subsec:ObvSMEFT}, at leading matching order (tree level in SMEFT), and then construct in \cref{subsec:Toml} a $T$ parameter generalization, $\Tom_l$, from them that serves as an indicator of custodial violation in general ($\nu$)SMEFT\@. Reparameterization invariance (RPI) plays an important role in our construction of $\Tom_l$, which we will explain in \cref{subsec:RPI}. Our ``observables'' here refers to quantities that we can calculate in the ($\nu$)SM/($\nu$)SMEFT that do not depend on the choice of operator basis, and they can in principle be measured by experiments. Some of these observables can be directly measured, such as $\hat{\alpha}$, ${\hat G}_F$, $\hat{m}_Z^2$, $\hat{m}_W^2$, while others need to be inferred from other measurements. In \cref{subsec:Measurements} we discuss how to extract these observables from experimental measurements. The observable set chosen in \cref{eqn:ExampleObservables} is only a demonstration example. There are more observables available in the canonical LEP choice, such as the hadronic branching ratio, the bottom quark branching ratio, or the total decay width of $Z$. We discuss these observables in \cref{appsec:Hadronic}.

\subsection{Observables in the SM}
\label{subsec:ObvSM}

In the SM, the observables in \cref{eqn:ExampleObservables} are given by the three Lagrangian parameters $g_1, g_2, v$ as\footnote{Throughout this paper, we neglect the lepton masses in $Z$ decay widths.}
\begin{subequations}\label{eqn:Obs7SM}
\begin{align}
{{\hat \alpha }_{{\,\text{SM}}}} &= \frac{{g_1^2g_2^2}}{{4\pi \left( {g_1^2 + g_2^2} \right)}} \,, \\[5pt]
{{\hat G}_{F{\text{, SM}}}} &= \frac{1}{{\sqrt 2 {v^2}}} \,, \\[5pt]
\hat m_{Z{\text{, SM}}}^2 &= \frac{1}{4}\left( {g_1^2 + g_2^2} \right){v^2} \,, \\[5pt]
\hat m_{W{\text{, SM}}}^2 &= \frac{1}{4}g_2^2{v^2} \,, \\[5pt]
{{\hat \Gamma }_{Z{\nu _L}{{\bar \nu }_L}{\text{, SM}}}} &= \frac{{{{\hat m}_{Z{\text{, SM}}}}}}{{96\pi }}\frac{{g_2^2}}{{c_\theta ^2}} \,, \\[5pt]
{{\hat \Gamma }_{Z{e_L}{{\bar e}_L}{\text{, SM}}}} &= \frac{{{{\hat m}_{Z{\text{, SM}}}}}}{{96\pi }}\frac{{g_2^2}}{{c_\theta ^2}} c_{2\theta}^2 \,, \\[5pt]
{{\hat \Gamma }_{Ze\bar e{\text{, SM}}}} &= \frac{{{{\hat m}_{Z{\text{, SM}}}}}}{{24\pi }}\frac{{g_2^2}}{{c_\theta ^2}}s_\theta ^4 \,,
\end{align}
\end{subequations}
where $\theta$ denotes the Weinberg angle
\begin{equation}
c_\theta = \cos\theta \equiv \frac{{{g_2}}}{{\sqrt {g_1^2 + g_2^2} }} \,,\quad
s_\theta = \sin\theta \equiv \frac{{{g_1}}}{{\sqrt {g_1^2 + g_2^2} }} \,.
\end{equation}

\subsection{Observables in SMEFT}
\label{subsec:ObvSMEFT}

Since the SM has only three inputs, the full set in \cref{eqn:Obs7SM} can be completely determined in terms of any subset of three observables. Typically, the most precisely measured subset is chosen, $\{\hat{\alpha}, \hat m^2_Z, \hat G_F\}$ or $\{\hat m^2_W, \hat m^2_Z, \hat G_F\}$. Once we include the contributions from SMEFT operators, three observables are no longer enough, as all of \cref{eqn:Obs7SM} will be polluted with different combinations of Wilson coefficients $C_i$. Said another way, it is still possible to swap out $g_1, g_2$ and $v$ for $\{\hat{\alpha}, \hat m^2_Z, \hat G_F\}$ or $\{\hat m^2_W, \hat m^2_Z, \hat G_F\}$, however, in the presence of SMEFT effects, $g_1, g_2$ and $v$ will be functions of $C_i$ rather than numbers fixed by experiment. This $C_i$ dependence is referred to as the ``electroweak input shifts'' in the literature. The exact form of the shifts depend on which three observables are used to solve for $g_1, g_2$ and $v$, either the $\hat{\alpha}$ scheme ($\{\hat{\alpha}, \hat m^2_Z, \hat G_F\}$) or the $\hat m_W$ scheme ($\{\hat m^2_W, \hat m^2_Z, \hat G_F\})$\footnote{Discussions of the strengths and weaknesses of the two schemes can be found in \cite{Brivio:2017bnu}.}. In this paper, we will exclusively use the $\hat \alpha$ scheme. Of course, the input shifts are only one place Wilson coefficients can enter, as every observable will also carry process-specific factors of $C_i$ depending on the fields and vertices involved.

Removing $\{\hat{\alpha}, \hat m^2_Z, \hat G_F\}$, we are left with four observables: $\left\{ \hat{m}_W^2,\, \hat\Gamma_{Z\nu_L\bar\nu_L},\, \hat\Gamma_{Ze_L\bar{e}_L},\, \hat\Gamma_{Ze\bar{e}} \right\}$. To make it easier to spot and quantify the effects from SMEFT, we swap out $\hat m^2_W$ for the Veltman $\hat\rho$, and divide all partial widths by their SM values.
\begin{subequations}\label{eqn:Obs4Def}
\begin{align}
\hat \rho &\equiv \frac{{\hat m_W^2}}{{\hat m_Z^2}}\frac{2}{{\hat x}} \left( {1 - \sqrt {1 - \hat x} } \right) \,, \\[8pt]
{{\hat r}_{Z{\nu _L}{{\bar \nu }_L}}} &\equiv \frac{{24\pi }}{{\sqrt 2 {{\hat G}_F}\hat m_Z^3}}\, {\hat \Gamma }_{Z{\nu _L}{{\bar \nu }_L}} \,, \\[8pt]
{{\hat r}_{Z{e_L}{{\bar e}_L}}} &\equiv \frac{{24\pi }}{{\sqrt 2 {{\hat G}_F}\hat m_Z^3\left(1 - \hat x\right)}}\, {\hat\Gamma}_{Z{e_L}{{\bar e}_L}} \,, \\[8pt]
{{\hat r}_{Ze\bar e}} &\equiv \frac{{24\pi }}{{\sqrt 2 {{\hat G}_F}\hat m_Z^3{{\left( {1 - \sqrt {1 - \hat x} } \right)}^2}}}\, {\hat \Gamma }_{Ze\bar e} \,,
\end{align}
\end{subequations}
where we have introduced the convenient combination
\begin{equation}\label{eqn:x}
\hat x \equiv \frac{2\sqrt 2 \pi \hat\alpha}{{\hat G}_F \hat m_Z^2} \,,\quad\text{with}\quad \hat x_\text{SM}^{} = s_{2\theta}^2 \,.
\end{equation}
The four observables in \cref{eqn:Obs4Def} are unity in SM, but are modified in SMEFT\@. Because we are only interested in the corrections from SMEFT at dim-6 level, we only need to keep up to the linear terms in the Wilson coefficients $C_i$ (see \cref{tbl:nuSMEFTWarsaw} for definitions of Warsaw basis operators). Assuming universality among fermion generations, we obtain
\begin{subequations}\label{eqn:Obs4nuSMEFTWarsaw}
\begin{align}
\hat \rho &= 1 + \frac{v^2}{c_{2\theta}} \Bigg[ - 2 s_\theta^2 \left( \frac{c_\theta}{s_\theta}\, C_{HWB} + C_{Hl}^{(3)} \right) + \12 s_\theta^2\, C_{12} - \12 c_\theta^2\, C_{HD} \Bigg] \,, \label{eq:rhohat-warsaw} \\[8pt]
{{\hat r}_{Z{\nu _L}{{\bar \nu }_L}}} &= 1 + v^2 \Bigg[ \12\, C_{12} - \12\, C_{HD} - 2\, C_{Hl}^{(1)} \Bigg] \,, \\[8pt]
{{\hat r}_{Z{e_L}{{\bar e}_L}}} &= 1 + \frac{v^2}{c_{2\theta}^2} \Bigg[ - 4s_\theta^2 \left( \frac{c_\theta}{s_\theta}\, C_{HWB} + C_{Hl}^{(3)} \right) + \12\, C_{12} - \12\, {C_{HD}} + 2c_{2\theta}\, C_{Hl}^{(1)} \Bigg] \,, \\[8pt]
{{\hat r}_{Ze\bar e}} &= 1 + \frac{v^2}{c_{2\theta}} \Bigg[ 2\left( \frac{c_\theta}{s_\theta}\, C_{HWB} + C_{Hl}^{(3)} \right) - \12\, C_{12} + \12\, {C_{HD}} - \frac{c_{2\theta}}{s_\theta^2}\, C_{He} \Bigg] \,.
\end{align}
\end{subequations}
More details of deriving these results are explained in \cref{appsec:Mapping}. We have checked that these results agree with Ref.~\cite{Brivio:2017vri}, (see also ~\cite{Elias-Miro:2013mua, Baglio:2018bkm}). Note that our expression for the Veltman $\hat{\rho}$ in \cref{eq:rhohat-warsaw} reduces to the Peskin-Takeuchi expression in \cref{eq:rhoSTU} in the special case of oblique corrections only where $C_{Hl}^{(3)} = C_{12} = 0$, upon identifying\footnote{Note that $S, T, U$ are already linear order in the Wilson coefficients, so the further difference between $\hat\alpha_\text{SM}$ and $\hat\alpha$ in their accompanying coefficients is beyond our SMEFT order. For this reason, we simply write the multiplying factor as $\alpha$. The same holds for our generalization $\Tom$ to be presented below.}
\begin{subequations}\label{eqn:STUnaive}
\begin{align}
\alpha S &= 2 v^2 s_{2\theta}\, C_{HWB} \,,\\[5pt]
\alpha T &= -\frac12 v^2\, C_{HD} \,, \label{eqn:PTTWarsaw} \\[5pt]
\alpha U &= 0 \,.
\end{align}
\end{subequations}
On the other hand, \cref{eqn:Obs4nuSMEFTWarsaw} holds for general SMEFT in the Warsaw basis. In addition, these results apply to SMEFT and $\nu$SMEFT alike, since we did not consider observables involving right-handed neutrinos.\footnote{In principle, one could also include the partial width $\hat\Gamma_{Z\nu_R\nu_R}$ in $\nu$SMEFT\@. However, we cannot construct a convenient ratio $\hat{r}_{Z\nu_R\nu_R}$, because $\hat\Gamma_{Z\nu_R\nu_R}$ vanishes in $\nu$SM\@. Furthermore, the existence of this partial width also relies on assuming the mass of the right-handed neutrino is below the electroweak scale.}

\subsection{Constructing $\Tom_l$ for ($\nu$)SMEFT to replace Peskin-Takeuchi $T$}
\label{subsec:Toml}

In order to work out a replacement of the $T$ parameter in the ($\nu$)SMEFT framework that serves to identify hard UV custodial symmetry violation, we rewrite the results in \cref{eqn:Obs4nuSMEFTWarsaw} into our custodial basis operators given in \cref{tbl:nuSMEFTCBasis}. This is straightforward, utilizing the translation relations provided in \cref{tbl:CFroma}:
\begin{subequations}\label{eqn:Obs4nuSMEFTCBasis}
\begin{align}
\hat \rho &= 1 + \frac{v^2}{c_{2\theta}} \Bigg[ 2s_\theta^2 \left( \frac{2c_\theta}{s_\theta}\, \WlC_{HWB} - \WlC_{Hl}^{(3)} \right) + \12 s_\theta ^2 \, \WlC_{12} - 2c_\theta^2\, \WlC_{HD} \Bigg] \,, \\[8pt]
{{\hat r}_{Z{\nu _L}{{\bar \nu }_L}}} &= 1 + v^2 \bigg[ \12\, \WlC_{12} - 2\, \WlC_{HD} + 2\, \WlC_{Hl}^{(1)} \bigg] \,, \\[8pt]
{{\hat r}_{Z{e_L}{{\bar e}_L}}} &= 1 + \frac{v^2}{c_{2\theta}^2} \Bigg[ 4s_\theta^2 \left( \frac{2c_\theta}{s_\theta}\, \WlC_{HWB} - \WlC_{Hl}^{(3)} \right) + \12\, \WlC_{12} - 2\, \WlC_{HD} - 2c_{2\theta} \, \WlC_{Hl}^{(1)} \Bigg] \,, \\[8pt]
{{\hat r}_{Ze\bar e}} &= 1 + \frac{v^2}{c_{2\theta}} \Bigg[ - 2 \left( \frac{2c_\theta}{s_\theta}\, \WlC_{HWB} - \WlC_{Hl}^{(3)} \right) - \12\, \WlC_{12} + 2\, \WlC_{HD} \notag\\
&\hspace{120pt} + \frac{c_{2\theta}}{s_\theta^2} \left( {\WlC_{H\lR}^{(1)+} - \WlC_{H{l_R}}^{(1)-} - \WlC_{H{l_R}}^{(3)+} + \WlC_{H{l_R}}^{(3)-}} \right) \Bigg] \,.
\end{align}
\end{subequations}
In the absence of custodial violation, these observables become
\begin{subequations}\label{eqn:Obs4MFCV}
\begin{align}
\hat \rho - 1 &\quad\longrightarrow\quad \frac{v^2}{c_{2\theta}} \Bigg[ 2s_\theta^2 \left( \frac{2c_\theta}{s_\theta}\, \WlC_{HWB} - \WlC_{Hl}^{(3)} \right) + \12s_\theta^2 \, \WlC_{12} \Bigg] \,, \\[8pt]
{\hat r}_{Z{\nu _L}{{\bar \nu }_L}} - 1 &\quad\longrightarrow\quad v^2 \left( \12\, \WlC_{12} \right) \,, \\[8pt]
{\hat r}_{Z{e_L}{{\bar e}_L}} - 1 &\quad\longrightarrow\quad \frac{v^2}{c_{2\theta}^2} \Bigg[ 4s_\theta^2 \left( \frac{2c_\theta}{s_\theta}\, \WlC_{HWB} - \WlC_{Hl}^{(3)} \right) + \12 \, \WlC_{12} \Bigg] \,, \\[8pt]
{\hat r}_{Ze\bar e} - 1 &\quad\longrightarrow\quad \frac{v^2}{c_{2\theta}} \Bigg[ - 2 \left( \frac{2c_\theta}{s_\theta}\, \WlC_{HWB} - \WlC_{Hl}^{(3)} \right) - \12 \, \WlC_{12} - \frac{c_{2\theta}}{s_\theta^2} \, \WlC_{H\lR}^{(3)+} \Bigg] \,.
\end{align}
\end{subequations}
While none of Eq.~\eqref{eqn:Obs4MFCV} vanish in the custodially symmetric limit, the first three observables are governed by only two independent combinations of (custodial symmetric) Wilson coefficients. Therefore, it is easy to identify the following linear combination that vanishes when there is no custodial violation:
\begin{equation}
\left( \hat\rho - 1 \right) + \12 \left( {\hat r}_{Z \nu_L {\bar\nu}_L} - 1 \right) - \12 c_{2\theta} \left( {\hat r}_{Z e_L {\bar e}_L} - 1 \right) \quad\longrightarrow\quad 0 \,.
\label{eqn:thefingerprint}
\end{equation}
Therefore, this combined observable can serve as an indicator of our custodial symmetry/violation. Going back to the general ($\nu$)SMEFT case where custodial violation is present, we see from \cref{eqn:Obs4nuSMEFTWarsaw,eqn:Obs4nuSMEFTCBasis} that this combined observable is given by
\begin{align}
&\left( \hat\rho - 1 \right) + \12 \left( {\hat r}_{Z \nu_L {\bar\nu}_L} - 1 \right) - \12 c_{2\theta} \left( {\hat r}_{Z e_L {\bar e}_L} - 1 \right) \notag\\[5pt]
&\hspace{40pt} = - \12 v^2 \left[ C_{HD} + 4\, C_{Hl}^{(1)} \right] = - 2 v^2 \left[ \WlC_{HD} - \WlC_{Hl}^{(1)} \right] \;\equiv\; \alpha \Tom_l \,\,.
\label{eqn:Toml}
\end{align}
We hence define the Wilson coefficients combination in the second line as $\alpha \Tom_l$ --- a generalization of the Peskin-Takeuchi $T$ parameter that is valid for general ($\nu$)SMEFT (written in appropriate basis, i.e.\ Warsaw or our custodial basis). Clearly, in the special case of just oblique corrections, $C_{Hl}^{(1)}=0$, our $\alpha \Tom_l$ reduces back to \cref{eqn:PTTWarsaw}.

We see from the above that if there is no custodial violation, $\WlC_{HD} = \WlC_{Hl}^{(1)} = 0$, then $\Tom_l=0$. However, the converse is not true. Custodial violation can conspire to yield a vanishing $\Tom_l$. This is a limitation of our example set of observables chosen in \cref{eqn:ExampleObservables}. As we will explain in \cref{subsec:RPI}, adding more observables does not resolve this issue until we move beyond $2 \to 2$ fermion experiments .

Our generalization above is named $\Tom_l$ and not just $\Tom$. This is because, in the presence of non-oblique corrections, one can in fact construct different generalizations of the $T$ parameter with different flavors of fermions. Our construction above used lepton partial widths of the $Z$ boson, and hence the name $\Tom_l$. We will discuss quark generalizations $\Tom_q$ and $\Tom_{\qR}$ in \cref{appsec:Hadronic}.

\subsection{The role of RPI in SMEFT}
\label{subsec:RPI}

The first three observables in \cref{eqn:Obs4MFCV} depend on three custodial symmetric Wilson coefficients $\WlC_{HWB}$, $\WlC_{Hl}^{(3)}$, and $\WlC_{12}$, so in general one would not expect a linear relation among them like \cref{eqn:thefingerprint}. From this point of view, it seemed that we were lucky to have the two Wilson coefficients $\WlC_{HWB}$ and $\WlC_{Hl}^{(3)}$ feeding into \cref{eqn:Obs4MFCV} only as a single linear combination
\begin{equation}
\left[ \dfrac{2c_\theta}{s_\theta}\, \WlC_{HWB} - \WlC_{Hl}^{(3)} \right] = - \left[\dfrac{c_\theta}{s_\theta}\, C_{HWB} + C_{Hl}^{(3)}\right] \,.
\label{eqn:RPICHWB}
\end{equation}

In fact, this grouping was inevitable due to an important property of the observables that we consider---the reparameterization invariance (RPI) when restricting to observables that only involve $2 \to 2$ fermion experiments \cite{Brivio:2017bnu}. Observables in $2 \to 2$ fermion experiments do not receive contributions from the following two operators \emph{outside} the Warsaw basis:
\begin{subequations}
\begin{align}
& ig_2\left(D^\mu H\right)^\dagger \tau^a \left(D^\nu H\right) W_{\mu\nu}^a \,, \\[3pt]
& ig_1\left(D^\mu H\right)^\dagger \left(D^\nu H\right) B_{\mu\nu} \,.
\end{align}
\end{subequations}
These two operators are equivalent to two linear combinations of Warsaw basis operators, which are hence two free directions that one can shift the Warsaw basis Wilson coefficients without affecting the $2 \to 2$ fermion observables. These are known as RPI shifts in SMEFT \cite{Brivio:2017bnu}.

In terms of the Wilson coefficients relevant for \cref{eqn:Obs4nuSMEFTWarsaw}, these RPI shifts are
\begin{subequations}\label{eqn:RPI}
\begin{align}
\mqty( C_{HWB} \\[5pt] C_{Hl}^{(3)} ) &\to \mqty( C_{HWB} \\[5pt] C_{Hl}^{(3)} ) + \epsilon_W \mqty( - \tan\theta \\[5pt] 1 ) \,, \label{eqn:RPIW} \\[10pt]
\mqty( C_{HWB} \\[5pt] C_{HD} \\[5pt] C_{Hl}^{(1)} \\[5pt] C_{He} ) &\to \mqty( C_{HWB} \\[5pt] C_{HD} \\[5pt] C_{Hl}^{(1)} \\[5pt] C_{He} ) + \epsilon_B \mqty( \cot\theta \\[5pt] -4 \\[5pt] 1 \\[5pt] 2 ) \,, \label{eqn:RPIB}
\end{align}
\end{subequations}
where $\epsilon_W$ and $\epsilon_B$ are arbitrary coefficients. The first shift above is especially strong, as there are only two Wilson coefficients involved. Staring at \cref{eqn:RPIW}, one can see that the only RPI combination of $C_{HWB}$ and $C_{Hl}^{(3)}$ is what appears in \cref{eqn:RPICHWB}. This explains why each of our observable's dependence on $C_{HWB}$ in \cref{eqn:Obs4nuSMEFTWarsaw} arises as this RPI combination.

Of course, one can also check that each of our observable's expression in \cref{eqn:Obs4nuSMEFTWarsaw} satisfies the second RPI above [\cref{eqn:RPIB}] as well. In fact, if one were to solve the six unknown Wilson coefficients in \cref{eqn:Obs4nuSMEFTWarsaw} from the four equations given measured values of the four observables, one would find precisely \cref{eqn:RPI} as the two undetermined directions. This means that our example set of observables chosen in \cref{eqn:ExampleObservables} saturates the resolving ability of $2 \to 2$ lepton observables---no undetermined directions remain beyond the RPI shifts. Therefore, adding more $2 \to 2$ fermion observables to \cref{eqn:Obs4nuSMEFTWarsaw}, such as $W$ decay widths, would not help pin down the Wilson coefficients.

Finally, we emphasize that our $\Tom_l$ is also an RPI combination, because it is constructed in \cref{eqn:Toml} with RPI observables. This can be readily checked against \cref{eqn:RPI}. On the other hand, $\Tom_l$ is not the only SMEFT RPI generalization of the Peskin-Takeuchi $T$ parameter; there are other $\Tom_f$ that can be constructed with hadronic widths of the $Z$ boson, as we will show in \cref{appsec:Hadronic}.

\subsection{Experimental measurements of our observables}
\label{subsec:Measurements}

We have presented our results in terms of the observables
\begin{align}
\left\{
\hat{\rho}, \,
\hat\Gamma_{Z\nu_L\bar\nu_L},\,
\hat\Gamma_{Ze_L\bar{e}_L},\,
\hat\Gamma_{Ze\bar{e}}
\right\} \,,
\end{align}
and an additional set of hadronic observables in \cref{appsec:Hadronic}. Let's now consider how to extract these observables from experimental measurements:
\begin{itemize}
\item We need the accurately measured $\hat{\alpha}$, $\hat{G}_F$, $\hat{m}_Z^2$ as basic inputs.
\item The observable $\hat{\rho}$ requires a measurement of $\hat{m}_W^2$.
\item The widths $\left\{ \hat\Gamma_{Ze_L\bar{e}_L}, \hat\Gamma_{Ze\bar{e}} \right\}$ are not directly measured in practice. Instead, we extract them from the measurements on the total partial width $\hat\Gamma_{Ze_L\bar{e}_L} + \hat\Gamma_{Ze\bar{e}}$ and the forward-backward asymmetry $\hat{A}_{FB}^{0,e}$. Direct measurements of the angular distributions of $e^+e^- \rightarrow e^+e^-$ on $Z$ resonance can determine $\hat{A}_{FB}^{0,e}$ \cite{Z-Pole}.
\item The total partial width into electrons $\hat\Gamma_{Ze_L\bar{e}_L} + \hat\Gamma_{Ze\bar{e}}$ is not directly measured either. Instead one uses measurements of the total rate $e^+e^- \rightarrow e^+e^-$ on $Z$ resonance as well as the total width of $Z$ boson, $\Gamma_Z$, determined by separate measurements scanning the lineshape of $e^+e^- \rightarrow \, \mbox{hadrons}$ \cite{Z-Pole}.
\item The partial width of $Z$ into neutrinos must be inferred by subtracting the measured contributions of the $Z$ partial widths from the measured total width $\Gamma_Z$ \cite{Z-Pole}. For this presentation, we assume flavor universality and neglect the masses of the quarks and leptons. The $Z$ partial width into neutrinos is
    \begin{equation}
    3 \hat\Gamma_{Z\nu_L\bar\nu_L} = \Gamma_Z - \hat{\Gamma}_{Zll} - \hat{\Gamma}_{Zqq} \,,
    \label{eqn:neutrinofromtotalandrest}
    \end{equation}
    where we emphasize the observable we have used throughout this paper, $\hat\Gamma_{Z\nu_L\bar\nu_L}$, is the width into just one generation of neutrinos, and $\hat{\Gamma}_{Zll}, \hat{\Gamma}_{Zqq}$ are the measured decay widths of $Z$ into leptons and hadrons respectively.
\end{itemize}

We are finally in a position to evaluate \cref{eqn:Toml} using experimental data on our observables as determined above. If one were to evaluate this expression using experimental measurements matched to just the tree-level relations, one would obtain a sizable numerical difference from zero. This is not surprising, since the Weinberg angle determined from the Veltman $\hat\rho$ differs substantially from the Weinberg angle determined from the charged lepton asymmetries \cite{PDG}. The main source of the discrepancy is the one-loop contribution from the top quark to the $W$ and $Z$ self-energies. Including this loop contribution to the Veltman $\hat\rho$ parameter will cause the numerical evaluation of $\Tom_l$ to be nearly $0$ within experimental errors. The more important quantity is thus the experimental error, i.e.\ \emph{sensitivity}, on $\Tom_l$. This is determined by including the errors on all of the experimental inputs $\hat\rho$, ${\hat r}_{Z{\nu _L}{{\bar \nu }_L}}$, and ${\hat r}_{Z{e_L}{{\bar e}_L}}$. The least well measured observable is ${\hat r}_{Z{e_L}{{\bar e}_L}}$, and thus the experimental error on this quantity dominates the constraint on the custodial violating contribution $- 2 v^2 \left[ a_{HD} - a_{Hl}^{(1)} \right]$. We find
\begin{align}
- 2 v^2 \left[ a_{HD} - a_{Hl}^{(1)} \right] \simeq 0 \pm 0.003 \,,
\end{align}
which implies, in the absence of an accidental cancellation,
\begin{align}
\Lambda_{CV} \equiv \frac{1}{\sqrt{2 \, |a_{HD} - a_{Hl}^{(1)}|}} \simeq 3.1 \; \text{ TeV} \,.
\end{align}
This is much smaller than the scale that would be deduced by doing a global fit to $S,T$ parameters under the assumption that the new physics contributes only to oblique corrections $\Lambda_{CV}^{\rm PDG} \simeq 6.6$~TeV \cite{PDG}.\footnote{The number was obtained by taking the upper and lower bounds on $T$ from the 90\% $S$-$T$ ellipse presented in PDG.}
This simple analysis illustrates that ``maximal'' custodial violation (tree-level contributions to $a_{HD} - a_{Hl}^{(1)}$) is allowed by precision electroweak data with a considerably lower scale of new physics than would be deduced under the assumption of oblique-only contributions.

\section{Custodial Violating Complications from EOM Redundancies}
\label{sec:EOM}

There is an intuitive but crucial assumption underlying our analysis in \cref{sec:Observables}, as was established in \cref{sec:CustodialBasis}:
\begin{itemize}
\item When the UV sector is custodial symmetric, any EFT operators generated by matching would preserve the identified custodial $SU(2)_R$ in the limit $g_1\to0$, and therefore all the custodial violating operators in our custodial basis (\Hvcolor~operators in \cref{tbl:nuSMEFTCBasis}) are absent.
\end{itemize}
Interestingly, this is NOT completely true. Only the first half of the above statement is true; while the second half can be invalidated by the EOM redundancies in ($\nu$)SMEFT\@.

After integrating out a custodial symmetric UV sector, the resulted EFT operators have to be custodial $SU(2)_R$-singlets (in the limit $g_1\to0$), but may lie outside of an arbitrarily chosen operator basis. In order to present the entire EFT in the Warsaw/custodial basis, one may need to apply redundancy relations to trade outside operators into linear combinations of Warsaw/custodial basis operators. While IBP and Fierz redundancies do not change the $SU(2)_R$ preserving/violating nature of operators, the EOM redundancies may mix operators that preserve the custodial $SU(2)_R$ with those that do not, because ($\nu$)SM Yukawa couplings break it. As a result, the linear combinations traded from outside operators may contain \Hvcolor~operators in our custodial basis in \cref{tbl:nuSMEFTCBasis}.

To better illustrate this issue, we take a closer look at a specific example that we will actually encounter later in some of our example UV theories in \cref{sec:UVTheories}. Consider the operator $Q_R \equiv |H|^2 |DH|^2$. Upon taking the limit $g_1\to0$, this operator preserves the global symmetry $SU(2)_{RH}$. Therefore, according to our discussions in \cref{sec:CustodialBasis}, it can be possibly generated by custodial symmetric UV sectors. For example, it is indeed generated at tree-level by integrating out a heavy $W'_L$ gauge boson, as we will see in \cref{subsec:Wprime}. The problem, however, is that $Q_R$ does not belong to the Warsaw/custodial basis; we need to use IBP and Higgs EOM redundancy relations to trade it into Warsaw/custodial basis operators. From the $\nu$SM Lagrangian given in \cref{eqn:LagSM,eqn:LagnuSM}, we obtain the Higgs EOM relation
\begin{equation}
{H^\dag }{D^2}H + \hc = 2\lambda {v^2}{\left| H \right|^2} - 4\lambda {\left| H \right|^4} - \left( {{Y_u}\bar q\tilde Hu + {Y_d}\bar qHd + {Y_\nu }\bar l\tilde H\nu  + {Y_e}\bar lHe + \hc} \right) \,.
\label{eqn:HiggsEOM}
\end{equation}
Note that in order to make the expression compact, we have multiplied the EOM by $H^\dagger$ from the left and also added its hermitian conjugate. Combining this with IBP, we can convert $Q_R$ into
\begin{equation}
{Q_R} = \left|H\right|^2 \left|DH\right|^2 \equiv 2\lambda {Q_H} + \frac{1}{2}{Q_{H\Box}} + \frac{1}{2} Q_Y - \lambda {v^2}{\left| H \right|^4} \,,
\label{eqn:QRtrading}
\end{equation}
where we have defined the operator combination
\begin{align}
Q_Y &\equiv {Y_u}{Q_{uH}} + {Y_d}{Q_{dH}} + {Y_\nu }{Q_{\nu H}} + {Y_e}{Q_{eH}} + \hc \notag\\[5pt]
&= \frac14 \left[ \left(Y_u+Y_d\right) \Op_{qH}^+ + \left(Y_u-Y_d\right) \Op_{qH}^- + \left(Y_\nu+Y_e\right) \Op_{lH}^+ + \left(Y_\nu-Y_e\right) \Op_{lH}^- \right] + \hc \,.
\label{eqn:QYDef}
\end{align}
Due to the Yukawa mismatch $Y_u\ne Y_d$ and $Y_\nu\ne Y_e$, this combination contains custodial violating operators $\showHv{\Op_{qH}^-}$ and $\showHv{\Op_{lH}^-}$. We see that once traded into Warsaw/custodial basis, the $SU(2)_{RH}$ preserving operator $Q_R$ corresponds to a mixture of custodial preserving and violating operators in our custodial basis. This is a consequence of applying Higgs EOM redundancies relations in \cref{eqn:HiggsEOM}, which breaks $SU(2)_{RH}$.

\subsection{Robustness of our observables and $\Tom_l$ parameter}
\label{subsec:EOMImplications}

Our observable results given in \cref{eqn:Obs4MFCV} assumed the presence of only custodial preserving operators in our custodial basis. Now, given that custodial violating operators could also appear from EOM redundancies, our analysis in \cref{sec:Observables} is potentially incomplete. In this section, we show that this EOM subtlety does not affect our results in \cref{sec:Observables}, provided we restrict to tree-level matching.

As was originally worked out in \cite{Arzt:1994gp, Einhorn:2013kja}, and recently emphasized and generalized by \cite{Craig:2019wmo}, only a small subset of dim-6 SMEFT operators can be generated by tree-level matching.\footnote{Note that this argument is not limited to the Warsaw basis operator set.} In particular, dim-6 operators with field strengths $X_{\mu\nu}$ cannot be generated at tree-level. This immediately removes the EOMs for gluons, $W$ boson, and $B$ boson out of consideration. So the only potentially problematic EOMs are those for the Higgs $H$ and the fermions $\psi$.

\begin{table}[t!]
\renewcommand{\arraystretch}{2.2}
\setlength{\arrayrulewidth}{.3mm}
\setlength{\tabcolsep}{1 em}
\begin{center}
\begin{tabular}{c|cccccc}
Total & $H^4 D^2$ & $\bar\psi\psi H^2 D$ & $H^2 D^4$ & $\bar\psi\psi D^3$ & $\bar\psi\psi X D$ & $\bar\psi\psi H D^2$ \\\hline
38    & 1         & 8                    & 1         & 4                  & 8                  & 16
\end{tabular}\vspace{0.2cm}
\caption{Custodial $SU(2)_R$ invariants outside of Warsaw basis, which could yield custodial violating operators in Warsaw basis upon using $H$ and $\psi$ EOM redundancies.}
\label{tbl:EOMCustodialInvariants}
\end{center}
\end{table}

Next, let us find all the $\nu$SMEFT dim-6 custodial $SU(2)_R$-singlet operators containing an EOM factor of $H$ or $\psi$, i.e.\ containing $D^2H$, $\slashed D \psi$, or $\slashed D\bar\psi$. Using the Hilbert series technique \cite{Jenkins:2009dy, Hanany:2010vu, Lehman:2015via, Henning:2015daa, Henning:2017fpj}, with these EOM redundancy relations relaxed,\footnote{This can be achieved by taking $H$, $\psi$, and $\bar\psi$ (and their descendants) as ``long representations'' of the conformal group, as opposed to ``short representations''. See \cite{Henning:2017fpj} for details.} we find that there are $38$ additional real custodial $SU(2)_R$-singlets outside of Warsaw basis. They can be divided into six classes according to the field content, as listed in \cref{tbl:EOMCustodialInvariants}.

When restricted to tree-level matching, again due to the argument given in \cite{Arzt:1994gp, Einhorn:2013kja, Craig:2019wmo}, only the first two classes in \cref{tbl:EOMCustodialInvariants}, i.e. $H^4 D^2$ and $\bar\psi\psi H^2 D$, can be generated. Let us examine what \Hvcolor~operators in our \cref{tbl:nuSMEFTCBasis} can be obtained from trading these two classes of operators into the Warsaw/custodial basis. They contain nine operators, which are nothing but the $\nu$SMEFT ``kinetic terms'' multiplied by $\left|H\right|^2$:
\begin{subequations}
\begin{align}
&\left|H\right|^2\left|DH\right|^2 \,,\\[5pt]
&\left|H\right|^2\bar\psi i\slashed D \psi+\hc \,,\qquad\text{with $\psi=q, l, \qR, \lR$} \,\,.
\end{align}
\end{subequations}
We already analyzed the first operator above, and showed the result of transforming it into the Warsaw basis through EOM in \cref{eqn:QRtrading,eqn:QYDef}.  The second operator transformed into the Warsaw basis becomes
\begin{align}
  {\left| H \right|^2}\bar qi\slashed{D}q &= {Y_u}{Q_{uH}} + {Y_d}{Q_{dH}} \,,\qquad\text{for $\psi=q$} \,,
\end{align}
and similarly for the others. We see that the custodial violating operators obtained through this procedure are all in ``class 5'' of \cref{tbl:nuSMEFTWarsaw}: $\bar\psi\psi H^3$. However, it is clear from \cref{eqn:Obs4nuSMEFTWarsaw} that none of these operators would feed into the observables discussed in \cref{sec:Observables}, even when they are present.\footnote{Recall that we have neglected the fermion masses in the decay widths.} Therefore, our results in \cref{eqn:Obs4MFCV} stand, and hence the subsequent analysis presented in \cref{sec:Observables}, provided that we limit ourselves to tree-level matching.\footnote{Amusingly, the argument here can also be recast into a (new) reparameterization invariance relation among $SU(2)_R$-singlet operators, in the same spirit as that in Ref.~\cite{Brivio:2017bnu} (see our discussion in \cref{subsec:RPI}). For example, let us rearrange the EOM relation in \cref{eqn:QRtrading} as
\begin{align}
Q_Y = 2{Q_R}  - 4\lambda {Q_H} - {Q_{H\Box}} + 2\lambda {v^2}{\left| H \right|^4} \,. \notag
\end{align}
When restricted to $Z$-pole observables and also neglecting the fermion masses as in \cref{sec:Observables}, the LHS does not contribute, and so neither does  the RHS\@. Therefore, the combination on the RHS is a free direction that can be viewed as a new set of RPI shifts among $SU(2)_R$-singlets. In this language, when the outside operator $Q_R$ is generated, one can use this new RPI shift to trade it for other $SU(2)_R$-singlets, which will then be in our Warsaw/custodial basis.}

\section{Application to UV Theories with Custodial Symmetry/Violation}
\label{sec:UVTheories}

In this section, we examine several UV theories and demonstrate that our $\Tom_l$ parameter is sensitive to (hard) custodial symmetry/violation. We consider in \cref{subsec:Triplet} a real triplet scalar; in \cref{subsec:Wprime} a heavy $W_L'$ from embedding $SU(2)_L$ into $SU(2)_A \times SU(2)_B$; in \cref{subsec:UVBL} a heavy $Z'$ from a spontaneously broken $U(1)_{B-L}$ theory; in \cref{subsec:UV221} heavy $W'$'s and $Z'$'s from embedding the electroweak group into $SU(2)_L \times SU(2)_R \times U(1)_{B-L}$; and finally in \cref{subsec:UVVLF} two heavy vector-like fermions transforming as $SU(2)_L$-singlets. Several highlights of the lessons that we will learn from these UV examples:
\begin{itemize}
  \item Our $\Tom_l$ parameter works perfectly for all of these examples. When the UV sector is custodial symmetric or violating, our $\Tom_l$ will be zero or nonzero accordingly.
  \item The heavy $W'_L$ example in \cref{subsec:Wprime} reminds us that the Veltman $\hat\rho$ can possibly deviate from unity in the case of custodial symmetric UV physics.
  \item The vector-like fermions theory discussed in \cref{subsec:UVVLF} serve as a striking example that our new $\Tom_l$ parameter captures non-oblique custodial violation of the UV theory while, unsurprisingly, the Peskin-Takeuchi $T$ parameter fails to do so.
\end{itemize}

\subsection{Triplet scalar extension}
\label{subsec:Triplet}

The first UV example we consider is the well studied SM extension by a real $SU(2)_L$-triplet scalar $\phi^a$; see e.g.~\cite{Skiba:2010xn, Khandker:2012zu, deBlas:2014mba, Chiang:2015ura, Ellis:2016enq, Dawson:2017vgm}. The most general renormalizable Lagrangian for this model is
\begin{equation}
\Lag_\text{UV} = \Lag_\text{SM} + \12 \left(D^\mu\phi^a\right)\left(D_\mu\phi^a\right) - \12M^2\phi^a\phi^a - A H^\dagger t^a H \phi^a - \kappa |H|^2 \phi^a\phi^a - \lambda_\phi (\phi^a\phi^a)^2 \,.
\end{equation}
This UV theory has (hard) custodial violation due to the interaction term $H^\dagger t^a H \phi^a$. It is well known that this custodial violation shows up already at tree-level in the EFT\@. In what follows below, we check that our new $\Tom_l$ parameter captures this effect.

Integrating out $\phi^a$ at tree level, we obtain a SMEFT up to dim-6
\begin{align}
\Lag_\text{SMEFT} &= \Lag_\text{SM} + \frac{A^2}{8M^2} \left|H\right|^4 - \frac{\kappa A^2}{4M^4} Q_H - \frac{A^2}{2M^4} \left( \frac14 Q_{H\Box} + Q_{HD} - Q_R \right) \notag\\[8pt]
&= \Lag_\text{SM} + \frac{A^2}{8M^2} \left( 1 - \frac{4\lambda v^2}{M^2} \right) \left|H\right|^4 \notag\\[3pt]
&\hspace{30pt} - \frac{A^2}{4M^4} \left( \kappa - 4\lambda \right) Q_H - \frac{A^2}{2M^4} \left( -\frac14 Q_{H\Box} + Q_{HD} - \12 Q_Y \right) \,.
\label{eqn:TripletSMEFT}
\end{align}
As expected, the custodial violating operator $Q_{HD}$ is generated. We also see the appearance of the operator $Q_R$ which is outside of the Warsaw basis. In the second line, we have traded it into combinations of Warsaw basis operators using \cref{eqn:QRtrading}, and hence obtained an additional custodial violating operator $Q_Y$ (see \cref{eqn:QYDef} for definition). Reading off the Warsaw basis Wilson coefficients $C_i$ from the above and translating to our custodial basis $\WlC_i$ using \cref{tbl:aFromC}, we obtain
\begin{subequations}
\begin{align}
\WlC_H &= - \frac{A^2}{32 M^4} \left( \kappa - 4\lambda \right) \,, \\[5pt]
\WlC_{H\Box} &= \frac{A^2}{4 M^4} \,, \\[5pt]
\WlC_{HD} &= - \frac{A^2}{8 M^4} \,, \\[5pt]
\WlC_{lH}^\pm &= \frac{A^2}{16 M^4} \left( Y_\nu \pm Y_e \right) \,, \\[5pt]
\WlC_{qH}^\pm &= \frac{A^2}{16 M^4} \left( Y_u \pm Y_d \right) \,.
\end{align}
\end{subequations}
Note that in addition to the $\showHv{\WlC_{HD}}$, the ``class 5'' (see \cref{tbl:nuSMEFTCBasis}) custodial violating operators $\showHv{\WlC_{lH}^-}$ and $\showHv{\WlC_{qH}^-}$ also show up due to the EOM subtlety discussed in \cref{sec:EOM}, but as explained in \cref{subsec:EOMImplications}, they do not invalidate our analysis.

Now using our definition in \cref{eqn:Toml}, we obtain
\begin{equation}
\alpha \Tom_l = - 2 v^2 \left[ \WlC_{HD} - \WlC_{Hl}^{(1)} \right] = \frac{v^2 A^2}{4M^4} \ne 0 \,.
\end{equation}
We see that our $\Tom_l$ parameter captures the hard custodial violation. Since there is no vertex correction in this example [see the first line of \cref{eqn:TripletSMEFT}], our $\Tom_l$ reduces to Peskin-Takeuchi $T$ parameter as explained before. So they work equally well in this case.

\subsection{A heavy $W_L'$ gauge boson}
\label{subsec:Wprime}

In this section, we consider a UV theory of embedding the $SU(2)_L$ of the SM into $SU(2)_A \times SU(2)_B$. Specifically, the gauge sector of the UV Lagrangian is
\begin{equation}
\Lag_\text{UV} \supset - \frac{1}{4}W_{A\mu \nu }^aW_A^{a\mu \nu } - \frac{1}{4}W_{B\mu \nu }^aW_B^{a\mu \nu } + \frac{1}{2}\tr \left[ {{{\left( {{D^\mu }\Phi } \right)}^\dag }\left( {{D_\mu }\Phi } \right)} \right] - V_\Phi \,,
\label{eqn:LagWprimeGauge}
\end{equation}
where the heavy scalar field $\Phi$ is a $2\times 2$ matrix that transforms as a bifundamental under $\left(U_A, U_B\right) \in SU(2)_A \times SU(2)_B$:
\begin{equation}
\Phi  \to U_A\, \Phi\, U_B^\dag \,\,.
\end{equation}
Therefore, the concrete form of its covariant derivative is
\begin{equation}
{D_\mu }\Phi  = {\partial _\mu }\Phi  - i{g_A}W_{A\mu}^a {t^a}\Phi  + i{g_B}\Phi W_{B\mu }^a{t^a} \,,
\end{equation}
with $t^a=\frac{1}{2}\sigma^a$ the $SU(2)$ generators in the fundamental representation.

The symmetry $SU(2)_A \times SU(2)_B$ is spontaneously broken by the vev of the heavy scalar field:
\begin{equation}
\Phi \supset \frac{v_\Phi}{\sqrt 2} \mqty(1 & 0\\0 & 1) \,.
\end{equation}
The unbroken group is the diagonal $SU(2)$ formed by the generators $t_A^a+t_B^a$, which we identify as our $SU(2)_L$ group in the SM\@. The corresponding gauge boson is the $W$ boson. For the broken generators, the corresponding gauge boson $W_L'$ acquire mass from $v_\Phi$:
\begin{align}
\frac{1}{2}\tr \left[ {{{\left( {{D^\mu }\Phi } \right)}^\dag }\left( {{D_\mu }\Phi } \right)} \right] &\supset \frac{1}{8}v_\Phi ^2\left( {{g_A}W_A^{a\mu } - {g_B}W_B^{a\mu }} \right)\left( {{g_A}W_{A\mu }^a - {g_B}W_{B\mu }^a} \right) \notag\\[5pt]
&= \frac{1}{8}v_\Phi ^2\left( {g_A^2 + g_B^2} \right){{W_L'}^{a\mu }}{W_L'}_\mu ^a \,\,.
\end{align}
We see that $m_{W_L'}^2=\frac{1}{4}\left(g_A^2+g_B^2\right)v_\Phi^2$, and
\begin{subequations}
\begin{align}
{W_L'}_\mu ^a &\equiv \frac{1}{{\sqrt {g_A^2 + g_B^2} }}\left( {{g_A}W_{A\mu }^a - {g_B}W_{B\mu }^a} \right) \,, \\[5pt]
W_\mu ^a &\equiv \frac{1}{{\sqrt {g_A^2 + g_B^2} }}\left( {{g_B}W_{A\mu }^a + {g_A}W_{B\mu }^a} \right) \,.
\end{align}
\end{subequations}
With the above rotation, we can rewrite the general covariant derivative as
\begin{align}
D_\mu &= \partial_\mu - i{g_A}W_{A\mu}^a t_A^a - i{g_B}W_{B\mu }^at_B^a \notag\\[5pt]
&= \partial_\mu - i{g_2}W_\mu ^a\left( {t_A^a + t_B^a} \right) - i{W_L'}_\mu ^a\left( {\frac{{g_A^2}}{{\sqrt {g_A^2 + g_B^2} }}t_A^a - \frac{{g_B^2}}{{\sqrt {g_A^2 + g_B^2} }}t_B^a} \right) \,,
\label{eqn:DBroken}
\end{align}
with the SM gauge coupling $g_2=\frac{g_Ag_B}{\sqrt{g_A^2+g_B^2}}$ identified.

For the UV interactions between the gauge sector in \cref{eqn:LagWprimeGauge} and the SM fields, we assume that $W_A$ plays the role of $W$ before the symmetry breaking, namely that the SM fields couple to $W_A$ exactly the way they couple to the $W$ boson in SM, and do not couple to $W_B$ at all.\footnote{While this is the simplest coupling scheme, it is also possible to split the left-handed fermion generations between coupling to $W_A$ and coupling to $W_B$. In this case, after the $SU(2)_A \times SU(2)_B \to SU(2)_L$ breaking all left handed fermions will couple as usual to $W_L$ but the interactions with $W'_L$ will be flavor-dependent.} This means that for nontrivially $SU(2)_L$-charged SM fields, $t_A^a\ne0$ but $t_B^a=0$. From \cref{eqn:DBroken}, we see that after the symmetry breaking, the SM fields couple to both $W$ and $W_L'$.

In the following, we will match this UV theory onto SMEFT by integrating out the heavy $W_L'$ gauge boson at the tree level. As is clear from the setup, the UV interactions in this example respect the symmetry $SU(2)_{RH}$ (as well as the other $SU(2)_R$ symmetries discussed in \cref{sec:CustodialSymmetries}), and hence are custodial symmetric by our definition. We therefore expect a vanishing $\Tom_l$ in the resulting EFT\@.

Up to linear power in $W_L'$, the UV interaction is
\begin{equation}
\Lag_\text{UV} \supset \frac{g_A^2}{\sqrt{g_A^2+g_B^2}} {W_L'}_\mu^a J_W^{a\mu} \,,
\end{equation}
where $J_W^{a\mu}$ denotes the SM $SU(2)_L$ current:
\begin{equation}
J_{W\mu}^a = \frac{1}{2} \left( H^\dagger i\lrD_{\text{SM, }\mu}^a H + \sum_\psi \bar\psi\gamma_\mu \tau^a\psi \right) \,.
\label{eqn:JWDef}
\end{equation}
Integrating out $W'_L$ at tree level, we obtain a SMEFT up to dim-6 as
\begin{equation}
\Lag_\text{SMEFT} =  - \frac{{g_A^4}}{{g_A^2 + g_B^2}}\frac{1}{{2m_{W_L'}^2}}J_{W\mu }^a J_W^{a\mu } = -\frac{2c_A^4}{v_\Phi^2}J_{W\mu}^a J_W^{a\mu} \,,
\label{eqn:LEFTJJ}
\end{equation}
where we have defined the mixing angle $c_A\equiv\frac{g_A}{\sqrt{g_A^2+g_B^2}}$. Clearly, this EFT Lagrangian preserves $SU(2)_{RH}$. Plugging in \cref{eqn:JWDef}, we obtain
\begin{align}
\Lag_\text{SMEFT} &= - \frac{{c_A^4}}{{v_\Phi ^2}}\left[ {\frac{1}{2}{Q_R} + \frac{1}{8}{Q_{H\Box}} + Q_{Hl}^{\left( 3 \right)} + Q_{Hq}^{\left( 3 \right)} + \frac{1}{2}{Q_{ll}} + \frac{1}{2}Q_{qq}^{\left( 3 \right)} + Q_{lq}^{\left( 3 \right)}} \right] \notag\\[5pt]
&= - \frac{{c_A^4}}{{v_\Phi ^2}}\Bigg[ \lambda {Q_H} + \frac{3}{8}{Q_{H\Box}} + \frac{1}{4}\left( {{Y_u}{Q_{uH}} + {Y_d}{Q_{dH}} + {Y_\nu }{Q_{\nu H}} + {Y_e}{Q_{eH}} + \hc} \right) \notag\\
&\hspace{40pt} + Q_{Hl}^{\left( 3 \right)} + Q_{Hq}^{\left( 3 \right)} + \frac{1}{2}{Q_{ll}} + \frac{1}{2}Q_{qq}^{\left( 3 \right)} + Q_{lq}^{\left( 3 \right)} \Bigg] \,.
\end{align}
From the first line above, we see that all the effective operators are $SU(2)_{RH}$ preserving, as expected from \cref{eqn:LEFTJJ}. However, in the second line, the $SU(2)_{RH}$ breaking operator $Q_Y$ (see \cref{eqn:QYDef} for definition) shows up, due to trading $Q_R$ for operators in the Warsaw basis using \cref{eqn:QRtrading}. Reading off the Warsaw basis Wilson coefficients $C_i$ from the above and translating to our custodial basis $\WlC_i$ using \cref{tbl:aFromC}, we obtain
\begin{subequations}\renewcommand\arraystretch{2.2}
	\begin{align}
	\left\{ \begin{array}{l}
	\WlC_H =  - \dfrac{{c_A^4}}{{v_\Phi ^2}} \dfrac{1}{8} \lambda \\
	\WlC_{H\Box} =  - \dfrac{{c_A^4}}{{v_\Phi ^2}}\dfrac{3}{8}
	\end{array} \right.  \quad&,\quad
	\left\{ \begin{array}{l}
	\WlC_{lH}^\pm = - \dfrac{{c_A^4}}{{v_\Phi ^2}} \dfrac{1}{16} \left(Y_\nu \pm Y_e\right) \\
	\WlC_{qH}^\pm = - \dfrac{{c_A^4}}{{v_\Phi ^2}} \dfrac{1}{16} \left(Y_u \pm Y_d\right)
	\end{array} \right. \,, \\[8pt]
	\WlC_{Hl}^{\left( 3 \right)} = \WlC_{Hq}^{\left( 3 \right)} =\,& 2{\WlC_{ll}} = 2\WlC_{qq}^{\left( 3 \right)} = \WlC_{lq}^{\left( 3 \right)} =  - \frac{{c_A^4}}{{v_\Phi ^2}} \,\,.
	\end{align}
\end{subequations}
Again, we find the appearance of the ``class 5'' (see \cref{tbl:nuSMEFTCBasis}) custodial violating operators $\showHv{\WlC_{lH}^-}$ and $\showHv{\WlC_{qH}^-}$, as expected from the EOM subtlety discussed in \cref{sec:EOM}. Nevertheless, they do not invalidate our analysis because they do not feed into our observables discussed in \cref{sec:Observables}, as we explained in \cref{subsec:EOMImplications}.

From the Wilson coefficients above and \cref{eqn:Toml}, it is straightforward to see that $\Tom_l$ vanishes:
\begin{equation}
\alpha \Tom_l = - 2 v^2 \left[ \WlC_{HD} - \WlC_{Hl}^{(1)} \right] = 0 \,,
\end{equation}
which demonstrates the consistency with the UV physics being custodial symmetric.

As a side note, this example also reminds us that the Veltman $\hat\rho$ can deviate from 1 in the presence of custodial symmetric UV physics.\footnote{This issue is unfortunately quite confusing in the PDG, which suggests $\hat\rho \not= 1$ implies custodial symmetry violation \cite{PDG}, which is not correct in general.}
To see this point, we can compute $\hat\rho$ with \cref{eqn:Obs4nuSMEFTCBasis}. However, we first need to extract the $\WlC_{12}$ [defined in \cref{eqn:C12WlC12}] from the $\WlC_{ll}$ result above. To do so, we restore the generation indices in $Q_{ll}$ from \cref{eqn:LEFTJJ}:
\begin{equation}
\Lag_\text{SMEFT} \supset -\frac{c_A^4}{2v_\Phi^2} \sum\limits_{p,r=1}^3 \left({\bar l}_p \gamma_\mu \tau^a l_p\right) \left({\bar l}_r \gamma^\mu \tau^a l_r\right) \,.
\end{equation}
To make this into the form of $Q_{\substack{ll \\ prst}}$, we need to also restore the $SU(2)_L$ indices being contracted, and use the group identity:
\begin{equation}
\tau_{ij}^a \tau_{kl}^a = 4 \left(\frac{1}{2}\delta_{il}\delta_{jk} - \frac{1}{4}\delta_{ij}\delta_{kl} \right) \,.
\end{equation}
Substituting this in, we get
\begin{align}
\Lag_\text{SMEFT} &\supset -\frac{c_A^4}{2v_\Phi^2} \sum\limits_{p,r=1}^3 \left({\bar l}_p^i \gamma_\mu \tau_{ij}^a l_p^j\right) \left({\bar l}_r^k \gamma^\mu \tau_{kl}^a l_r^l\right) \notag\\[3pt]
&= -\frac{c_A^4}{2v_\Phi^2} \sum\limits_{p,r=1}^3 \left[ 2\left({\bar l}_p^i \gamma_\mu l_p^j\right) \left({\bar l}_r^j \gamma^\mu l_r^i\right) - \left({\bar l}_p^i \gamma_\mu l_p^i\right) \left({\bar l}_r^j \gamma^\mu l_r^j\right) \right] \notag\\[3pt]
&= -\frac{c_A^4}{2v_\Phi^2} \sum\limits_{p,r=1}^3 \left[ 2\left({\bar l}_p \gamma_\mu l_r\right) \left({\bar l}_r \gamma^\mu l_p\right) - \left({\bar l}_p \gamma_\mu l_p\right) \left({\bar l}_r \gamma^\mu l_r\right) \right] \,.
\end{align}
To obtain the last line above, we have used Fierz identity for the first term in the square brackets, and then suppressed the $SU(2)_L$ indices as usual. Now we can read off the Wilson coefficient with generation indices:
\begin{equation}
\WlC_{\substack{ll \\ prst}} = -\frac{c_A^4}{2v_\Phi^2} \left( 2\delta_{pt}\delta_{rs} - \delta_{pr}\delta_{st} \right) \,.
\end{equation}
Now from \cref{eqn:C12WlC12} we get
\begin{equation}
\WlC_{12} = \WlC_{\substack{ll \\ 1221}} + \WlC_{\substack{ll \\ 2112}} = -2 \frac{c_A^4}{v_\Phi^2} \,.
\end{equation}
Plugging all the relevant Wilson coefficients into \cref{eqn:Obs4nuSMEFTCBasis}, we obtain $\hat\rho$ as
\begin{equation}
\hat \rho = 1 + \frac{v^2}{c_{2\theta}} \Bigg[ 2s_\theta^2 \left( \frac{2c_\theta}{s_\theta}\, \WlC_{HWB} - \WlC_{Hl}^{(3)} \right) + \12 s_\theta ^2 \, \WlC_{12} - 2c_\theta^2\, \WlC_{HD} \Bigg] = 1 + \frac{s_\theta^2}{c_{2\theta}} \frac{{c_A^4 v^2}}{{v_\Phi ^2}} \ne 1 \,.
\end{equation}
Note that $\hat\rho\ne1$ in this example is from the non-oblique corrections $\WlC_{Hl}^{(3)}$ and $\WlC_{12}$. On the other hand, $\WlC_{HD} = \WlC_{HWB} = 0$, so a naive implementation of \cref{eq:rhoSTU,eqn:STUnaive} would misleadingly predict Veltman $\hat\rho=1$. This highlights one  limitation of the oblique framework (although this is not about
custodial symmetry).

In fact, the UV theory in this example is actually a ``universal theory'', in the sense that one can find an operator basis (different from Warsaw/custodial basis) in which all the effective operators are oblique corrections. Concretely, the SMEFT Lagrangian we obtained in \cref{eqn:LEFTJJ} can be fully written into a single effective operator $\left(D^\mu W_{\mu\nu}^a\right)^2$ by using the SM $W$ boson EOM, which is then obviously oblique (only contributing to two-point function of the $W$ boson). However, even in case of a universal theory, finding the desired basis and working out the oblique parameters in that basis requires additional effort, and must again be done on a case-by-case basis. On the other hand, restricting to the Warsaw basis and accommodating the non-oblique corrections provides a more systematic approach.

\subsection{A heavy $Z'$ associated with the $U(1)_{B-L}$ symmetry}
\label{subsec:UVBL}

In this section, we consider a UV model with a heavy $Z'$ gauge boson, associated with the $U(1)_{B-L}$ symmetry in SM (see, e.g., Ref.~\cite{Heeck:2014zfa}). This classical symmetry can be broken at the quantum level through triangle anomalies. To consistently gauge the symmetry, one has to ensure that the triangle anomaly contributions from different fermion species are cancelled. This can be simply achieved by introducing three SM-singlet right-handed neutrinos $\nu$, a requirement that is satisfied automatically by $\nu$SM and $\nu$SMEFT\@.

Assuming that this $U(1)_{B-L}$ gauge boson $Z'$ couples to the $B-L$ current $j_{B-L} \equiv j_B - j_L$ through a coupling $\frac{1}{2}g_Z$, our UV Lagrangian is\footnote{In principle, our $U(1)_{B-L}$ gauge boson $Z'_{B-L}$ can also mix with the hypercharge gauge boson $B$ through a coupling $\12 \epsilon B^{\mu\nu} Z'_{\mu\nu}$. We set this coupling to zero for simplicity in this UV theory example. This is legitimate in our analysis as we only focuse on the tree-level matching and neglect radiative effects.}
\begin{equation}
\Lag_\text{UV} = \Lag_\text{SM} - \frac{1}{4}{Z'_{\mu \nu }}{{Z'}^{\mu \nu }} + \frac{1}{2}{M^2}{Z'_\mu}{{Z'}^\mu } + g_Z{Z'_\mu}\sum\limits_{\psi=q, u, d, l, \nu, e}  {\bar \psi {\gamma ^\mu }{{\hyp}'_\psi} \psi } \,.
\label{eqn:LUVZprime}
\end{equation}
Here the specific values of the charge ${\hyp}'_\psi = \frac{1}{2}(B-L)$ are
\begin{subequations}\label{eqn:hypprime}
\begin{align}
{\hyp}'_q = {\hyp}'_u   = {\hyp}'_d =  \frac{1}{6} &\equiv {\hyp}'_1  \qquad\text{for quarks} \,, \\[5pt]
{\hyp}'_l = {\hyp}'_\nu = {\hyp}'_e = -\frac{1}{2} &\equiv {\hyp}'_2  \qquad\text{for leptons} \,.
\end{align}
\end{subequations}
We have also assumed that the $Z'$ has a large mass $M\gg v$. This can be acquired through the Higgsing from a heavy scalar in the UV which only couples to $Z'$, or via a \emph{St{\"u}eckelberg} mechanism which allows $M$ to be a free parameter in the model.

This example is trivially custodial symmetric by our definition, because the UV interactions do not involve the SM Higgs, and hence the $SU(2)_{RH}$ symmetry is trivially preserved. The $\nu$SMEFT side of the story is similarly trivial. Only operators not involving the Higgs field can be generated at the tree level (the four-fermion operators in this case, as we will see); they are custodial preserving operators due to trivially respecting $SU(2)_{RH}$. In particular, no \Hvcolor~operators can be possibly generated by this example, so we can already get $\Tom_l=0$ without carrying out the matching calculation.

Nevertheless, this example is still interesting, because apart from the custodial symmetry (which must involve $SU(2)_{RH}$; see \cref{subsec:CustodialDef}), our custodial basis also helps making manifest the operator structure under the isospin symmetries $SU(2)_{R\qR}$ and $SU(2)_{R\lR}$. The UV interactions in \cref{eqn:LUVZprime} with the charges given in \cref{eqn:hypprime} clearly also preserve these two isospin symmetries. Below, we will check that no \qvcolor~or \lvcolor~operators will be generated in the resulting $\nu$SMEFT\@.

Integrating out the $Z'$ at tree-level, we obtain the $\nu$SMEFT Lagrangian
\begin{equation}
\Lag_{\nu\text{SMEFT}} = - \frac{{g_Z^2}}{{2{M^2}}}\left( {\sum\limits_{\psi=q, u, d, l, \nu, e}  {\bar \psi {\gamma _\mu }{{\hyp}'_\psi} \psi } } \right)\left( {\sum\limits_{\psi=q, u, d, l, \nu, e}  {\bar \psi {\gamma ^\mu }{{\hyp}'_\psi}\psi } } \right) \,.
\label{eqn:LEFTZprime}
\end{equation}
We see that only four-fermion operators of the type $(\bar LL)(\bar LL)$, $(\bar RR)(\bar RR)$, and $(\bar LL)(\bar RR)$ are generated. In Warsaw basis, the Wilson coefficients can be summarized as
\begin{subequations}\label{eqn:WlCZprime}
\begin{align}
C_{ud}^{\left( 1 \right)} = C_{qu}^{\left( 1 \right)} = C_{qd}^{\left( 1 \right)} = 2C_{qq}^{\left( 1 \right)} = 2{C_{uu}} = 2{C_{dd}} &= - \frac{{{g_Z^2}}}{{2{M^2}}}2({\hyp}'_1)^2 \,, \\[8pt]
{C_{\nu e}} = {C_{l\nu }} = {C_{le}} = 2{C_{ll}} = 2{C_{\nu \nu }} = 2{C_{ee}} &= - \frac{{{g_Z^2}}}{{2{M^2}}}2({\hyp}'_2)^2 \,, \\[8pt]
C_{lq}^{\left( 1 \right)} = {C_{\nu u}} = {C_{\nu d}} = {C_{eu}} = {C_{ed}} = {C_{lu}} = {C_{ld}} = {C_{q\nu }} = {C_{qe}} &= - \frac{{{g_Z^2}}}{{2{M^2}}}2{{\hyp}'_1}{{\hyp}'_2} \,.
\end{align}
\end{subequations}
Transforming to our custodial basis defined in \cref{tbl:nuSMEFTCBasis} (again by applying the dictionary in \cref{tbl:aFromC}), we see that the only nonzero Wilson coefficients are those preserving both of the isospin symmetries $SU(2)_{R\qR}\times SU(2)_{R\lR}$:
\begin{equation}\renewcommand\arraystretch{2.5}
\left\{ \begin{array}{l}
{\WlC_{ll}} = - \dfrac{{{g_Z^2}}}{{2{M^2}}}({\hyp}'_2)^2\\
\WlC_{qq}^{\left( 1 \right)} =  - \dfrac{{{g_Z^2}}}{{2{M^2}}}({\hyp}'_1)^2\\
\WlC_{lq}^{\left( 1 \right)} =  - \dfrac{{{g_Z^2}}}{{2{M^2}}}2{{\hyp}'_1}{{\hyp}'_2}
\end{array} \right.  \,,\quad
\left\{ \begin{array}{l}
\WlC_{\lR\lR}^{ +  + } =  - \dfrac{{{g_Z^2}}}{{2{M^2}}}({\hyp}'_2)^2\\
\WlC_{\qR\qR}^{\left( 1 \right) +  + } =  - \dfrac{{{g_Z^2}}}{{2{M^2}}}({\hyp}'_1)^2\\
\WlC_{\lR\qR}^{\left( 1 \right) +  + } =  - \dfrac{{{g_Z^2}}}{{2{M^2}}}2{{\hyp}'_1}{{\hyp}'_2}
\end{array} \right.  \,,\quad
\left\{ \begin{array}{l}
\WlC_{l\lR}^ +  =  - \dfrac{{{g_Z^2}}}{{2{M^2}}}2({\hyp}'_2)^2\\
\WlC_{l\qR}^ +  =  - \dfrac{{{g_Z^2}}}{{2{M^2}}}2{{\hyp}'_1}{{\hyp}'_2}\\
\WlC_{q\lR}^ +  =  - \dfrac{{{g_Z^2}}}{{2{M^2}}}2{{\hyp}'_1}{{\hyp}'_2}\\
\WlC_{q\qR}^{\left( 1 \right) + } =  - \dfrac{{{g_Z^2}}}{{2{M^2}}}2({\hyp}'_1)^2
\end{array} \right. \,.
\end{equation}
No \qvcolor~or \lvcolor~operators in our \cref{tbl:nuSMEFTCBasis} is generated, consistent with what we expected from the UV physics.

\subsection{Heavy $W'$s and $Z'$s from a UV  theory with $SU(2)_L \times SU(2)_R \times U(1)_{B-L}$}
\label{subsec:UV221}

In this section, we consider a simple custodial symmetric UV embedding of the electroweak sector, by promoting the electroweak gauge symmetry to $SU(2)_L \times SU(2)_R \times U(1)_{B-L}$, which we hence refer to as the 2-2-1 model. The covariant derivative is now
\begin{equation}
D_\mu = \partial_\mu - i g W_\mu^a t^a - i g_R R_\mu^a t_R^a - i g_K {\hyp}' K_\mu \,.
\end{equation}
with $R_\mu^a, K_\mu$ the gauge bosons and $t_R^a, {\hyp}'$ the corresponding generators for $SU(2)_R$ and $U(1)_{B-L}$. Note that the gauge coupling $g_R$ forces the three different $SU(2)_R$ symmetries in $\nu$SM to be the same; or in other words, it breaks them down to the diagonal subgroup of $SU(2)_{RH}\times SU(2)_{R\qR}\times SU(2)_{R\lR}$.

In order to break the enlarged symmetry $SU(2)_L \times SU(2)_R \times U(1)_{B-L}$ down to electroweak symmetry at low energy, we introduce a new heavy scalar field $\Phi$, which is an $SU(2)_R$-doublet with ${\hyp}'_\Phi=\frac{1}{2}$ and $SU(2)_L$-singlet. Upon acquiring a vev
\begin{equation}
\Phi \supset \frac{1}{\sqrt 2}\mqty( 0 \\ v_\phi ) \,,
\end{equation}
it breaks $SU(2)_R \times U(1)_{B-L}$ to $U(1)_Y$, with the hypercharge $\hyp=t_R^3+{\hyp}'$.\footnote{The story is completely in parallel with how the SM Higgs $H$ breaks $SU(2)_L \times U(1)_Y$ to $U(1)_{EM}$, with electric charge $Q=t^3+\hyp$.} In this example, the custodial symmetry is an exact symmetry respected by the UV theory at the high energy scale. However, it is spontaneously broken at the scale $v_\phi$. Once we integrate out the heavy gauge bosons and $\Phi$, this $v_\phi$ gives rise to (all) the custodial violating effects in the resulting SMEFT, putting the hypercharge part of the dim-4 custodial violations and those at higher mass dimensions onto the same footing. This is in analogy with the case of MFV \cite{DAmbrosio:2002vsn}.\footnote{Note that this example would not account for the Yukawa-induced custodial violation in the SM\@.}

The UV sector in this example is
\begin{equation}
\Lag_\text{UV} \supset - \frac{1}{4}W_{\mu \nu }^a{W^{a,\mu \nu }} - \frac{1}{4}R_{\mu \nu }^a{R^{a,\mu \nu }} - \frac{1}{4}{K_{\mu \nu }}{K^{\mu \nu }} + {\left| {D\Phi } \right|^2} - {V_\Phi } + {\left| {DH} \right|^2} + \bar \psi i\slashed D\psi \,.
\label{eqn:LUV221}
\end{equation}
Here we have switched off any possible interactions between $\Phi$ and $H$ for simplicity, and hence focus on the effects of integrating out the heavy gauge bosons. After the symmetry breaking, we can identify the mass eigenstates of the gauge bosons
\begin{equation}
\left( {R_\mu ^a,{K_\mu }} \right) \to \left( {R_\mu ^ \pm ,{X_\mu },{B_\mu }} \right) \,,
\end{equation}
among which $B_\mu$ remains massless, but $R_\mu^\pm$ and $X_\mu$ obtain masses
\begin{subequations}\label{eqn:mRmX}
\begin{align}
m_R^2 &= \frac{1}{4}g_R^2v_\phi^2 \,, \\[5pt]
m_X^2 &= \frac{1}{4} \left( g_R^2 + g_K^2 \right)v_\phi^2 \,.
\end{align}
\end{subequations}
We then integrate out these heavy gauge bosons (together with the heavy scalar $\Phi$) at tree level, and obtain the EFT Lagrangian up to dim-6
\begin{align}
\Lag_\text{EFT} &= {\Lag_{{\text{SM}}}} + \frac{{g_R^2}}{{2m_R^2}}\left[ \left( iD_\text{SM}^\mu {\tilde H}^\dag \right) H + \sum\limits_{\psi=\qR,\lR} \bar\psi \gamma^\mu t_R^- \psi \right] \left[ H^\dag \left( iD_{\text{SM,}\mu}\tilde H \right) - \sum\limits_{\psi=\qR,\lR} \bar\psi \gamma_\mu t_R^+ \psi \right] \notag\\[5pt]
&\hspace{35pt} -\frac{{g_R^2}}{{2m_X^2c_R^2}}\left[ {\frac{{c_R^2}}{2}\left( {{H^\dag }i\lrD_{{\text{SM}}}^\mu H} \right) + \sum\limits_{\psi=\qR,\lR} \bar \psi {\gamma ^\mu }\left( {t_R^3 - s_R^2\hyp} \right)\psi } \right] \notag\\[3pt]
&\hspace{100pt} \times \left[ {\frac{{c_R^2}}{2}\left( H^\dag i{\lrD}_{\text{SM,}\mu} H \right) + \sum\limits_{\psi=\qR,\lR} \bar \psi {\gamma _\mu }\left( {t_R^3 - s_R^2\hyp} \right)\psi } \right] \,.
\label{eqn:LEFT221}
\end{align}
Here the mixing angle is defined as usual
\begin{equation}
c_R = \cos\theta_R \equiv \frac{g_R}{\sqrt{g_R^2+g_K^2}} \,,
\label{eqn:cRdef}
\end{equation}
and the SM gauge coupling for hypercharge is recovered as
\begin{equation}
g_1^2 = \frac{g_R^2 g_K^2}{g_R^2+g_K^2} \,.
\label{eqn:g1221}
\end{equation}

We see from the result in \cref{eqn:LEFT221} that there are generically custodial violating operators, such as $Q_{HD}$ appearing in the following combinations
\begin{subequations}
\begin{align}
\left[ \left( iD_\text{SM}^\mu {\tilde H}^\dagger \right) H \right] \left[ H^\dagger \left( iD_{\text{SM,}\mu} {\tilde H} \right) \right] &= Q_{HD} - Q_R \,, \\[5pt]
\left( H^\dag i\lrD_\text{SM}^\mu H \right) \left( H^\dag i\lrD_{\text{SM,}\mu} H \right) &= Q_{H\Box} + 4 Q_{HD} \,.
\end{align}
\end{subequations}
This is simply a reflection that the $SU(2)_R$ is spontaneously broken by $v_\phi$. Next, we carry out the standard routine of expanding the EFT Lagrangian, trading operators outside of our desired basis (such as $Q_R$ above) into the Warsaw basis, reading off the Wilson coefficients $C_i$ and translating them into our custodial basis $\WlC_i$. The end result contains a large set of Wilson coefficients. The coefficients relevant for computing our observables in \cref{eqn:Obs4nuSMEFTCBasis} are:
\begin{subequations}\label{eqn:221general}
\begin{align}
\WlC_{HD} = -\12\, \WlC_{H\lR}^{(1)-} &= \frac{1}{2v_\phi^2} \left( 1 - c_R^4 \right) \,, \\[5pt]
\WlC_{Hl}^{(1)} = \WlC_{H\lR}^{(1)+} &= \frac{1}{v_\phi^2}\, s_R^2\, c_R^2 \,, \\[5pt]
\WlC_{H\lR}^{(3)+} &= \frac{1}{v_\phi^2} \,.
\end{align}
\end{subequations}

First we notice the appearance of $\WlC_{H\lR}^{(3)+}$. It is generated because in the 2-2-1 model, the gauging of $SU(2)_R$ reduces the three independent global $SU(2)_{RH} \times SU(2)_{R\qR} \times SU(2)_{R\lR}$ down to one single gauged $SU(2)_R$. In fact, $\WlC_{H\qR}^{(3)+}$ and $\WlC_{\lR\qR}^{(3)++}$ are generated as well, but not listed above as they do not enter into our observables.

Next, a non-zero $\WlC_{HD}$ and $\WlC_{Hl}^{(1)}$ indeed means that the UV theory violates custodial symmetry. Plugging \cref{eqn:221general} into \cref{eqn:Toml}, our $\Tom_l$ parameter serves as a good indicator of the custodial symmetry/violation:
\begin{equation}
\alpha\Tom_l = - 2 v^2 \left[ a_{HD} - a_{Hl}^{(1)} \right] = - \frac{v^2}{v_\phi^2}\, s_R^4 \,\,.
\end{equation}
Clearly, $\Tom_l$ is generically nonzero in the 2-2-1 model.  However, fixing $g_1$ and recalling the relations between $g_1$ and the primordial couplings $g_R, g_K$ in \cref{eqn:cRdef,eqn:g1221}, there are two interesting limits: $g_R \gg g_K$ and $g_R \ll g_K$.
\begin{itemize}
  \item In the limit $g_R \gg g_K$, we have $g_K \to g_1$, $g_R \gg g_1$, and $s_R \to 0$. Then, $\Tom_l \to 0$ in this limit. More specifically, the behavior of the custodial violating operators in $\Tom_l$ are $\WlC_{HD}\,, \WlC_{Hl}^{(1)} \to \frac{g_1^2}{v_\phi^2 \, g_R^2} \to 0$. Not only are they small but also proportional to the SM hypercharge coupling $g_1$. This limit is asymptotically custodial symmetric.
  \item In the limit of $g_R \ll g_K$, we have $g_R \to g_1$, $g_K\gg g_1$, and $s_R \to 1$. We see that $\alpha\Tom_l \to -\frac{v^2}{v_\phi^2}$ approaching its maximum size allowed. More specifically, the Wilson coefficient $\WlC_{Hl}^{(1)} \to \frac{g_1^2}{v_\phi^2 \, g_K^2} \to 0$, while $\WlC_{HD} \to \frac{1}{2 v_\phi^2} \ne 0$. One may naively think this limit ought to work precisely as the $Z'_{B-L}$ boson model discussed in \cref{subsec:UVBL}. However, this is NOT the case. Although $g_R$ is small compared to $g_K$, it has not been completely switched off, and the EFT does not necessarily imply a light $m_R$, because $g_1 v_\phi$ should be viewed as parametrically larger compared to electroweak scale (as $v_\phi \gg v$). So in this limit, we actually decouple $m_X$ instead of $m_R$, resulting in a custodial violating UV theory, as indicated by the nonzero $\Tom_l$ parameter.
\end{itemize}

\subsection{Heavy vector-like fermions}
\label{subsec:UVVLF}

In this section, we illustrate an example of integrating out a UV sector with heavy vector-like fermions that interact with the Standard Model charged leptons and neutrinos \cite{delAguila:2000rc, delAguila:2008pw, Crivellin:2020ebi}. Such a UV model does not belong to the category of universal theories \cite{Barbieri:2004qk, Wells:2015uba, Wells:2015cre}, thus the oblique assumption is not admissible. Below, we will see explicitly that the Peskin-Takeuchi $T$ parameter fails to detect the hard custodial violation in the UV interactions, while our new $\Tom_l$ parameter works perfectly.

Consider a UV model with two vector-like fermions $N$ and $E$ that are SM $SU(2)_L$-singlets. They share a common mass $M\gg v$ and interact with the SM in the same way as the $\nu$SM right-handed leptons $\nu$ and $e$,
\begin{equation}
\Lag_\text{UV} = \Lag_\text{SM} + {\bar N} (i\slashed{D} - M) N + {\bar E} (i\slashed{D} - M) E - \left( Y_N\, {\bar l} \tilde{H} N + Y_E\, {\bar l} H E + \hc \right) \,.
\label{eqn:LUVVF1}
\end{equation}
The new UV Yukawa interactions can be rewritten following the same way shown in \cref{eqn:YukawaCombine}:
\begin{equation}
Y_N\, \bar{l} {\tilde H} N + Y_E\, \bar{l} H E  = \bar{l}\, \Sigma\, \mqty( Y_N & 0 \\ 0 & Y_E )\, \mqty( N \\ E ) \,.
\end{equation}
We see that if $|Y_N| = |Y_E|$, the $SU(2)_{RH}$ symmetry can be preserved by the UV sector\footnote{The phase mismatch between $Y_N$ and $Y_E$ can be absorbed by redefining the field $N$ or $E$.} in the limit $g_1\to 0$ ($N$ and $E$ have different hypercharges). In this case, the UV sector is custodial symmetric. Otherwise, it has hard custodial violation. Let us now check if our $\Tom_l$ parameter can distinguish these two scenarios.

Integrating the heavy vector-like fermions out at tree level, we obtain a SMEFT Lagrangian at dim-6 as\footnote{As is well known, the dim-5 ``neutrino mass'' operator is also generated by this UV theory, but it is irrelevant for our current discussion.}
\begin{subequations}
\begin{align}
\label{eqn:LEFTVF1}
\Lag_\text{SMEFT} \supset \left( Y_N \bar l \tilde{H} \right) \frac{i\slashed{D}}{M^2} \left( Y_N^* \tilde{H}^\dag l \right) + \left( Y_E \bar l H \right) \frac{i\slashed{D}}{M^2} \left( Y_E^* H^\dag l \right)\,.
\end{align}
\end{subequations}
Expanding this SMEFT Lagrangian and trading operators into Warsaw and custodial basis, we obtain the Wilson coefficients
\begin{subequations}
\begin{align}
\WlC_{Hl}^{(3)} &= - \frac{1}{4M^2} \left( |Y_N|^2 + |Y_E|^2 \right) \,, \\[5pt]
\WlC_{Hl}^{(1)} &= - \frac{1}{4M^2} \left( |Y_N|^2 - |Y_E|^2 \right) \,, \\[5pt]
\WlC_{lH}^\pm &= \frac{1}{8M^2} \left( Y_\nu |Y_N|^2 \pm Y_e |Y_E|^2 \right) \,.
\end{align}
\end{subequations}
Plugging these into \cref{eqn:Toml}, we get
\begin{equation}
\alpha \Tom_l = - 2 v^2 \left[ \WlC_{HD} - \WlC_{Hl}^{(1)} \right] = - \frac{v^2}{2 M^2} \left( |Y_N|^2 - |Y_E|^2 \right) \,.
\end{equation}
We see that indeed our $\Tom_l$ parameter vanishes only if $\left|Y_N\right|=\left|Y_E\right|$, and does not vanish in general. Thus, $\Tom_l$ serves as a perfect indicator of the UV custodial violation. In addition, we notice that in this example, the SMEFT framework captures the UV custodial violation through the Wilson coefficient $\WlC_{Hl}^{(1)}$, a non-oblique correction, while $\WlC_{HD}=0$. Therefore, the Peskin-Takeuchi $T$ parameter fails to capture custodial violation in this UV theory. Explicitly,
\begin{equation}
\alpha T = - \12 v^2\, C_{HD} = - 2 v^2\, \WlC_{HD} = 0 \,.
\end{equation}
This example demonstrates the utility of our new $\Tom_l$ parameter for indicating both oblique as well as non-oblique custodial violation arising in $(\nu)$SMEFT.

\section{Discussion}
\label{sec:Discussion}

We have investigated how to faithfully detect hard custodial symmetry/violation in the UV physics beyond the SM, where hard refers to violations that persist in the limit of vanishing $U(1)_Y$ gauge coupling $g_1\to 0$. Working with dim-6 ($\nu$)SMEFT, we introduced a new basis---the custodial basis, which is simply a rewriting of the Warsaw basis operators to make manifest the symmetric/breaking structures of the various $SU(2)_R$ symmetries in ($\nu$)SM\@. This custodial basis facilitates the recognition of operators that can/cannot be generated at tree level by custodial symmetric UV physics. With the help of electroweak precision observables, we then identified several example RPI combinations of dim-6 SMEFT Wilson coefficients $\Tom_l$ (as well as $\Tom_q$ and $\Tom_{\qR}$, in \cref{appsec:Hadronic})  that serve as generalizations of the Peskin-Takeuchi $T$ parameter to accommodate non-oblique corrections from general UV physics.

Given measurements of $\hat\alpha,\, \hat{G}_F,\, \hat{m}_Z^2$, we showed that the electroweak precision observables
\begin{equation}
\Big\{
\hat\rho,\,
{\hat r}_{Z{\nu _L}{{\bar\nu}_L}},\,
{\hat r}_{Z{e_L}{{\bar e}_L}} \Big\}
\label{eqn:observablesindiscussion}
\end{equation}
can be used to construct $\Tom_l$ [see \cref{eqn:Toml}]:
\begin{align}
&\left( \hat\rho - 1 \right) + \12 \left( {\hat r}_{Z \nu_L {\bar\nu}_L} - 1 \right) - \12 c_{2\theta} \left( {\hat r}_{Z e_L {\bar e}_L} - 1 \right) \notag\\[5pt]
&\hspace{40pt} = - \12 v^2 \left[ C_{HD} + 4\, C_{Hl}^{(1)} \right] = - 2 v^2 \left[ \WlC_{HD} - \WlC_{Hl}^{(1)} \right] \;\equiv\; \alpha \Tom_l \,\,.
\end{align}
The measurement $\Tom_l \ne 0$ implies the UV sector violates custodial symmetry at tree-level. Importantly, the converse is not true: $\Tom_l = 0$ does \emph{not} immediately imply no custodial violation in the UV sector. There are several exceptions that we have highlighted throughout the paper. For example, our observable has been demonstrated to capture just hard breaking of custodial symmetry, and is not sensitive to soft custodial violations arising from the gauging of hypercharge. In addition, $\Tom_l$ is unable to rule out the accidental cancellation $a_{HD} = a_{Hl}^{(1)}$. Furthermore, we have emphasized in \cref{sec:EOM} that $\Tom_l$ is not sensitive to all custodial violating ($\nu$)SMEFT operators, e.g., it is not sensitive to $\Op_{lH}^{-}$ or $\Op_{qH}^{-}$. As we argued in \cref{sec:EOM}, this is a good thing since they may be faked by the EOM redundancy in rewriting custodial symmetric operators outside our custodial basis. Finally, our $\Tom_l$ is also not sensitive to custodial violation that appears only at loop level at leading matching order. Here we should distinguish between two possibilities:  there are well-known loop corrections to our observables purely from the SM physics, such as the contribution to $\hat{\rho}$ from the custodial-violating difference between the top and bottom quark Yukawa couplings.  These effects could be incorporated into the framework by redefining our observables to include the SM loop effects (e.g. the PDG provides a prescription to do this for the Veltman $\rho$ parameter \cite{PDG}). However, additional contributions to our observables that arise from radiative corrections from ($\nu$)SMEFT operators are not included.  For some theories, radiative corrections are known, for example the singlet scalar model \cite{Jiang:2018pbd,Haisch:2020ahr}. In future work we will investigate if there are persistent patterns that bely a UV theory with custodial symmetry even after radiative corrections are included.

Following the same logic used to construct $\Tom_l$, one can use the hadronic \emph{pseudo-observables} discussed in \cref{appsec:Hadronic} to construct two additional parameters, $\Tom_q$ and $\Tom_{\qR}$ (see \cref{eqn:Tomq,eqn:TomqR}):
\begin{subequations}
\begin{align}
&\hspace{-10pt} (\hat\rho - 1) - \12 (3 - 4 s_\theta^2) ({\hat r}_{Zu_L{\bar u}_L} - 1) + \12 (3 - 2 s_\theta^2) ({\hat r}_{Zd_L{\bar d}_L} - 1) \notag\\[5pt]
&\hspace{20pt} = -\12 v^2 \left[ C_{HD} - 12\, C_{Hq}^{(1)} \right] = -2v^2 \left[ a_{HD} + 3\, a_{Hq}^{(1)} \right] \;\equiv\; \alpha\Tom_q \,, \\[15pt]
&\hspace{-10pt} (\hat\rho - 1) + 2s_\theta^2 ({\hat r}_{Zu{\bar u}} - 1) - s_\theta^2 ({\hat r}_{Zd{\bar d}} - 1) \notag\\[5pt]
&\hspace{20pt} = -\12 v^2 \bigg[ C_{HD} - 6 \left( C_{Hu} + C_{Hd} \right) \bigg] = -2v^2 \left[ a_{HD} + 3\, a_{H\qR}^{(1)+} + 3\, a_{H\qR}^{(3)-} \right] \;\equiv\; \alpha\Tom_{\qR} \,.
\end{align}
\end{subequations}
Of course we cannot separately measure the partial widths into left-handed or right-handed up and down quarks, so unfortunately there is no actual utility of these results.

We demonstrated the viability and usefulness of our results by calculating $\Tom_l$ for several example UV theories. In some cases, the result is trivial. For example for the heavy $Z'$ associated with $U(1)_{B-L}$ [\cref{subsec:UVBL}], the prediction is $\Tom_l = 0$, and more specifically $\hat \rho = {\hat r}_{Z{\nu _L}{{\bar \nu }_L}} = {\hat r}_{Z{e_L}{{\bar e}_L}} = {\hat r}_{Ze\bar e} = 1$. By itself, this is uninformative, since predicting these observables do not deviate from unity is indistinguishable from the SM\@. However, when combined with other observables that deviate from the SM prediction, e.g., a new/modified four-fermion interaction, measuring $\Tom_l$ consistent with zero provides evidence that the UV physics is custodial symmetric and consistent with a $U(1)_{B-L}$ interpretation. Similar arguments applies to the heavy $W'_L$ boson example [\cref{subsec:Wprime}], which is also custodial symmetric. In addition, this example also reminded us that the Veltman $\hat\rho\ne 1$ is possible for custodial symmetric UV physics.

We also considered UV sectors that (generically) possess hard custodial violations. In the real $SU(2)_L$-triplet scalar example [\cref{subsec:Triplet}], our $\Tom_l$ works exactly the same as the Peskin-Takeuchi $T$ parameter. Both parameters indicate the presence of hard custodial violation in the UV sector. In the 2-2-1 example [\cref{subsec:UV221}], we embedded the SM into a larger gauge symmetry, $SU(2)_L \times SU(2)_R \times U(1)_{B-L}$. In this case, the gauging of all the three global symmetries $SU(2)_{RH} \times SU(2)_{R\qR} \times SU(2)_{R\lR}$ marries them into one single $SU(2)_R$. Then the spontaneous breaking of $SU(2)_R \times U(1)_{B-L} \rightarrow U(1)_Y$ down to the SM generically leads to hard custodial violation, where we found that $\Tom_l$ is in general nonzero. Finally, the heavy vector-like fermions [\cref{subsec:UVVLF}] is a striking example where the UV physics is not a  ``universal'' theory and \emph{requires} our replacement parameter $\Tom_l$. Hard custodial violation feeds into SMEFT via the non-oblique correction $\WlC_{lH}^{(1)}$ (but not $\WlC_{HD}$). As a result, the Peskin-Takeuchi $T$ parameter fails to detect it, but our new $\Tom_l$ parameter works perfectly.

In this paper, we have assumed flavor universality in constructing our $\Tom_f$ parameters. One can certainly generalize our analysis in \cref{sec:Observables,appsec:Hadronic} to include flavor-dependent deviations to the observables. This would allow for a construction of flavor sensitive $\Tom_f$ parameters, which could be used for probing non-trivial flavor structure in the UV custodial violation. The SM Yukawa couplings are examples of flavor-dependent custodial violation. Unlike the hypercharge coupling, they have no direct linkage with a general UV sector. However, for UV theories with ``minimal flavor violation'' \cite{DAmbrosio:2002vsn}, couplings in the UV sector are proportional to (powers of) the SM Yukawa couplings. Such UV sectors are necessarily custodial violating (as well as flavor violating). Nevertheless, with the aforementioned generalization, our $\Tom_f$ parameters are capable of capturing this custodial violation, provided that custodial violating dim-6 operators (beyond those in ``class 5'' of our \cref{tbl:nuSMEFTCBasis}) are generated at tree level.

\section*{Acknowledgments}

We thank Spencer Chang for very helpful discussions as this work was being completed. The work of GDK, XL, and TT was supported in part by the U.S. Department of Energy under Grant Number DE-SC0011640. The work of AM was supported in part by the National Science Foundation under Grant Number PHY-1820860.

%----------------------------------------------------------------------
\appendix

\section{Details of Mapping onto Observables}
\label{appsec:Mapping}

In this Appendix, we provide some details on the intermediate steps that lead to our results in \cref{eqn:Obs4nuSMEFTWarsaw}. We work with the Warsaw basis of dim-6 $\nu$SMEFT shown in \cref{tbl:nuSMEFTWarsaw}, assuming flavor universality. We will perform tree-level mapping, and only up to dim-6.

First, we find the corrections to the two-point functions of electroweak gauge bosons
\begin{subequations}\label{eqn:2ptPiVV}
\begin{align}
\Pi_{WW} \left( p^2 \right) &= 2 p^2 v^2\, C_{HW} \,, \\[8pt]
\Pi_{ZZ} \left( p^2 \right) &= \12 {\hat m}_{Z\text{, SM}}^2 v^2\, C_{HD} + 2 p^2 v^2 \left( c_\theta^2\, C_{HW} + s_\theta^2\, C_{HB} + c_\theta s_\theta\, C_{HWB} \right) \,, \\[8pt]
\Pi_{\gamma\gamma}\left( p^2 \right) &= 2 p^2 v^2\left( s_\theta^2\, C_{HW} + c_\theta^2\, C_{HB} - c_\theta s_\theta\, C_{HWB} \right) \,, \\[8pt]
\Pi_{\gamma Z} \left( p^2 \right) &= p^2 v^2 \left[ 2 c_\theta s_\theta \left( C_{HW} - C_{HB} \right) - \left( c_\theta^2 - s_\theta^2 \right) C_{HWB} \right] \,,
\end{align}
\end{subequations}
where as usual $\Pi_{VV}\left(p^2\right)$ denotes the transverse part of the full two-point function of the gauge bosons:
\begin{equation}
i\Pi _{VV}^{\mu \nu } \left(p^2\right) = i\Pi_{VV}\left(p^2\right) \left( {{\eta ^{\mu \nu }} - \frac{{{p^\mu }{p^\nu }}}{{{p^2}}}} \right) + \left( {i\frac{{{p^\mu }{p^\nu }}}{{{p^2}}}}\,\, \text{term} \right) \,.
\end{equation}
Next, we move on to the three-point vertices. For the observables considered in \cref{sec:Observables}, the relevant vertex corrections between the electroweak gauge bosons and the leptons are
\begin{subequations}\label{eqn:3ptVll}
\begin{align}
{V_{Z{\nu_L}{\bar\nu}_L}} &= 1 - {v^2} \bigg[ C_{Hl}^{(1)} - C_{Hl}^{(3)} \bigg] \,, \\[5pt]
{V_{Z{e_L}{{\bar e}_L}}} &= 1 + \frac{v^2}{c_{2\theta}} \bigg[ C_{Hl}^{(1)} + C_{Hl}^{(3)} \bigg] \,, \\[5pt]
{V_{Ze\bar e}} &= 1 - \frac{v^2}{{2s_\theta ^2}}\, C_{He} \,, \\[5pt]
{V_{Wl\bar l}} &= 1 + v^2\, C_{Hl}^{\left( 3 \right)} \,.
\end{align}
\end{subequations}
Note that corrections to the four-fermion vertices would not feed into $\hat\alpha$ due to lack of pole structure. The only four-fermion vertex needs to be considered in our analysis is $C_{12}$ mentioned in \cref{subsec:a12}, which will feed into $\hat{G}_F$.

With the above, we would like to find the modifications to \cref{eqn:Obs7SM}. The first four observables are relatively simple:
\begin{subequations}\label{eqn:First4Rev}
\begin{align}
\hat \alpha &= \frac{{g_1^2g_2^2}}{{4\pi \left( {g_1^2 + g_2^2} \right)}}\left[ {{{\left. {\frac{{{p^2}}}{{{p^2} - {\Pi _{\gamma \gamma }}\left( {{p^2}} \right)}}} \right|}_{{p^2} \to 0}}} \right] \notag\\[5pt]
 &= {{\hat\alpha}_\text{SM}} \bigg[ 1 + 2{v^2}\left( s_\theta^2\, C_{HW} + c_\theta^2\, C_{HB} - c_\theta s_\theta\, C_{HWB} \right) \bigg] \,, \\[10pt]
{{\hat G}_F} &= \frac{{\sqrt 2 g_2^2}}{8}V_{Wl\bar l}^2\left[ {{{\left. {\frac{{ - 1}}{{{p^2} - {\hat m}_{W{\text{, SM}}}^2 - {\Pi _{WW}}\left( {{p^2}} \right)}}} \right|}_{{p^2} \to 0}}} \right] - \frac{1}{2\sqrt{2}}\, C_{12} \notag\\[3pt]
 &= {{\hat G}_{F{\text{, SM}}}}\left[ 1 + 2v^2\, C_{Hl}^{(3)} - \12 v^2\, C_{12} \right] \,, \\[10pt]
\hat m_Z^2 &= {\hat m}_{Z{\text{, SM}}}^2 + {\Pi _{ZZ}}\left( {{\hat m}_{Z{\text{, SM}}}^2} \right) \notag\\[5pt]
 &= \hat m_{Z{\text{, SM}}}^2 \left[ 1 + \12 v^2\, C_{HD} + 2 v^2 \left( c_\theta^2\, C_{HW} + s_\theta^2\, C_{HB} + c_\theta s_\theta\, C_{HWB} \right) \right] \,, \\[10pt]
\hat m_W^2 &= {\hat m}_{W{\text{, SM}}}^2 + \Pi_{WW} \left( {\hat m}_{W{\text{, SM}}}^2 \right) = \hat m_{W{\text{, SM}}}^2 \left( 1 + 2{v^2}\, C_{HW} \right) \,.
\end{align}
\end{subequations}
These will lead us to the $\hat\rho$ expression in \cref{eqn:Obs4nuSMEFTWarsaw}.

For the decay widths in \cref{eqn:Obs7SM}, we need a bit more setup. We define the amplitude $i\hat M$ as the strength $\hat\kappa$ multiplied by the polarization kinematics:
\begin{equation}
i{{\hat M}_{Z\psi \bar \psi }} \equiv i\hat \kappa \left( {\epsilon_\mu {{\bar u}_\psi }{\gamma ^\mu }{P_{L/R}}{v_{\bar \psi }}} \right) \,,
\end{equation}
with $\epsilon_\mu$ denoting the polarization vectors for $Z$ boson, $u$ and $v$ denoting the Dirac spinors for the fermion legs, and $P_{L/R}=\frac{1\mp\gamma^5}{2}$ denoting the projector depending on the chirality of the fermion $\psi$. With this, one can compute the decay width
\begin{equation}
{{\hat \Gamma }_{Z\psi \bar \psi }} = \frac{1}{{16\pi {{\hat m}_Z}}}\overline {{{\left| {{{\hat M}_{Z\psi \bar \psi }}} \right|}^2}}  = \frac{{{{\hat m}_Z}}}{{24\pi }}{{\hat \kappa }^2} \,,
\end{equation}
where fermion masses are neglected. The $\hat r$ observables defined in \cref{eqn:Obs4Def} can then be expressed as
\begin{subequations}\label{eqn:rReExpress}
\begin{align}
{{\hat r}_{Z{\nu _L}{{\bar \nu }_L}}} &= \frac{{\hat \kappa _{Z{\nu _L}{{\bar \nu }_L}}^2}}{{\sqrt 2 {{\hat G}_F}\hat m_Z^2}} \,, \\[5pt]
{{\hat r}_{Z{e_L}{{\bar e}_L}}} &= \frac{{\hat \kappa _{Z{e_L}{{\bar e}_L}}^2}}{{\sqrt 2 {{\hat G}_F}\hat m_Z^2\left( {1 - \hat x} \right)}} \,, \\[5pt]
{{\hat r}_{Ze\bar e}} &= \frac{{\hat \kappa _{Ze\bar e}^2}}{{\sqrt 2 {{\hat G}_F}\hat m_Z^2{{\left( {1 - \sqrt {1 - \hat x} } \right)}^2}}} \,,
\end{align}
\end{subequations}
where $\hat x$ is defined as before by \cref{eqn:x}. In SM, these ratios are unity. In SMEFT dim-6 Warsaw basis, the above strengths are modified
\begin{subequations}\label{eqn:Last3Rev}
\begin{align}
\hspace{-5pt} {\hat \kappa }_{Z{\nu _L}{{\bar \nu }_L}} &= \hat\kappa _{Z{\nu _L}{{\bar \nu }_L}{\text{, SM}}} \left( R_Z \right)^{1/2}\, V_{Z{\nu _L}{{\bar \nu }_L}} \notag\\[5pt]
 &= {\hat\kappa _{Z{\nu _L}{{\bar \nu }_L}{\text{, SM}}}}\Bigg[ 1 + v^2 \left( c_\theta^2\, C_{HW} + s_\theta^2\, C_{HB} + c_\theta s_\theta\, C_{HWB} \right) - v^2 \left( C_{Hl}^{(1)} - C_{Hl}^{(3)} \right) \Bigg] \,, \\[15pt]
{\hat\kappa}_{Z{e_L}{{\bar e}_L}} &= \hat\kappa _{Z{e_L}{{\bar e}_L}{\text{, SM}}} \left( {{R_Z}} \right)^{1/2} \left[ V_{Z{e_L}{{\bar e}_L}} + \frac{s_{2\theta}}{c_{2\theta}}\, \frac{1}{p^2}\, \Pi_{\gamma Z} \left( p^2 \right) \right] \notag\\[5pt]
 &= \hat\kappa_{Z{e_L}{{\bar e}_L}{\text{, SM}}} \Bigg[ 1 + v^2 \left( c_\theta^2\, C_{HW} + s_\theta^2\, C_{HB} + c_\theta s_\theta\, C_{HWB} \right) + v^2 \frac{1}{c_{2\theta}} \left( C_{Hl}^{(1)} + C_{Hl}^{(3)} \right) \notag\\
 &\hspace{100pt} + v^2 \frac{s_{2\theta}^2}{c_{2\theta}} \left( C_{HW} - C_{HB} \right) - 2 v^2 c_\theta s_\theta\, C_{HWB} \Bigg] \,, \\[15pt]
{\hat \kappa }_{Ze\bar e} &= \hat\kappa_{Ze\bar e{\text{, SM}}} \left( R_Z \right)^{1/2} \left[ V_{Ze\bar e} - \frac{c_\theta}{s_\theta }\, \frac{1}{p^2}\, \Pi_{\gamma Z} \left( p^2 \right) \right] \notag\\[5pt]
 &= \hat\kappa _{Ze\bar e{\text{, SM}}} \Bigg[ 1 + v^2 \left( c_\theta^2\, C_{HW} + s_\theta^2\, C_{HB} + c_\theta s_\theta\, C_{HWB} \right) - v^2 \frac{1}{2s_\theta^2}\, C_{He} \notag\\
 &\hspace{100pt} - v^2\, 2c_\theta^2 \left( C_{HW} - C_{HB} \right) + v^2\, \frac{c_\theta}{s_\theta}\, c_{2\theta}\, C_{HWB} \Bigg] \,,
\end{align}
\end{subequations}
where $R_Z$ is the residue of the $Z$ boson at the pole mass:
\begin{align}
\label{eqn:Residue}
R_Z &= 1 + \left. \left[ \frac{d}{dp^2} \Pi_{ZZ} \left( p^2 \right) \right] \right|_{{p^2} = \hat m_{Z{\text{, SM}}}^2} = 1 + 2 v^2 \left( c_\theta^2\, C_{HW} + s_\theta^2\, C_{HB} + c_\theta s_\theta\, C_{HWB} \right) \,.
\end{align}
Plugging \cref{eqn:Last3Rev} and \cref{eqn:First4Rev} into \cref{eqn:rReExpress} will lead us to the expressions for the partial widths in \cref{eqn:Obs4nuSMEFTWarsaw}.

\section{Hadronic pseudo-Observables}
\label{appsec:Hadronic}

In this Appendix, we consider a set of four quark partial widths \emph{pseudo-observables} in addition to those listed in \cref{eqn:ExampleObservables}:
\begin{equation}
\left\{
\hat\Gamma_{Zu_L\bar{u}_L},\,
\hat\Gamma_{Zd_L\bar{d}_L},\,
\hat\Gamma_{Zu\bar{u}},\,
\hat\Gamma_{Zd\bar{d}} \right\} \,.
\label{eqn:hadronicobservables}
\end{equation}
In order, these denote the partial decay widths of the $Z$ boson to left-handed up-type quarks, left-handed down-type quarks, right-handed up-type quarks, and right-handed down-type quarks. Note that in $Z$ decay measurements, the first two generations of quarks are essentially indistinguishable. The measurable observables in practice are $\hat\Gamma_{Zqq}$ (which will be needed in measuring $\hat\Gamma_{Z\nu_L\bar\nu_L}$, see discussions in \cref{subsec:Measurements}) and measurements involving the $b$-quark. For this reason, we refer to these hadronic partial widths of $Z$ as \emph{pseudo-observables} to distinguish them from the \emph{observables} discussed in \cref{sec:Observables}.

We present our results in terms of definite parity hadronic final states in order to most easily compare with the results in \cref{sec:Observables}. In SM they are given by the three Lagrangian parameters $g_1, g_2, v$:
\begin{subequations}\label{eqn:ObsHad}
\begin{align}
{{\hat \Gamma }_{Z{u _L}{{\bar u }_L}{\text{, SM}}}} &= \frac{{{{\hat m}_{Z{\text{, SM}}}}}}{{288\pi }}\frac{{g_2^2}}{{c_\theta ^2}} \left( 3 - 4s_\theta^2 \right)^2 \,, \\[8pt]
{{\hat \Gamma }_{Z{d _L}{{\bar d }_L}{\text{, SM}}}} &= \frac{{{{\hat m}_{Z{\text{, SM}}}}}}{{288\pi }}\frac{{g_2^2}}{{c_\theta ^2}} \left( 3 - 2s_\theta^2 \right)^2 \,, \\[8pt]
{{\hat \Gamma }_{Z{u}{{\bar u }}{\text{, SM}}}} &= \frac{{{{\hat m}_{Z{\text{, SM}}}}}}{{18\pi }}\frac{{g_2^2}}{{c_\theta ^2}} s_\theta^4 \,, \\[8pt]
{{\hat \Gamma }_{Z{d}{{\bar d }}{\text{, SM}}}} &= \frac{{{{\hat m}_{Z{\text{, SM}}}}}}{{72\pi }}\frac{{g_2^2}}{{c_\theta ^2}} s_\theta^4 \,.
\end{align}
\end{subequations}
We then construct the following ratios (similar with \cref{eqn:Obs4Def}) to keep track of the deviations from SM
\begin{subequations}
\begin{align}\label{eqn:ObsHadDef}
{{\hat r}_{Z{u _L}{{\bar u }_L}}} &\equiv \frac{{72\pi }}{{\sqrt 2 {{\hat G}_F}\hat m_Z^3 (1+2\sqrt{1-\hat x})^2}}\, {\hat \Gamma }_{Z{u _L}{{\bar u }_L}} \,, \\[8pt]
{{\hat r}_{Z{d _L}{{\bar d }_L}}} &\equiv \frac{{72\pi }}{{\sqrt 2 {{\hat G}_F}\hat m_Z^3 (2+\sqrt{1-\hat x})^2}}\, {\hat \Gamma }_{Z{d _L}{{\bar d }_L}} \,, \\[8pt]
{{\hat r}_{Z{u}{{\bar u}}}} &\equiv \frac{{18\pi }}{{\sqrt 2 {{\hat G}_F}\hat m_Z^3 (1-\sqrt{1-\hat x})^2}}\, {\hat \Gamma }_{Z{u}{{\bar u}}} \,, \\[8pt]
{{\hat r}_{Z{d}{{\bar d}}}} &\equiv \frac{{72\pi }}{{\sqrt 2 {{\hat G}_F}\hat m_Z^3 (1-\sqrt{1-\hat x})^2}}\, {\hat \Gamma }_{Z{d}{{\bar d}}} \,,
\end{align}
\end{subequations}
where $\hat x$ is defined as before by \cref{eqn:x}. These four ratios are unity in SM, but will get modified in SMEFT\@. Following the same procedure shown in \cref{appsec:Mapping}, we obtain their general Warsaw basis corrections as
\begin{subequations}\label{eqn:ObsHadWarsaw}
\begin{align}
{\hat r}_{Zu_L{\bar u}_L} &= 1 + \frac{v^2}{c_{2\theta}(3 - 4s_\theta^2)} \Bigg[ -8s_\theta^2 \left( \frac{c_\theta}{s_\theta}\, C_{HWB} + C_{Hl}^{(3)} - \frac14\, C_{12} \right) -6c_{2\theta} \left( C_{Hl}^{(3)} - C_{Hq}^{(3)} - \frac14\, C_{12} \right) \nn[3pt]
&\hspace{110pt} - \12 (3 - 2s_\theta^2)\, C_{HD} - 6c_{2\theta}\, C_{Hq}^{(1)} \Bigg] \,, \\[10pt]
{\hat r}_{Zd_L{\bar d}_L} &= 1 + \frac{v^2}{c_{2\theta}(3 - 2s_\theta^2)} \Bigg[ -4s_\theta^2 \left( \frac{c_\theta}{s_\theta}\, C_{HWB} + C_{Hl}^{(3)} - \frac14\, C_{12} \right) -6c_{2\theta} \left( C_{Hl}^{(3)} - C_{Hq}^{(3)} - \frac14\, C_{12} \right) \nn[3pt]
&\hspace{110pt} -\12 (3-4s_\theta^2)\, C_{HD} + 6c_{2\theta}\, C_{Hq}^{(1)} \Bigg] \,, \\[10pt]
{\hat r}_{Zu{\bar u}} &= 1 + \frac{v^2}{c_{2\theta}} \Bigg[ 2\left( \frac{c_\theta}{s_\theta}\, C_{HWB} + C_{Hl}^{(3)} - \frac14\, C_{12} \right) + \12\, C_{HD} + \frac{3c_{2\theta}}{{2s_\theta^2}}\, C_{Hu} \Bigg] \,, \\[10pt]
{\hat r}_{Zd{\bar d}} &= 1 + \frac{v^2}{c_{2\theta}} \Bigg[ 2\left( \frac{c_\theta}{s_\theta}\, C_{HWB} + C_{Hl}^{(3)} - \frac14\, C_{12} \right) + \12\, {C_{HD}} - \frac{3c_{2\theta}}{s_\theta^2}\, C_{Hd}  \Bigg] \,.
\end{align}
\end{subequations}
Although there are eight Wilson coefficients $C_i$ involved in the above, they only come with six different combinations. Furthermore, recall from \cref{eq:rhohat-warsaw} that we have
\begin{equation}
\hat \rho = 1 + \frac{v^2}{c_{2\theta}} \Bigg[ - 2 s_\theta^2 \left( \frac{c_\theta}{s_\theta}\, C_{HWB} + C_{Hl}^{(3)} - \frac14\, C_{12} \right) - \12 c_\theta^2\, C_{HD} \Bigg] \,.
\end{equation}
We see that in the three quantities $\left\{\hat\rho, {\hat r}_{Zu_L{\bar u}_L}, {\hat r}_{Zd_L{\bar d}_L}\right\}$, only two independent combinations of custodial preserving operators show up. Therefore, analogous to the procedure of constructing $\Tom_l$ from \cref{eqn:Obs4MFCV,eqn:thefingerprint,eqn:Toml}, we can construct a new $T$ parameter generalization $\Tom_q$ using $\hat\rho$ and the left-handed partial widths:
\begin{align}
&(\hat\rho - 1) - \12 (3 - 4 s_\theta^2) ({\hat r}_{Zu_L{\bar u}_L} - 1) + \12 (3 - 2 s_\theta^2) ({\hat r}_{Zd_L{\bar d}_L} - 1) \notag\\[8pt]
&\hspace{30pt} = -\12 v^2 \left[ C_{HD} - 12\, C_{Hq}^{(1)} \right] = -2v^2 \left[ a_{HD} + 3\, a_{Hq}^{(1)} \right] \;\equiv\; \alpha\Tom_q \,.
\label{eqn:Tomq}
\end{align}
Similarly, a $\Tom_{\qR}$ can be constructed using $\hat\rho$ and the right-handed partial widths:
\begin{align}
\hspace{-20pt} &(\hat\rho - 1) + 2s_\theta^2 ({\hat r}_{Zu{\bar u}} - 1) - s_\theta^2 ({\hat r}_{Zd{\bar d}} - 1) \notag\\[8pt]
&\hspace{30pt} = -\12 v^2 \bigg[ C_{HD} - 6 \left( C_{Hu} + C_{Hd} \right) \bigg] = -2v^2 \left[ a_{HD} + 3\, a_{H\qR}^{(1)+} + 3\, a_{H\qR}^{(3)-} \right] \;\equiv\; \alpha\Tom_{\qR} \,.
\label{eqn:TomqR}
\end{align}
In the second lines of \cref{eqn:Tomq,eqn:TomqR}, we have used \cref{tbl:CFroma} to write them in terms of our custodial basis Wilson coefficients $\WlC_i$, where it becomes manifest that $\Tom_q$ and $\Tom_{\qR}$ receive contributions only from custodial violating operators (\Hvcolor~operators) in our \cref{tbl:nuSMEFTCBasis}. Unfortunately, we cannot separately measure the $Z$ partial widths into left- and/or right-handed up and down quarks, so these two results have no practical value. Nevertheless, it is worth mentioning that parameters analogous to $\Tom_l$ can at least in principle be constructed.

\section{Tables of Operators, Coefficients, and Translations}
\label{appsec:Tables}

In this Appendix, we gather tables of operator bases and relevant translation relations. \cref{tbl:nuSMEFTWarsaw} summarizes all the independent baryon-preserving operators in the Warsaw basis for dim-6 $\nu$SMEFT (suppressing flavor indices). These operators are recombined to form our custodial basis summarized in \cref{tbl:nuSMEFTCBasis}. \cref{tbl:BasisTranslation} provides an explicit translation dictionary between the operators in these two operator bases. Translation dictionaries between the Wilson coefficients $C_i$ and $\WlC_i$, in both directions, are further provided in \cref{tbl:aFromC,tbl:CFroma}. \cref{tbl:SMEFTfromnuSMEFT} summarizes the restrictions on the Wilson coefficients $C_i$ and $\WlC_i$ to reduce $\nu$SMEFT back to SMEFT\@.

Our notation and color scheme for the operators in the two bases are
\begin{align}
\mbox{Warsaw basis operators} &\; \left\{
\begin{array}{cl}
Q_i           & ~~\mbox{SMEFT} \\[3pt]
\shownu{Q_i}  & ~~\mbox{additional operators in }\nu\mbox{SMEFT}
\end{array} \right. \label{eqn:ColorSchemeWarsaw} \\[5pt]
\mbox{custodial basis operators} &\; \left\{
\begin{array}{cl}
\showHv{\Op_i}  & ~~\mbox{custodial violating} \\[3pt]
\showqv{\Op_i}  & ~~SU(2)_{RH}\,\mbox{-preserving}\,,\; SU(2)_{R\qR}\,\mbox{-violating} \\[3pt]
\showlv{\Op_i}  & ~~SU(2)_{RH}\,\mbox{-preserving}\,,\; SU(2)_{R\lR}\,\mbox{-violating} \\[3pt]
\Op_i           & ~~\mbox{all other custodial preserving} \\
\end{array} \right. \label{eqn:ColorSchemeCustodial}
\end{align}
We emphasize here that in \cref{tbl:nuSMEFTCBasis} only red operators $\showHv{\Op_i}$ are custodial violating. See text in \cref{sec:CustodialBasis} for details.

Our notations in \cref{tbl:nuSMEFTCBasis} are also a bit compact. For example, we sometimes use $\Op^\cpmH$ to group custodial preserving/violating operators together, which respectively involves $P_\cpmH$. Such examples include $\Op_{lH}^\cpmH$, $\Op_{qH}^\cpmH$, $\Op_{lW}^\cpmH$, $\Op_{qG}^\cpmH$, $\Op_{qW}^\cpmH$, $\Op_{H\lR}^{(3)\cpmH}$, and $\Op_{H\qR}^{(3)\cpmH}$. A similar kind of notation is also applied to some custodial preserving four-fermion operators that break the isospin $SU(2)_{R\qR}$ or $SU(2)_{R\lR}$. In particular, the notation $O^{\textbf{\hspace{0.7pt}\showlv{-}\showqv{-}}}$ implies that the operator violates both the lepton and quark isospin.

\setcounter{table}{0}
\renewcommand{\thetable}{\Alph{section}.\arabic{table}}

\begin{table}[tbp]
\centering\footnotesize
\vspace{-0.8cm}
\hspace{-0.3cm}
\begin{minipage}[t]{4.1cm}
\renewcommand{\arraystretch}{1.3}
\begin{tabular}[t]{c|c}
\multicolumn{2}{c}{$1:X^3$} \\
\hline
$Q_G$   & $f^{ABC} G_\mu^{A\nu} G_\nu^{B\rho} G_\rho^{C\mu}$ \\
$Q_\tG$ & $f^{ABC} \tG_\mu^{A\nu} G_\nu^{B\rho} G_\rho^{C\mu}$ \\
$Q_W$   & $\epsilon^{abc}\, W_\mu^{a\nu} W_\nu^{b\rho} W_\rho^{c\mu}$ \\
$Q_\tW$ & $\epsilon^{abc}\, \tW_\mu^{a\nu} W_\nu^{b\rho} W_\rho^{c\mu}$ \\
\end{tabular}
\end{minipage}
\begin{minipage}[t]{2.0cm}
\renewcommand{\arraystretch}{1.3}
\begin{tabular}[t]{c|c}
\multicolumn{2}{c}{$2:H^6$} \\
\hline
$Q_H$ & $\left|H\right|^6$ \\
\end{tabular}
\end{minipage}
\begin{minipage}[t]{5.2cm}
\renewcommand{\arraystretch}{1.3}
\begin{tabular}[t]{c|c}
\multicolumn{2}{c}{$3:H^4 D^2$} \\
\hline
$Q_{H\Box}$ & $-\left(\partial_\mu |H|^2\right)\left(\partial^\mu |H|^2\right)$ \\
$Q_{HD}$    & $\left[\left(D_\mu H^\dag\right)H\right] \left[H^\dag \left(D^\mu H\right)\right]$ \\
\end{tabular}
\end{minipage}
\begin{minipage}[t]{3.2cm}
\renewcommand{\arraystretch}{1.3}
\begin{tabular}[t]{c|c}
\multicolumn{2}{c}{$5: \bar\psi\psi H^3 + \hc$} \\
\hline
\shownu{$Q_{\nu H}$} & \shownu{$\left|H\right|^2 (\bar{l}\tH\nu)$} \\
$Q_{eH}$             & $\left|H\right|^2 (\bar{l}He)$ \\
$Q_{uH}$             & $\left|H\right|^2 (\bar{q}\tH u)$ \\
$Q_{dH}$             & $\left|H\right|^2 (\bar{q}Hd)$ \\
\end{tabular}
\end{minipage}
\vspace{0.2cm}

\begin{minipage}[t]{4.4cm}
\renewcommand{\arraystretch}{1.4}
\begin{tabular}[t]{c|c}
\multicolumn{2}{c}{$4:X^2H^2$} \\
\hline
$Q_{H G}$    & $\left|H\right|^2 G^A_{\mu\nu} G^{A\mu\nu}$ \\
$Q_{H\tG}$   & $\left|H\right|^2 \tG^A_{\mu\nu} G^{A\mu\nu}$ \\
$Q_{H W}$    & $\left|H\right|^2 W^a_{\mu\nu} W^{a\mu\nu}$ \\
$Q_{H\tW}$   & $\left|H\right|^2 \tW^a_{\mu\nu} W^{a\mu\nu}$ \\
$Q_{H B}$    & $\left|H\right|^2 B_{\mu\nu} B^{\mu\nu}$ \\
$Q_{H\tB}$   & $\left|H\right|^2 \tB_{\mu\nu} B^{\mu\nu}$ \\
$Q_{H WB}$   & $H^\dag\tau^a H\, W^a_{\mu\nu} B^{\mu\nu}$ \\
$Q_{H\tW B}$ & $H^\dag\tau^a H\, \tW^a_{\mu\nu} B^{\mu\nu}$ \\
\end{tabular}
\end{minipage}
\begin{minipage}[t]{4.2cm}
\renewcommand{\arraystretch}{1.4}
\begin{tabular}[t]{c|c}
\multicolumn{2}{c}{$6:\bar\psi\psi XH + \hc$} \\
\hline
\shownu{$Q_{\nu W}$} & \shownu{$(\bar{l} \sigma^{\mu\nu} \nu) \tau^a \tH W_{\mu\nu}^a$} \\
$Q_{eW}$             & $(\bar{l} \sigma^{\mu\nu} e) \tau^a H W_{\mu\nu}^a$ \\
\shownu{$Q_{\nu B}$} & \shownu{$(\bar{l} \sigma^{\mu\nu} \nu) \tH B_{\mu\nu}$} \\
$Q_{eB}$             & $(\bar{l} \sigma^{\mu\nu} e) H B_{\mu\nu}$ \\
$Q_{uG}$             & $(\bar{q} \sigma^{\mu\nu} T^A u) \tH G_{\mu\nu}^A$ \\
$Q_{dG}$             & $(\bar{q} \sigma^{\mu\nu} T^A d) H G_{\mu\nu}^A$ \\
$Q_{uW}$             & $(\bar{q} \sigma^{\mu\nu} u) \tau^a \tH W_{\mu\nu}^a$ \\
$Q_{dW}$             & $(\bar{q} \sigma^{\mu\nu} d) \tau^a H W_{\mu\nu}^a$ \\
$Q_{uB}$             & $(\bar{q} \sigma^{\mu\nu} u) \tH B_{\mu\nu}$ \\
$Q_{dB}$             & $(\bar{q} \sigma^{\mu\nu} d) H B_{\mu\nu}$ \\
\end{tabular}
\end{minipage}
\begin{minipage}[t]{5.8cm}
\renewcommand{\arraystretch}{1.4}
\begin{tabular}[t]{c|c}
\multicolumn{2}{c}{$7:\bar\psi\psi H^2 D$} \\
\hline
$Q_{Hl}^{(1)}$              & $(H^\dagger i \lrD_\mu H)(\bar{l} \gamma^\mu l)$ \\
$Q_{Hl}^{(3)}$              & $(H^\dagger i \lrD_\mu^a H)(\bar{l} \gamma^\mu \tau^a l)$ \\
$Q_{Hq}^{(1)}$              & $(H^\dagger i \lrD_\mu H)(\bar{q} \gamma^\mu q)$ \\
$Q_{Hq}^{(3)}$              & $(H^\dagger i \lrD_\mu^a H)(\bar{q} \gamma^\mu \tau^a q)$ \\
\shownu{$Q_{H\nu}$}         & \shownu{$(H^\dagger i \lrD_\mu H)(\bar{\nu} \gamma^\mu \nu)$} \\
$Q_{He}$                    & $(H^\dagger i \lrD_\mu H)(\bar{e} \gamma^\mu e)$ \\
\shownu{$Q_{H\nu e} + \hc$} & \shownu{$(\tH^\dagger iD_\mu H)(\bar{\nu} \gamma^\mu e)$} \\
$Q_{Hu}$                    & $(H^\dagger i \lrD_\mu H)(\bar{u} \gamma^\mu u)$ \\
$Q_{Hd}$                    & $(H^\dagger i \lrD_\mu H)(\bar{d} \gamma^\mu d)$ \\
$Q_{Hud} + \hc$             & $(\tH^\dagger iD_\mu H)(\bar{u} \gamma^\mu d)$ \\
\end{tabular}
\end{minipage}
\vspace{0.2cm}

\begin{minipage}[t]{4.1cm}
\renewcommand{\arraystretch}{1.3}
\begin{tabular}[t]{c|c}
\multicolumn{2}{c}{$8:(\bar LL)(\bar LL)$} \\
\hline
$Q_{ll}$       & $(\bar l \gamma_\mu l)(\bar l \gamma^\mu l)$ \\
$Q_{qq}^{(1)}$ & $(\bar q \gamma_\mu q)(\bar q \gamma^\mu q)$ \\
$Q_{qq}^{(3)}$ & $(\bar q \gamma_\mu \tau^a q)(\bar q \gamma^\mu \tau^a q)$ \\
$Q_{lq}^{(1)}$ & $(\bar l \gamma_\mu l)(\bar q \gamma^\mu q)$ \\
$Q_{lq}^{(3)}$ & $(\bar l \gamma_\mu \tau^a l)(\bar q \gamma^\mu \tau^a q)$ \\
\end{tabular}
\end{minipage}
\begin{minipage}[t]{5.6cm}
\renewcommand{\arraystretch}{1.3}
\begin{tabular}[t]{c|c}
\multicolumn{2}{c}{$8:(\bar RR)(\bar RR)$} \\
\hline
\shownu{$Q_{\nu\nu}$}      & \shownu{$(\bar{\nu}\gamma_\mu\nu)(\bar{\nu}\gamma^\mu\nu)$} \\
$Q_{ee}$                   & $(\bar{e}\gamma_\mu e)(\bar{e}\gamma^\mu e)$ \\
\shownu{$Q_{\nu e}$}       & \shownu{$(\bar{\nu}\gamma_\mu\nu)(\bar{e}\gamma^\mu e)$} \\
$Q_{uu}$                   & $(\bar{u}\gamma_\mu u)(\bar{u}\gamma^\mu u)$ \\
$Q_{dd}$                   & $(\bar{d}\gamma_\mu d)(\bar{d}\gamma^\mu d)$ \\
$Q_{ud}^{(1)}$             & $(\bar{u}\gamma_\mu u)(\bar{d}\gamma^\mu d)$ \\
$Q_{ud}^{(8)}$             & $(\bar{u}\gamma_\mu T^A u)(\bar{d}\gamma^\mu T^A d)$ \\
\shownu{$Q_{\nu u}$}       & \shownu{$(\bar{\nu}\gamma_\mu\nu)(\bar{u}\gamma^\mu u)$} \\
\shownu{$Q_{\nu d}$}       & \shownu{$(\bar{\nu}\gamma_\mu\nu)(\bar{d}\gamma^\mu d)$} \\
$Q_{eu}$                   & $(\bar{e}\gamma_\mu e)(\bar{u}\gamma^\mu u)$ \\
$Q_{ed}$                   & $(\bar{e}\gamma_\mu e)(\bar{d}\gamma^\mu d)$ \\
\shownu{$Q_{\nu edu}+\hc$} & \shownu{$(\bar{\nu}\gamma_\mu e)(\bar{d}\gamma^\mu u)$} \\
\end{tabular}
\end{minipage}
\begin{minipage}[t]{4.5cm}
\renewcommand{\arraystretch}{1.3}
\begin{tabular}[t]{c|c}
\multicolumn{2}{c}{$8:(\bar LL)(\bar RR)$} \\
\hline
\shownu{$Q_{l\nu}$} & \shownu{$(\bar{l}\gamma_\mu l)(\bar{\nu}\gamma^\mu\nu)$} \\
$Q_{le}$            & $(\bar{l}\gamma_\mu l)(\bar{e}\gamma^\mu e)$ \\
$Q_{lu}$            & $(\bar{l}\gamma_\mu l)(\bar{u}\gamma^\mu u)$ \\
$Q_{ld}$            & $(\bar{l}\gamma_\mu l)(\bar{d}\gamma^\mu d)$ \\
\shownu{$Q_{q\nu}$} & \shownu{$(\bar{q}\gamma_\mu q)(\bar{\nu}\gamma^\mu\nu)$} \\
$Q_{qe}$            & $(\bar{q}\gamma_\mu q)(\bar{e}\gamma^\mu e)$ \\
$Q_{qu}^{(1)}$      & $(\bar{q}\gamma_\mu q)(\bar{u}\gamma^\mu u)$ \\
$Q_{qd}^{(1)}$      & $(\bar{q}\gamma_\mu q)(\bar{d}\gamma^\mu d)$ \\
$Q_{qu}^{(8)}$      & $(\bar{q}\gamma_\mu T^A q)(\bar{u}\gamma^\mu T^A u)$ \\
$Q_{qd}^{(8)}$      & $(\bar{q}\gamma_\mu T^A q)(\bar{d}\gamma^\mu T^A d)$ \\
\end{tabular}
\end{minipage}
\vspace{0.2cm}

\begin{minipage}[t]{4.0cm}
\renewcommand{\arraystretch}{1.3}
\begin{tabular}[t]{c|c}
\multicolumn{2}{c}{$8:(\bar LR)(\bar RL)+\hc$} \\
\hline
\shownu{$Q_{l\nu uq}$} & \shownu{$(\bar{l}^i\nu)(\bar{u}q^i)$} \\
$Q_{ledq}$             & $(\bar{l}^i e)(\bar{d}q^i)$ \\
\end{tabular}
\end{minipage}
\begin{minipage}[t]{5.0cm}
\renewcommand{\arraystretch}{1.3}
\begin{tabular}[t]{c|c}
\multicolumn{2}{c}{$8:(\bar LR)(\bar L R)+\hc$} \\
\hline
\shownu{$Q_{l\nu le}$} & \shownu{$(\bar{l}^i\nu)\epsilon_{ij}(\bar{l}^j e)$} \\
$Q_{quqd}^{(1)}$ & $(\bar{q}^i u)\epsilon_{ij}(\bar{q}^j d)$ \\
$Q_{quqd}^{(8)}$ & $(\bar{q}^i T^A u)\epsilon_{ij}(\bar{q}^j T^A d)$ \\
\shownu{$Q_{l\nu qd}^{(1)}$} & \shownu{$(\bar{l}^i\nu)\epsilon_{ij}(\bar{q}^j d)$} \\
$Q_{le qu}^{(1)}$ & $(\bar{l}^i e)\epsilon_{ij}(\bar{q}^j u)$ \\
\shownu{$Q_{l\nu qd}^{(3)}$} & \shownu{$(\bar{l}^i\sigma_{\mu\nu}\nu)\epsilon_{ij}(\bar{q}^j \sigma^{\mu\nu} d)$} \\
$Q_{le qu}^{(3)}$ & $(\bar{l}^i \sigma_{\mu\nu} e)\epsilon_{ij}(\bar{q}^j \sigma^{\mu\nu} u)$ \\
\end{tabular}
\end{minipage}
\vspace{0.2cm}

\caption{$\nu$SMEFT dim-6 baryon-preserving operators in Warsaw basis. In addition to the $76=42+(17+\hc)$ SMEFT operators, there are $25=7+(9+\hc)$ new operators involving right-handed neutrinos $\nu$, which are colored in \nucolor.}
\label{tbl:nuSMEFTWarsaw}
\end{table}

\begin{table}[tbp]
\centering\footnotesize
\hspace{-0.4cm}
\begin{minipage}[t]{4.1cm}
\renewcommand{\arraystretch}{1.6}
\begin{tabular}[t]{c|c}
\multicolumn{2}{c}{$1:X^3$} \\
\hline
{$\Op_G$}   & {$f^{ABC} G_\mu^{A\nu} G_\nu^{B\rho} G_\rho^{C\mu}$} \\
{$\Op_\tG$} & {$f^{ABC} \tG_\mu^{A\nu} G_\nu^{B\rho} G_\rho^{C\mu}$} \\
{$\Op_W$}   & {$\epsilon^{abc}\, W_\mu^{a\nu} W_\nu^{b\rho} W_\rho^{c\mu}$} \\
{$\Op_\tW$} & {$\epsilon^{abc}\, \tW_\mu^{a\nu} W_\nu^{b\rho} W_\rho^{c\mu}$} \\
\end{tabular}
\end{minipage}
\begin{minipage}[t]{3.0cm}
\renewcommand{\arraystretch}{1.6}
\begin{tabular}[t]{c|c}
\multicolumn{2}{c}{$2:H^6$} \\
\hline
{$\Op_H$} & {$\left[\tr\left(\Sigma^\dagger\Sigma\right)\right]^3$} \\
\end{tabular}
\end{minipage}
\begin{minipage}[t]{4.1cm}
\renewcommand{\arraystretch}{1.6}
\begin{tabular}[t]{c|c}
\multicolumn{2}{c}{$3:H^4 D^2$} \\
\hline
{$\Op_{H\Box}$} & {$\left[\tr\left(\Sigma^\dagger iD_\mu\Sigma\right)\right]^2$} \\
\showHv{$\Op_{HD}$}               & \showHv{$\left[\tr\left(\Sigma^\dagger iD_\mu\Sigma\tau_R^3\right)\right]^2$} \\
\end{tabular}
\end{minipage}
\begin{minipage}[t]{4.3cm}
\renewcommand{\arraystretch}{1.6}
\begin{tabular}[t]{c|c}
\multicolumn{2}{c}{$5: \bar\psi\psi H^3 + \hc$} \\
\hline
$\Op_{lH}^\cpmH$ & $\tr(\Sigma^\dagger\Sigma)\left(\bar{l}\Sigma P_\cpmH \lR\right)$ \\
$\Op_{qH}^\cpmH$ & $\tr(\Sigma^\dagger\Sigma)\left(\bar{q}\Sigma P_\cpmH \qR\right)$ \\
\end{tabular}
\end{minipage}
%\vspace{0.3cm}

\hspace{-0.4cm}
\begin{minipage}[t]{5.0cm}
\renewcommand{\arraystretch}{1.6}
\begin{tabular}[t]{c|c}
\multicolumn{2}{c}{$4:X^2H^2$} \\
\hline
{$\Op_{HG}$}   & {$\tr\left(\Sigma^\dagger\Sigma\right) G^A_{\mu\nu} G^{A\mu\nu}$} \\
{$\Op_{H\tG}$} & {$\tr\left(\Sigma^\dagger\Sigma\right) \tG^A_{\mu\nu} G^{A\mu\nu}$} \\
{$\Op_{HW}$}   & {$\tr\left(\Sigma^\dagger\Sigma\right) W^a_{\mu\nu} W^{a\mu\nu}$} \\
{$\Op_{H\tW}$} & {$\tr\left(\Sigma^\dagger\Sigma\right) \tW^a_{\mu\nu} W^{a\mu\nu}$} \\
$\Op_{HB}$              & {$\tr\left(\Sigma^\dagger\Sigma\right) B_{\mu\nu} B^{\mu\nu}$} \\
$\Op_{H\tB}$            & {$\tr\left(\Sigma^\dagger\Sigma\right) \tB_{\mu\nu} B^{\mu\nu}$} \\
$\Op_{HWB}$             & $\tr\left(\Sigma^\dagger\tau^a\Sigma\tau_R^3\right) W^a_{\mu\nu} B^{\mu\nu}$ \\
$\Op_{H\tW B}$          & $\tr\left(\Sigma^\dagger\tau^a\Sigma\tau_R^3\right) \tW^a_{\mu\nu} B^{\mu\nu}$ \\
\end{tabular}
\end{minipage}
\begin{minipage}[t]{4.5cm}
\renewcommand{\arraystretch}{1.6}
\begin{tabular}[t]{c|c}
\multicolumn{2}{c}{$6:\bar\psi\psi XH + \hc$} \\
\hline
{$\Op_{lW}^\cpmH$} & {$(\bar{l} \sigma^{\mu\nu} \tau^a \Sigma P_\cpmH \lR) W_{\mu\nu}^a$} \\
$\Op_{lB}^\pm$             & {$(\bar{l} \sigma^{\mu\nu} \Sigma P_\mp \lR) B_{\mu\nu}$} \\
{$\Op_{qG}^\cpmH$} & {$(\bar{q} \sigma^{\mu\nu} T^A \Sigma P_\cpmH \qR) G_{\mu\nu}^A$} \\
{$\Op_{qW}^\cpmH$} & {$(\bar{q} \sigma^{\mu\nu} \tau^a \Sigma P_\cpmH \qR) W_{\mu\nu}^a$} \\
$\Op_{qB}^\pm$             & {$(\bar{q} \sigma^{\mu\nu} \Sigma P_\mp \qR) B_{\mu\nu}$} \\
\end{tabular}
\end{minipage}
\begin{minipage}[t]{6.1cm}
\renewcommand{\arraystretch}{1.6}
\begin{tabular}[t]{c|c}
\multicolumn{2}{c}{$7:\bar\psi\psi H^2 D$} \\
\hline
\showHv{$\Op_{Hl}^{(1)}$}                  & $\showHv{\tr\left(\Sigma^\dagger iD_\mu\Sigma\tau_R^3\right)}\left(\bar{l} \gamma^\mu l\right)$ \\
{$\Op_{Hl}^{(3)}$}       & {$\tr\left(\Sigma^\dagger \tau^a iD_\mu\Sigma\right)\left(\bar{l} \gamma^\mu \tau^a l\right)$} \\
\showHv{$\Op_{Hq}^{(1)}$}                  & $\showHv{\tr\left(\Sigma^\dagger iD_\mu\Sigma\tau_R^3\right)}\left(\bar{q} \gamma^\mu q\right)$ \\
{$\Op_{Hq}^{(3)}$}       & {$\tr\left(\Sigma^\dagger \tau^a iD_\mu\Sigma\right)\left(\bar{q} \gamma^\mu \tau^a q\right)$} \\
\showHv{$\Op_{H\lR}^{(1)\pm}$}             & $\showHv{\tr\left(\Sigma^\dagger iD_\mu\Sigma\tau_R^3\right)}\left(\lRbar \gamma^\mu P_\cpmH \lR\right)$ \\
{$\Op_{H\lR}^{(3)\cpmH}$} & {$\tr\left(\Sigma^\dagger iD_\mu\Sigma\tau_R^a\right)\left(\lRbar \gamma^\mu \tau_R^a P_\cpmH \lR\right)$} \\
\showHv{$\Op_{H\qR}^{(1)\pm}$}             & $\showHv{\tr\left(\Sigma^\dagger iD_\mu\Sigma\tau_R^3\right)}\left(\qRbar \gamma^\mu P_\cpmH \qR\right)$ \\
{$\Op_{H\qR}^{(3)\cpmH}$} & {$\tr\left(\Sigma^\dagger iD_\mu\Sigma\tau_R^a\right)\left(\qRbar \gamma^\mu \tau_R^a P_\cpmH \qR\right)$} \\
\end{tabular}
\end{minipage}
%\vspace{0.3cm}

\hspace{-0.2cm}
\begin{minipage}[t]{4.0cm}
\renewcommand{\arraystretch}{1.6}
\begin{tabular}[t]{c|c}
\multicolumn{2}{c}{$8:(\bar LL)(\bar LL)$} \\
\hline
{$\Op_{ll}$}       & {$(\bar l \gamma_\mu l)(\bar l \gamma^\mu l)$} \\
{$\Op_{qq}^{(1)}$} & {$(\bar q \gamma_\mu q)(\bar q \gamma^\mu q)$} \\
{$\Op_{qq}^{(3)}$} & {$(\bar q \gamma_\mu \tau^a q)(\bar q \gamma^\mu \tau^a q)$} \\
{$\Op_{lq}^{(1)}$} & {$(\bar l \gamma_\mu l)(\bar q \gamma^\mu q)$} \\
{$\Op_{lq}^{(3)}$} & {$(\bar l \gamma_\mu \tau^a l)(\bar q \gamma^\mu \tau^a q)$} \\
\end{tabular}
\end{minipage}
\begin{minipage}[t]{5.6cm}
\renewcommand{\arraystretch}{1.6}
\begin{tabular}[t]{c|c}
\multicolumn{2}{c}{$8:(\bar RR)(\bar RR)$} \\
\hline
$\Op_{\lR\lR}^{\cpml\cpml}$     & $(\lRbar\gamma_\mu P_\cpml\lR)(\lRbar\gamma^\mu P_\cpml\lR)$ \\
\showlv{$\Op_{\lR\lR}^{+-}$}    & $(\lRbar\gamma_\mu P_+\lR)\showlv{(\lRbar\gamma^\mu P_-\lR)}$ \\
$\Op_{\qR\qR}^{(1)\cpmq\cpmq}$  & $(\qRbar\gamma_\mu P_\cpmq\qR)(\qRbar\gamma^\mu P_\cpmq\qR)$ \\
\showqv{$\Op_{\qR\qR}^{(1)+-}$} & $(\qRbar\gamma_\mu P_+\qR)\showqv{(\qRbar\gamma^\mu P_-\qR)}$ \\
$\Op_{\qR\qR}^{(3)++}$          & $(\qRbar\gamma_\mu\tau_R^a\qR)(\qRbar\gamma^\mu\tau_R^a\qR)$ \\
$\Op_{\lR\qR}^{(1)\cpml\cpmq}$  & $(\lRbar\gamma_\mu P_\cpml\lR)(\qRbar\gamma^\mu P_\cpmq\qR)$ \\
$\Op_{\lR\qR}^{(1)\cpml\cmpq}$  & $(\lRbar\gamma_\mu P_\cpml\lR)(\qRbar\gamma^\mu P_\cmpq\qR)$ \\
$\Op_{\lR\qR}^{(3)+\cpmq}$      & $(\lRbar\gamma_\mu\tau_R^a\lR)(\qRbar\gamma^\mu\tau_R^a P_\cpmq\qR)$ \\
\end{tabular}
\end{minipage}
\begin{minipage}[t]{5.3cm}
\renewcommand{\arraystretch}{1.6}
\begin{tabular}[t]{c|c}
\multicolumn{2}{c}{$8:(\bar LL)(\bar RR)$} \\
\hline
$\Op_{l\lR}^\cpml$      & {$(\bar{l}\gamma_\mu l)(\lRbar\gamma^\mu P_\cpml\lR)$} \\
$\Op_{l\qR}^\cpmq$      & {$(\bar{l}\gamma_\mu l)(\qRbar\gamma^\mu P_\cpmq\qR)$} \\
$\Op_{q\lR}^\cpml$      & {$(\bar{q}\gamma_\mu q)(\lRbar\gamma^\mu P_\cpml\lR)$} \\
$\Op_{q\qR}^{(1)\cpmq}$ & {$(\bar{q}\gamma_\mu q)(\qRbar\gamma^\mu P_\cpmq\qR)$} \\
$\Op_{q\qR}^{(8)\cpmq}$ & {$(\bar{q}\gamma_\mu T^A q)(\qRbar\gamma^\mu T^A P_\cpmq\qR)$} \\
\end{tabular}
\end{minipage}
%\vspace{0.3cm}

\begin{minipage}[t]{5.0cm}
\renewcommand{\arraystretch}{1.6}
\begin{tabular}[t]{c|c}
\multicolumn{2}{c}{$8:(\bar LR)(\bar RL)+\hc$} \\
\hline
{$\Op_{l\lR\qR q}^\cpmdouble$} & {$(\bar{l}^i \lR^j) P_\cpmdouble^{jk} (\qRbar^k q^i)$} \\
\end{tabular}
\end{minipage}
\begin{minipage}[t]{6.4cm}
\renewcommand{\arraystretch}{1.6}
\begin{tabular}[t]{c|c}
\multicolumn{2}{c}{$8:(\bar LR)(\bar L R)+\hc$} \\
\hline
{$\Op_{l\lR l\lR}$}           & {$(\bar{l}^i \lR^k) \epsilon_{ij}\epsilon_{kl} (\bar{l}^j \lR^l)$} \\
{$\Op_{q\qR q\qR}^{(1)}$}     & {$(\bar{q}^i \qR^k) \epsilon_{ij}\epsilon_{kl} (\bar{q}^j \qR^l)$} \\
{$\Op_{q\qR q\qR}^{(8)}$}     & {$(\bar{q}^i T^A \qR^k) \epsilon_{ij}\epsilon_{kl} (\bar{q}^j T^A \qR^l)$} \\
{$\Op_{l\lR q\qR}^{(1)\cpmdouble}$} & {$(\bar{l}^i \lR^k) \epsilon_{ij}\left(\epsilon P_\cpmdouble\right)_{kl} (\bar{q}^j \qR^l)$} \\
{$\Op_{l\lR q\qR}^{(3)\cpmdouble}$} & {$(\bar{l}^i \sigma_{\mu\nu} \lR^k) \epsilon_{ij}\left(\epsilon P_\cpmdouble\right)_{kl} (\bar{q}^j \sigma^{\mu\nu} \qR^l)$} \\
\end{tabular}
\end{minipage}
\vspace{0.2cm}

\caption{$\nu$SMEFT dim-6 baryon-preserving operators in our custodial basis. Custodial violating operators are colored in \Hvcolor, and all other operators are custodial preserving. Operators preserving $SU(2)_{RH}$ trivially but violating the isospin $SU(2)_{R\qR}$ or $SU(2)_{R\lR}$ or both are colored \qvcolor~or \lvcolor~or both.}
\label{tbl:nuSMEFTCBasis}
\end{table}

\begin{table}[tbp]
\centering\footnotesize
\begin{minipage}[t]{2.5cm}
\renewcommand{\arraystretch}{1.6}
\begin{tabular}[t]{c|c}
\multicolumn{2}{c}{$1:X^3$} \\
\hline
{$\Op_G$}   & {$Q_G$} \\
{$\Op_\tG$} & {$Q_\tG$} \\
{$\Op_W$}   & {$Q_W$} \\
{$\Op_\tW$} & {$Q_\tW$} \\
\end{tabular}
\end{minipage}
\begin{minipage}[t]{2.6cm}
\renewcommand{\arraystretch}{1.6}
\begin{tabular}[t]{c|c}
\multicolumn{2}{c}{$2:H^6$} \\
\hline
$\Op_H$ & $8 Q_H$ \\
\end{tabular}
\end{minipage}
\begin{minipage}[t]{4.1cm}
\renewcommand{\arraystretch}{1.6}
\begin{tabular}[t]{c|c}
\multicolumn{2}{c}{$3:H^4 D^2$} \\
\hline
$\Op_{H\Box}$       & $Q_{H\Box}$ \\
\showHv{$\Op_{HD}$} & $Q_{H\Box} + 4 Q_{HD}$ \\
\end{tabular}
\end{minipage}
\begin{minipage}[t]{3.6cm}
\renewcommand{\arraystretch}{1.6}
\begin{tabular}[t]{c|c}
\multicolumn{2}{c}{$5: \bar\psi\psi H^3 + \hc$} \\
\hline
{$\Op_{lH}^\cpmH$} & $2 \left(\shownu{Q_{\nu H}} \cpmH Q_{eH}\right)$ \\
{$\Op_{qH}^\cpmH$} & $2 \left(Q_{uH} \cpmH Q_{dH}\right)$ \\
\end{tabular}
\end{minipage}
\vspace{0.3cm}

\begin{minipage}[t]{3.6cm}
\renewcommand{\arraystretch}{1.6}
\begin{tabular}[t]{c|c}
\multicolumn{2}{c}{$4:X^2H^2$} \\
\hline
{$\Op_{HG}$}   & {$2 Q_{HG}$} \\
{$\Op_{H\tG}$} & {$2 Q_{H\tG}$} \\
{$\Op_{HW}$}   & {$2 Q_{HW}$} \\
{$\Op_{H\tW}$} & {$2 Q_{H\tW}$} \\
$\Op_{HB}$              & $2 Q_{HB}$ \\
$\Op_{H\tB}$            & $2 Q_{H\tB}$ \\
$\Op_{HWB}$             & $-2 Q_{HWB}$ \\
$\Op_{H\tW B}$          & $-2 Q_{H\tW B}$ \\
\end{tabular}
\end{minipage}
\begin{minipage}[t]{3.6cm}
\renewcommand{\arraystretch}{1.6}
\begin{tabular}[t]{c|c}
\multicolumn{2}{c}{$6:\bar\psi\psi XH + \hc$} \\
\hline
$\Op_{lW}^\cpmH$ & $\shownu{Q_{\nu W}} \cpmH Q_{eW}$ \\
$\Op_{lB}^\pm$   & $\shownu{Q_{\nu B}} \mp Q_{eB}$ \\
$\Op_{qG}^\cpmH$ & $Q_{uG} \cpmH Q_{dG}$ \\
$\Op_{qW}^\cpmH$ & $Q_{uW} \cpmH Q_{dW}$ \\
$\Op_{qB}^\pm$   & $Q_{uB} \mp Q_{dB}$ \\
\end{tabular}
\end{minipage}
\begin{minipage}[t]{6.0cm}
\renewcommand{\arraystretch}{1.6}
\begin{tabular}[t]{c|c}
\multicolumn{2}{c}{$7:\bar\psi\psi H^2 D$} \\
\hline
\showHv{$\Op_{Hl}^{(1)}$}      & $-Q_{Hl}^{(1)}$ \\
$\Op_{Hl}^{(3)}$               & $Q_{Hl}^{(3)}$ \\
\showHv{$\Op_{Hq}^{(1)}$}      & $-Q_{Hq}^{(1)}$ \\
$\Op_{Hq}^{(3)}$               & $Q_{Hq}^{(3)}$ \\
\showHv{$\Op_{H\lR}^{(1)\pm}$} & $-\left(\shownu{Q_{H\nu}} \showHv{\pm} Q_{He}\right)$ \\
$\Op_{H\lR}^{(3)\cpmH}$        & $\cpmH 2 \left(\shownu{Q_{H\nu e}} \cpmH \hc\right) - \shownu{Q_{H\nu}} \cpmH Q_{He}$ \\
\showHv{$\Op_{H\qR}^{(1)\pm}$} & $-\left(Q_{Hu} \showHv{\pm} Q_{Hd}\right)$ \\
$\Op_{H\qR}^{(3)\cpmH}$        & $\cpmH 2 \left(Q_{Hud} \cpmH \hc\right) - Q_{Hu} \cpmH Q_{Hd}$ \\
\end{tabular}
\end{minipage}
\vspace{0.3cm}

\begin{minipage}[t]{2.5cm}
\renewcommand{\arraystretch}{1.6}
\begin{tabular}[t]{c|c}
\multicolumn{2}{c}{$8:(\bar LL)(\bar LL)$} \\
\hline
{$\Op_{ll}$}       & {$Q_{ll}$} \\
{$\Op_{qq}^{(1)}$} & {$Q_{qq}^{(1)}$} \\
{$\Op_{qq}^{(3)}$} & {$Q_{qq}^{(3)}$} \\
{$\Op_{lq}^{(1)}$} & {$Q_{lq}^{(1)}$} \\
{$\Op_{lq}^{(3)}$} & {$Q_{lq}^{(3)}$} \\
\end{tabular}
\end{minipage}
\begin{minipage}[t]{8.6cm}
\renewcommand{\arraystretch}{1.6}
\begin{tabular}[t]{c|c}
\multicolumn{2}{c}{$8:(\bar RR)(\bar RR)$} \\
\hline
$\Op_{\lR\lR}^{\cpml\cpml}$     & $\shownu{Q_{\nu\nu}} + Q_{ee} \cpml 2 \shownu{Q_{\nu e}}$ \\
\showlv{$\Op_{\lR\lR}^{+-}$}    & $\shownu{Q_{\nu\nu}} \showlv{-} Q_{ee}$ \\
$\Op_{\qR\qR}^{(1)\cpmq\cpmq}$  & {$Q_{uu} + Q_{dd} \cpmq 2 Q_{ud}^{(1)}$} \\
\showqv{$\Op_{\qR\qR}^{(1)+-}$} & $Q_{uu} \showqv{-} Q_{dd}$ \\
$\Op_{\qR\qR}^{(3)++}$          & $8 Q_{ud}^{(8)} - \frac{2N_c-4}{N_c} Q_{ud}^{(1)} + Q_{uu} + Q_{dd} $ \\
$\Op_{\lR\qR}^{(1)\cpml\cpmq}$  & $\left(\shownu{Q_{\nu u}} + Q_{ed}\right) \cpmdouble \left(\shownu{Q_{\nu d}} + Q_{eu}\right)$ \\
$\Op_{\lR\qR}^{(1)\cpml\cmpq}$  & $\left(\shownu{Q_{\nu u}} - Q_{ed}\right) \cmpql \left(\shownu{Q_{\nu d}} - Q_{eu}\right)$ \\
$\Op_{\lR\qR}^{(3)+\cpmq}$      & $2\left(\shownu{Q_{\nu edu}} \cpmq \hc\right) + \left(\shownu{Q_{\nu u}} - Q_{eu}\right) \cmpqtwo \left(\shownu{Q_{\nu d}} - Q_{ed}\right)$ \\
\end{tabular}
\end{minipage}
\begin{minipage}[t]{3.2cm}
\renewcommand{\arraystretch}{1.6}
\begin{tabular}[t]{c|c}
\multicolumn{2}{c}{$8:(\bar LL)(\bar RR)$} \\
\hline
$\Op_{l\lR}^\cpml$      & $\shownu{Q_{l\nu}} \cpml Q_{le}$ \\
$\Op_{l\qR}^\cpmq$      & $Q_{lu} \cpmq Q_{ld}$ \\
$\Op_{q\lR}^\cpml$      & $\shownu{Q_{q\nu}} \cpml Q_{qe}$ \\
$\Op_{q\qR}^{(1)\cpmq}$ & $Q_{qu}^{(1)} \cpmq Q_{qd}^{(1)}$ \\
$\Op_{q\qR}^{(8)\cpmq}$ & $Q_{qu}^{(8)} \cpmq Q_{qd}^{(8)}$ \\
\end{tabular}
\end{minipage}
\vspace{0.3cm}

\begin{minipage}[t]{4.5cm}
\renewcommand{\arraystretch}{1.6}
\begin{tabular}[t]{c|c}
\multicolumn{2}{c}{$8:(\bar LR)(\bar RL)+\hc$} \\
\hline
$\Op_{l\lR\qR q}^\cpmdouble$ & $\shownu{Q_{l\nu uq}} \cpmdouble Q_{ledq}$ \\
\end{tabular}
\end{minipage}
\begin{minipage}[t]{4.5cm}
\renewcommand{\arraystretch}{1.6}
\begin{tabular}[t]{c|c}
\multicolumn{2}{c}{$8:(\bar LR)(\bar L R)+\hc$} \\
\hline
$\Op_{l\lR l\lR}$                 & $2 \shownu{Q_{l\nu le}}$ \\
$\Op_{q\qR q\qR}^{(1)}$           & $2 Q_{quqd}^{(1)}$ \\
$\Op_{q\qR q\qR}^{(8)}$           & $2 Q_{quqd}^{(8)}$ \\
$\Op_{l\lR q\qR}^{(1)\cpmdouble}$ & $-Q_{lequ}^{(1)} \cpmdouble \shownu{Q_{l\nu qd}^{(1)}}$ \\
$\Op_{l\lR q\qR}^{(3)\cpmdouble}$ & $-Q_{lequ}^{(3)} \cpmdouble \shownu{Q_{l\nu qd}^{(3)}}$ \\
\end{tabular}
\end{minipage}
\vspace{0.5cm}

\caption{A dictionary of the custodial basis operators $\Op_i$ in terms of Warsaw basis operators $Q_i$. The color scheme is the same as given in \cref{eqn:ColorSchemeWarsaw,eqn:ColorSchemeCustodial}.}
\label{tbl:BasisTranslation}
\end{table}

\begin{table}[tbp]
\centering\footnotesize
\begin{minipage}[t]{2.5cm}
\renewcommand{\arraystretch}{1.6}
\begin{tabular}[t]{c|c}
\multicolumn{2}{c}{$1:X^3$} \\
\hline
{$\WlC_G$}   & {$C_G$} \\
{$\WlC_\tG$} & {$C_\tG$} \\
{$\WlC_W$}   & {$C_W$} \\
{$\WlC_\tW$} & {$C_\tW$} \\
\end{tabular}
\end{minipage}
\begin{minipage}[t]{2.6cm}
\renewcommand{\arraystretch}{1.6}
\begin{tabular}[t]{c|c}
\multicolumn{2}{c}{$2:H^6$} \\
\hline
{$\WlC_H$} & {$\frac{1}{8} C_H$} \\
\end{tabular}
\end{minipage}
\begin{minipage}[t]{4.1cm}
\renewcommand{\arraystretch}{1.6}
\begin{tabular}[t]{c|c}
\multicolumn{2}{c}{$3:H^4 D^2$} \\
\hline
{$\WlC_{H\Box}$} & {$C_{H\Box} - \frac{1}{4}C_{HD}$} \\
$\WlC_{HD}$               & $\frac{1}{4}C_{HD}$ \\
\end{tabular}
\end{minipage}
\begin{minipage}[t]{3.5cm}
\renewcommand{\arraystretch}{1.6}
\begin{tabular}[t]{c|c}
\multicolumn{2}{c}{$5: \bar\psi\psi H^3 + \hc$} \\
\hline
{$\WlC_{lH}^\pm$} & {$\frac{1}{4} \left(C_{\nu H} \pm C_{eH}\right)$} \\
{$\WlC_{qH}^\pm$} & {$\frac{1}{4} \left(C_{uH} \pm C_{dH}\right)$} \\
\end{tabular}
\end{minipage}
\vspace{0.3cm}

\begin{minipage}[t]{3.5cm}
\renewcommand{\arraystretch}{1.6}
\begin{tabular}[t]{c|c}
\multicolumn{2}{c}{$4:X^2H^2$} \\
\hline
{$\WlC_{HG}$}   & {$\frac{1}{2} C_{HG}$} \\
{$\WlC_{H\tG}$} & {$\frac{1}{2} C_{H\tG}$} \\
{$\WlC_{HW}$}   & {$\frac{1}{2} C_{HW}$} \\
{$\WlC_{H\tW}$} & {$\frac{1}{2} C_{H\tW}$} \\
$\WlC_{HB}$              & $ \frac{1}{2} C_{HB}$ \\
$\WlC_{H\tB}$            & $ \frac{1}{2} C_{H\tB}$ \\
$\WlC_{HWB}$             & $-\frac{1}{2} C_{HWB}$ \\
$\WlC_{H\tW B}$          & $-\frac{1}{2} C_{H\tW B}$ \\
\end{tabular}
\end{minipage}
\begin{minipage}[t]{3.9cm}
\renewcommand{\arraystretch}{1.6}
\begin{tabular}[t]{c|c}
\multicolumn{2}{c}{$6:\bar\psi\psi XH + \hc$} \\
\hline
{$\WlC_{lW}^\pm$} & {$\frac{1}{2} \left( C_{\nu W} \pm C_{eW} \right)$} \\
$\WlC_{lB}^\pm$             & $\frac{1}{2} \left( C_{\nu B} \pm C_{eB} \right)$ \\
{$\WlC_{qG}^\pm$} & {$\frac{1}{2} \left( C_{uG} \pm C_{dG} \right)$} \\
{$\WlC_{qW}^\pm$} & {$\frac{1}{2} \left( C_{uW} \pm C_{dW} \right)$} \\
$\WlC_{qB}^\pm$             & $\frac{1}{2} \left( C_{uB} \pm C_{dB} \right)$ \\
\end{tabular}
\end{minipage}
\begin{minipage}[t]{7.2cm}
\renewcommand{\arraystretch}{1.6}
\begin{tabular}[t]{c|c}
\multicolumn{2}{c}{$7:\bar\psi\psi H^2 D$} \\
\hline
$\WlC_{Hl}^{(1)}$                  & $-C_{Hl}^{(1)}$ \\
{$\WlC_{Hl}^{(3)}$}       & {$C_{Hl}^{(3)}$} \\
$\WlC_{Hq}^{(1)}$                  & $-C_{Hq}^{(1)}$ \\
{$\WlC_{Hq}^{(3)}$}       & {$C_{Hq}^{(3)}$} \\
$\WlC_{H\lR}^{(1)\pm}$             & $-\frac{1}{2}\left(C_{H\nu}\pm C_{He}\right) + \frac{1}{4}\left(\pm C_{H\nu e} - C_{H\nu e}^*\right)$ \\
{$\WlC_{H\lR}^{(3)\pm}$} & {$\frac{1}{4}\left(\pm C_{H\nu e} + C_{H\nu e}^*\right)$} \\
$\WlC_{H\qR}^{(1)\pm}$             & $-\frac{1}{2}\left(C_{Hu}\pm C_{Hd}\right) + \frac{1}{4}\left(\pm C_{Hud} - C_{Hud}^*\right)$ \\
{$\WlC_{H\qR}^{(3)\pm}$} & {$\frac{1}{4}\left(\pm C_{Hud} + C_{Hud}^*\right)$} \\
\end{tabular}
\end{minipage}
\vspace{0.3cm}

\begin{minipage}[t]{2.3cm}
\renewcommand{\arraystretch}{1.6}
\begin{tabular}[t]{c|c}
\multicolumn{2}{c}{$8:(\bar LL)(\bar LL)$} \\
\hline
{$\WlC_{ll}$}       & {$C_{ll}$} \\
{$\WlC_{qq}^{(1)}$} & {$C_{qq}^{(1)}$} \\
{$\WlC_{qq}^{(3)}$} & {$C_{qq}^{(3)}$} \\
{$\WlC_{lq}^{(1)}$} & {$C_{lq}^{(1)}$} \\
{$\WlC_{lq}^{(3)}$} & {$C_{lq}^{(3)}$} \\
\end{tabular}
\end{minipage}
\begin{minipage}[t]{9.0cm}
\renewcommand{\arraystretch}{1.6}
\begin{tabular}[t]{c|c}
\multicolumn{2}{c}{$8:(\bar RR)(\bar RR)$} \\
\hline
{$\WlC_{\lR\lR}^{\pm\pm}$}    & {$\frac{1}{4}\left( C_{\nu\nu} + C_{ee} \pm C_{\nu e} \right)$} \\
$\WlC_{\lR\lR}^{+-}$                     & $\frac{1}{2}\left(C_{\nu\nu} - C_{ee}\right)$ \\
{$\WlC_{\qR\qR}^{(1)\pm\pm}$} & {$\frac{1}{4}\left[ \left(C_{uu}+C_{dd}\right)\pm C_{ud}^{(1)} -\frac{1}{4}C_{ud}^{(8)} \pm \left(\frac{1}{4}-\frac{1}{2N_c}\right)C_{ud}^{(8)} \right] $} \\
$\WlC_{\qR\qR}^{(1)+-}$                  & $\frac{1}{2}\left(C_{uu} - C_{dd}\right)$ \\
{$\WlC_{\qR\qR}^{(3)++}$}       & {$\frac{1}{8} C_{ud}^{(8)}$} \\
{$\WlC_{\lR\qR}^{(1)+\pm}$}    & {$\frac{1}{4}\left[ \left(C_{\nu u} + C_{eu}\right) \pm \left(C_{\nu d} + C_{ed}\right) \right]$} \\
$\WlC_{\lR\qR}^{(1)-\pm}$                & $\frac{1}{4} \left[ \left(C_{\nu u} - C_{eu}\right) \pm \left(C_{\nu d} - C_{ed}\right) + \left( -C_{\nu edu} \pm C_{\nu edu}^* \right) \right]$ \\
{$\WlC_{\lR\qR}^{(3)+\pm}$}    & {$\frac{1}{4}\left( C_{\nu edu} \pm C_{\nu edu}^* \right)$} \\
\end{tabular}
\end{minipage}
\begin{minipage}[t]{3.7cm}
\renewcommand{\arraystretch}{1.6}
\begin{tabular}[t]{c|c}
\multicolumn{2}{c}{$8:(\bar LL)(\bar RR)$} \\
\hline
{$\WlC_{l\lR}^\pm$}      & {$\frac{1}{2}\left( C_{l\nu} \pm C_{le} \right)$} \\
{$\WlC_{l\qR}^\pm$}      & {$\frac{1}{2}\left( C_{lu} \pm C_{ld} \right)$} \\
{$\WlC_{q\lR}^\pm$}      & {$\frac{1}{2}\left( C_{q\nu} \pm C_{qe} \right)$} \\
{$\WlC_{q\qR}^{(1)\pm}$} & {$\frac{1}{2}\left[ C_{qu}^{(1)} \pm C_{qd}^{(1)} \right]$} \\
{$\WlC_{q\qR}^{(8)\pm}$} & {$\frac{1}{2}\left[ C_{qu}^{(8)} \pm C_{qd}^{(8)} \right]$} \\
\end{tabular}
\end{minipage}
\vspace{0.3cm}

\begin{minipage}[t]{5.0cm}
\renewcommand{\arraystretch}{1.6}
\begin{tabular}[t]{c|c}
\multicolumn{2}{c}{$8:(\bar LR)(\bar RL)+\hc$} \\
\hline
{$\WlC_{l\lR\qR q}^\pm$} & {$\frac{1}{2}\left( C_{l\nu uq} \pm C_{ledq} \right)$} \\
\end{tabular}
\end{minipage}
\begin{minipage}[t]{5.0cm}
\renewcommand{\arraystretch}{1.6}
\begin{tabular}[t]{c|c}
\multicolumn{2}{c}{$8:(\bar LR)(\bar L R)+\hc$} \\
\hline
{$\WlC_{l\lR l\lR}$}           & {$\frac{1}{2} C_{l\nu le}$} \\
{$\WlC_{q\qR q\qR}^{(1)}$}     & {$\frac{1}{2} C_{quqd}^{(1)}$} \\
{$\WlC_{q\qR q\qR}^{(8)}$}     & {$\frac{1}{2} C_{quqd}^{(8)}$} \\
{$\WlC_{l\lR q\qR}^{(1)\pm}$} & {$\frac{1}{2}\left[ -C_{lequ}^{(1)} \pm C_{l\nu qd}^{(1)} \right]$} \\
{$\WlC_{l\lR q\qR}^{(3)\pm}$} & {$\frac{1}{2}\left[ -C_{lequ}^{(3)} \pm C_{l\nu qd}^{(3)} \right]$} \\
\end{tabular}
\end{minipage}
\vspace{0.5cm}

\caption{A translation dictionary: the custodial basis Wilson coefficients $\WlC_i$ in terms of the Warsaw basis Wilson coefficients $C_i$.}
\label{tbl:aFromC}
\end{table}

\begin{table}[tbp]
\centering\footnotesize
\begin{minipage}[t]{3.8cm}
\renewcommand{\arraystretch}{1.3}
\begin{tabular}[t]{c|c}
\multicolumn{2}{c}{$1:X^3$} \\
\hline
$C_G\,,\, C_\tG$ & $\WlC_G\,,\, \WlC_\tG$ \\
$C_W\,,\, C_\tW$ & $\WlC_W\,,\, \WlC_\tW$ \\
\end{tabular}
\end{minipage}
\begin{minipage}[t]{2.3cm}
\renewcommand{\arraystretch}{1.3}
\begin{tabular}[t]{c|c}
\multicolumn{2}{c}{$2:H^6$} \\
\hline
$C_H$ & $8\,\WlC_H$ \\
\end{tabular}
\end{minipage}
\begin{minipage}[t]{3.6cm}
\renewcommand{\arraystretch}{1.3}
\begin{tabular}[t]{c|c}
\multicolumn{2}{c}{$3:H^4 D^2$} \\
\hline
$C_{H\Box}$ & $\WlC_{H\Box} + \WlC_{HD}$ \\
$C_{HD}$    & $4\,\WlC_{HD}$ \\
\end{tabular}
\end{minipage}
\begin{minipage}[t]{4.5cm}
\renewcommand{\arraystretch}{1.3}
\begin{tabular}[t]{c|c}
\multicolumn{2}{c}{$5: \bar\psi\psi H^3 + \hc$} \\
\hline
$C_{\nu H}\,,\, C_{eH}$ & $2 \left(\WlC_{lH}^+ \pm \WlC_{lH}^-\right)$ \\
$C_{uH}\,,\, C_{dH}$    & $2 \left(\WlC_{qH}^+ \pm \WlC_{qH}^-\right)$ \\
\end{tabular}
\end{minipage}
\vspace{0.3cm}

\begin{minipage}[t]{3.4cm}
\renewcommand{\arraystretch}{1.6}
\begin{tabular}[t]{c|c}
\multicolumn{2}{c}{$4:X^2H^2$} \\
\hline
$C_{HG}$     & $ 2\,\WlC_{HG}$ \\
$C_{H\tG}$   & $ 2\,\WlC_{H\tG}$ \\
$C_{HW}$     & $ 2\,\WlC_{HW}$ \\
$C_{H\tW}$   & $ 2\,\WlC_{H\tW}$ \\
$C_{HB}$     & $ 2\,\WlC_{HB}$ \\
$C_{H\tB}$   & $ 2\,\WlC_{H\tB}$ \\
$C_{HWB}$    & $-2\,\WlC_{HWB}$ \\
$C_{H\tW B}$ & $-2\,\WlC_{H\tW B}$ \\
\end{tabular}
\end{minipage}
\begin{minipage}[t]{4.3cm}
\renewcommand{\arraystretch}{1.6}
\begin{tabular}[t]{c|c}
\multicolumn{2}{c}{$6:\bar\psi\psi XH + \hc$} \\
\hline
$C_{\nu W}\,,\, C_{eW}$ & $\WlC_{lW}^+ \pm \WlC_{lW}^-$ \\
$C_{\nu B}\,,\, C_{eB}$ & $\WlC_{lB}^+ \pm \WlC_{lB}^-$ \\
$C_{uG}\,,\, C_{dG}$    & $\WlC_{qG}^+ \pm \WlC_{qG}^-$ \\
$C_{uW}\,,\, C_{dW}$    & $\WlC_{qW}^+ \pm \WlC_{qW}^-$ \\
$C_{uB}\,,\, C_{dB}$    & $\WlC_{qB}^+ + \WlC_{qB}^-$ \\
\end{tabular}
\end{minipage}
\begin{minipage}[t]{6.8cm}
\renewcommand{\arraystretch}{1.6}
\begin{tabular}[t]{c|c}
\multicolumn{2}{c}{$7:\bar\psi\psi H^2 D$} \\
\hline
$C_{Hl}^{(1)}$ & $-\,\WlC_{Hl}^{(1)}$ \\
$C_{Hl}^{(3)}$ & $   \WlC_{Hl}^{(3)}$ \\
$C_{Hq}^{(1)}$ & $-\,\WlC_{Hq}^{(1)}$ \\
$C_{Hq}^{(3)}$ & $   \WlC_{Hq}^{(3)}$ \\
$C_{H\nu}\,,\, C_{He}$     & $-\,\WlC_{H\lR}^{(1)+} \mp \WlC_{H\lR}^{(1)-} \mp \WlC_{H\lR}^{(3)+} -\WlC_{H\lR}^{(3)-}$ \\
$C_{H\nu e}$   & $2\left[\WlC_{H\lR}^{(3)+} - \WlC_{H\lR}^{(3)-}\right]$ \\
$C_{Hu}\,,\, C_{Hd}$       & $-\,\WlC_{H\qR}^{(1)+} \mp \WlC_{H\qR}^{(1)-} \mp \WlC_{H\qR}^{(3)+} -\WlC_{H\qR}^{(3)-}$ \\
$C_{Hud}$      & $2\left[\WlC_{H\qR}^{(3)+} - \WlC_{H\qR}^{(3)-}\right]$ \\
\end{tabular}
\end{minipage}
\vspace{0.3cm}

\begin{minipage}[t]{2.3cm}
\renewcommand{\arraystretch}{1.6}
\begin{tabular}[t]{c|c}
\multicolumn{2}{c}{$8:(\bar LL)(\bar LL)$} \\
\hline
$C_{ll}$       & $\WlC_{ll}$ \\
$C_{qq}^{(1)}$ & $\WlC_{qq}^{(1)}$ \\
$C_{qq}^{(3)}$ & $\WlC_{qq}^{(3)}$ \\
$C_{lq}^{(1)}$ & $\WlC_{lq}^{(1)}$ \\
$C_{lq}^{(3)}$ & $\WlC_{lq}^{(3)}$ \\
\end{tabular}
\end{minipage}
\begin{minipage}[t]{9.2cm}
\renewcommand{\arraystretch}{1.6}
\begin{tabular}[t]{c|c}
\multicolumn{2}{c}{$8:(\bar RR)(\bar RR)$} \\
\hline
$C_{\nu\nu}$   & $\WlC_{\lR\lR}^{++} + \WlC_{\lR\lR}^{--} + \WlC_{\lR\lR}^{+-}$ \\
$C_{ee}$       & $\WlC_{\lR\lR}^{++} + \WlC_{\lR\lR}^{--} - \WlC_{\lR\lR}^{+-}$ \\
$C_{\nu e}$    & $2\left(\WlC_{\lR\lR}^{++} - \WlC_{\lR\lR}^{--}\right)$ \\
$C_{uu}$       & $\WlC_{\qR\qR}^{(1)++} + \WlC_{\qR\qR}^{(1)--} + \WlC_{\qR\qR}^{(1)+-} + \WlC_{\qR\qR}^{(3)++}$ \\
$C_{dd}$       & $\WlC_{\qR\qR}^{(1)++} + \WlC_{\qR\qR}^{(1)--} - \WlC_{\qR\qR}^{(1)+-} + \WlC_{\qR\qR}^{(3)++}$ \\
$C_{ud}^{(1)}$ & $2\left[\WlC_{\qR\qR}^{(1)++} - \WlC_{\qR\qR}^{(1)--}\right] + \left(\frac{4}{N_c}-2\right)\WlC_{\qR\qR}^{(3)++}$ \\
$C_{ud}^{(8)}$ & $8\,\WlC_{\qR\qR}^{(3)++}$ \\
$C_{\nu u}$    & $\WlC_{\lR\qR}^{(1)++} + \WlC_{\lR\qR}^{(1)--} + \WlC_{\lR\qR}^{(1)+-} + \WlC_{\lR\qR}^{(1)-+} + \WlC_{\lR\qR}^{(3)++} + \WlC_{\lR\qR}^{(3)+-}$ \\
$C_{\nu d}$    & $\WlC_{\lR\qR}^{(1)++} - \WlC_{\lR\qR}^{(1)--} - \WlC_{\lR\qR}^{(1)+-} + \WlC_{\lR\qR}^{(1)-+} - \WlC_{\lR\qR}^{(3)++} + \WlC_{\lR\qR}^{(3)+-}$ \\
$C_{eu}$       & $\WlC_{\lR\qR}^{(1)++} - \WlC_{\lR\qR}^{(1)--} + \WlC_{\lR\qR}^{(1)+-} - \WlC_{\lR\qR}^{(1)-+} - \WlC_{\lR\qR}^{(3)++} - \WlC_{\lR\qR}^{(3)+-}$ \\
$C_{ed}$       & $\WlC_{\lR\qR}^{(1)++} + \WlC_{\lR\qR}^{(1)--} - \WlC_{\lR\qR}^{(1)+-} - \WlC_{\lR\qR}^{(1)-+} + \WlC_{\lR\qR}^{(3)++} - \WlC_{\lR\qR}^{(3)+-}$ \\
$C_{\nu edu}$  & $2\left[\WlC_{\lR\qR}^{(3)++} + \WlC_{\lR\qR}^{(3)+-}\right]$ \\
\end{tabular}
\end{minipage}
\begin{minipage}[t]{3.3cm}
\renewcommand{\arraystretch}{1.6}
\begin{tabular}[t]{c|c}
\multicolumn{2}{c}{$8:(\bar LL)(\bar RR)$} \\
\hline
$C_{l\nu}$     & $\WlC_{l\lR}^+ + \WlC_{l\lR}^-$ \\
$C_{le}$       & $\WlC_{l\lR}^+ - \WlC_{l\lR}^-$ \\
$C_{lu}$       & $\WlC_{l\qR}^+ + \WlC_{l\qR}^-$ \\
$C_{ld}$       & $\WlC_{l\qR}^+ - \WlC_{l\qR}^-$ \\
$C_{q\nu}$     & $\WlC_{q\lR}^+ + \WlC_{q\lR}^-$ \\
$C_{qe}$       & $\WlC_{q\lR}^+ - \WlC_{q\lR}^-$ \\
$C_{qu}^{(1)}$ & $\WlC_{q\qR}^{(1)+} + \WlC_{q\qR}^{(1)-}$ \\
$C_{qd}^{(1)}$ & $\WlC_{q\qR}^{(1)+} - \WlC_{q\qR}^{(1)-}$ \\
$C_{qu}^{(8)}$ & $\WlC_{q\qR}^{(8)+} + \WlC_{q\qR}^{(8)-}$ \\
$C_{qd}^{(8)}$ & $\WlC_{q\qR}^{(8)+} - \WlC_{q\qR}^{(8)-}$ \\
\end{tabular}
\end{minipage}
\vspace{0.3cm}

\begin{minipage}[t]{6.0cm}
\renewcommand{\arraystretch}{1.6}
\begin{tabular}[t]{c|c}
\multicolumn{2}{c}{$8:(\bar LR)(\bar RL)+\hc$} \\
\hline
$C_{l\nu uq}\,,\, C_{ledq}$ & $\WlC_{l\lR\qR q}^+ \pm \WlC_{l\lR\qR q}^-$ \\
\end{tabular}
\end{minipage}
\begin{minipage}[t]{6.0cm}
\renewcommand{\arraystretch}{1.6}
\begin{tabular}[t]{c|c}
\multicolumn{2}{c}{$8:(\bar LR)(\bar L R)+\hc$} \\
\hline
$C_{l\nu le}$                            & $2\,\WlC_{l\lR l\lR}$ \\
$C_{quqd}^{(1)}\,,\, C_{quqd}^{(8)}$     & $2\,\WlC_{q\qR q\qR}^{(1)}\,,\, 2\,\WlC_{q\qR q\qR}^{(8)}$ \\
%$C_{quqd}^{(8)}$                         & $2\,\WlC_{q\qR q\qR}^{(8)}$ \\
$C_{l\nu qd}^{(1)}\,,\, C_{le qu}^{(1)}$ & $\mp\,\WlC_{l\lR q\qR}^{(1)+}-\WlC_{l\lR q\qR}^{(1)-}$ \\
$C_{l\nu qd}^{(3)}\,,\, C_{le qu}^{(3)}$ & $\mp\,\WlC_{l\lR q\qR}^{(3)+}-\WlC_{l\lR q\qR}^{(3)-}$ \\
\end{tabular}
\end{minipage}
\vspace{0.2cm}

\caption{A translation dictionary: the Warsaw basis Wilson coefficients $C_i$ in terms of the custodial basis Wilson coefficients $\WlC_i$.}
\label{tbl:CFroma}
\end{table} 

\begin{table}
\centering
\renewcommand{\arraystretch}{1.8}
\setlength{\arrayrulewidth}{.2mm}
\setlength{\tabcolsep}{1 em}
\begin{tabular}{|c|c|}
\hline
  $\nu$SMEFT $\to$ SMEFT in Warsaw basis & $\nu$SMEFT $\to$ SMEFT in custodial basis \\ [1ex]
\hline
$C_{\nu H} = 0$ & $\WlC_{lH}^+ = - \, \WlC_{lH}^-$ \\ [1ex]
\hline
$C_{\nu W} = C_{\nu B} = 0$ & $\WlC_{lW}^+ = - \, \WlC_{lW}^-  \quad,\quad  \WlC_{lB}^+ = - \, \WlC_{lB}^-$ \\ [1ex]
\hline
$C_{H \nu e} = C_{H \nu e}^* = 0$ & $\WlC_{H\lR}^{(3)+} = \WlC_{H\lR}^{(3)-} = 0$ \\ [1ex]
\hline
$C_{H \nu} = 0$ & $\WlC_{H\lR}^{(1)+} = - \, \WlC_{H\lR}^{(1)-}$ \\ [1ex]
\hline
$C_{\nu \nu} = C_{\nu e} = 0$ & $\WlC_{\lR\lR}^{++} = \WlC_{\lR\lR}^{--} = -\12 \, \WlC_{\lR\lR}^{+-}$ \\ [1ex]
\hline
$C_{\nu e d u} = C_{\nu e d u}^* = 0$ & $\WlC_{\lR \qR}^{(3)++} = \WlC_{\lR \qR}^{(3)+-} = 0$ \\ [1ex]
\hline
$C_{\nu u} = C_{\nu d} = 0$ & $\WlC_{\lR \qR}^{(1)++} = - \, \WlC_{\lR \qR}^{(1)-+}  \quad,\quad  \WlC_{\lR \qR}^{(1)+-} = - \, \WlC_{\lR \qR}^{(1)--} $ \\ [1ex]
\hline
$C_{l \nu} = C_{q \nu} = 0$ & $\WlC_{l \lR}^+ = - \, \WlC_{l \lR}^-  \quad,\quad   \WlC_{q \lR}^+ = - \, \WlC_{q \lR}^-$ \\ [1ex]
\hline
$C_{l \nu u q} = 0$ & $\WlC_{l \lR \qR q}^+ = - \, \WlC_{l \lR \qR q}^-$ \\ [1ex]
\hline
$C_{l \nu l e} = 0$ & $\WlC_{l \lR l \lR} = 0$ \\ [1ex]
\hline
$C_{l \nu q d}^{(1)} = C_{l \nu q d}^{(3)} = 0$ & $\WlC_{l \lR q \qR}^{(1)+} = - \, \WlC_{l \lR q \qR}^{(1)-}  \quad,\quad  \WlC_{l \lR q \qR}^{(3)+} = - \, \WlC_{l \lR q \qR}^{(3)-}$ \\ [1ex]
\hline
\end{tabular}
\vspace{3mm}
\caption{Reducing $\nu$SMEFT to SMEFT:  the left (right) column shows the constraints on the Wilson coefficients in Warsaw (custodial) basis.}
\label{tbl:SMEFTfromnuSMEFT}
\end{table}

\vfill
\newpage
\vfill
\newpage
\bibliographystyle{utphys}
\bibliography{ref_custodial}

\providecommand{\href}[2]{#2}\begingroup\raggedright\begin{thebibliography}{10}

\bibitem{Z-Pole}
{The ALEPH, DELPHI, L3, OPAL, SLD Collaborations, the LEP Electroweak Working
  Group, the SLD Electroweak and Heavy Flavour Groups}, ``{Precision
  Electroweak Measurements on the Z Resonance},'' {\em Phys. Rept.} {\bf 427}
  (2006)  257,
\href{http://arxiv.org/abs/hep-ex/0509008}{{\tt hep-ex/0509008}}.
%%CITATION = HEP-EX 0509008;%%.

\bibitem{LEP-2}
{The ALEPH, DELPHI, L3, OPAL Collaborations, the LEP Electroweak Working
  Group}, ``{Electroweak Measurements in Electron-Positron Collisions at
  W-Boson-Pair Energies at LEP},'' {\em Phys. Rept.} {\bf 532} (2013)  119,
\href{http://arxiv.org/abs/1302.3415}{{\tt arXiv:1302.3415 [hep-ex]}}.
%%CITATION = ARXIV:1302.3415;%%.

\bibitem{Peskin:1991sw}
M.~E. Peskin and T.~Takeuchi, ``{Estimation of oblique electroweak
  corrections},''
\href{http://dx.doi.org/10.1103/PhysRevD.46.381}{{\em Phys.Rev.} {\bf D46}
  (1992)  381--409}.
%%CITATION = PHRVA,D46,381;%%.

\bibitem{Sikivie:1980hm}
P.~Sikivie, L.~Susskind, M.~B. Voloshin, and V.~I. Zakharov, ``{Isospin
  Breaking in Technicolor Models},''
\href{http://dx.doi.org/10.1016/0550-3213(80)90214-X}{{\em Nucl. Phys.} {\bf
  B173} (1980)  189--207}.
%%CITATION = NUPHA,B173,189;%%.

\bibitem{Hill:2002ap}
C.~T. Hill and E.~H. Simmons, ``{Strong Dynamics and Electroweak Symmetry
  Breaking},'' \href{http://dx.doi.org/10.1016/S0370-1573(03)00140-6}{{\em
  Phys. Rept.} {\bf 381} (2003)  235--402},
  \href{http://arxiv.org/abs/hep-ph/0203079}{{\tt arXiv:hep-ph/0203079
  [hep-ph]}}.
[Erratum: Phys. Rept.390,553(2004)].
%%CITATION = HEP-PH/0203079;%%.

\bibitem{Georgi:1984af}
H.~Georgi and D.~B. Kaplan, ``{Composite Higgs and Custodial SU(2)},''
\href{http://dx.doi.org/10.1016/0370-2693(84)90341-1}{{\em Phys. Lett.} {\bf
  145B} (1984)  216--220}.
%%CITATION = PHLTA,145B,216;%%.

\bibitem{Dugan:1984hq}
M.~J. Dugan, H.~Georgi, and D.~B. Kaplan, ``{Anatomy of a Composite Higgs
  Model},''
\href{http://dx.doi.org/10.1016/0550-3213(85)90221-4}{{\em Nucl. Phys.} {\bf
  B254} (1985)  299--326}.
%%CITATION = NUPHA,B254,299;%%.

\bibitem{Agashe:2004rs}
K.~Agashe, R.~Contino, and A.~Pomarol, ``{The Minimal composite Higgs model},''
  \href{http://dx.doi.org/10.1016/j.nuclphysb.2005.04.035}{{\em Nucl. Phys.}
  {\bf B719} (2005)  165--187},
\href{http://arxiv.org/abs/hep-ph/0412089}{{\tt arXiv:hep-ph/0412089
  [hep-ph]}}.
%%CITATION = HEP-PH/0412089;%%.

\bibitem{Agashe:2006at}
K.~Agashe, R.~Contino, L.~Da~Rold, and A.~Pomarol, ``{A Custodial symmetry for
  $Zb \bar b$},'' \href{http://dx.doi.org/10.1016/j.physletb.2006.08.005}{{\em
  Phys. Lett.} {\bf B641} (2006)  62--66},
\href{http://arxiv.org/abs/hep-ph/0605341}{{\tt arXiv:hep-ph/0605341
  [hep-ph]}}.
%%CITATION = HEP-PH/0605341;%%.

\bibitem{Giudice:2007fh}
G.~F. Giudice, C.~Grojean, A.~Pomarol, and R.~Rattazzi, ``{The
  Strongly-Interacting Light Higgs},''
  \href{http://dx.doi.org/10.1088/1126-6708/2007/06/045}{{\em JHEP} {\bf 06}
  (2007)  045},
\href{http://arxiv.org/abs/hep-ph/0703164}{{\tt arXiv:hep-ph/0703164
  [hep-ph]}}.
%%CITATION = HEP-PH/0703164;%%.

\bibitem{Csaki:2002qg}
C.~Csaki, J.~Hubisz, G.~D. Kribs, P.~Meade, and J.~Terning, ``{Big corrections
  from a little Higgs},''
  \href{http://dx.doi.org/10.1103/PhysRevD.67.115002}{{\em Phys. Rev.} {\bf
  D67} (2003)  115002},
\href{http://arxiv.org/abs/hep-ph/0211124}{{\tt arXiv:hep-ph/0211124
  [hep-ph]}}.
%%CITATION = HEP-PH/0211124;%%.

\bibitem{Hewett:2002px}
J.~L. Hewett, F.~J. Petriello, and T.~G. Rizzo, ``{Constraining the littlest
  Higgs},'' \href{http://dx.doi.org/10.1088/1126-6708/2003/10/062}{{\em JHEP}
  {\bf 10} (2003)  062},
\href{http://arxiv.org/abs/hep-ph/0211218}{{\tt arXiv:hep-ph/0211218
  [hep-ph]}}.
%%CITATION = HEP-PH/0211218;%%.

\bibitem{Chang:2003zn}
S.~Chang, ``{A 'Littlest Higgs' model with custodial SU(2) symmetry},''
  \href{http://dx.doi.org/10.1088/1126-6708/2003/12/057}{{\em JHEP} {\bf 12}
  (2003)  057},
\href{http://arxiv.org/abs/hep-ph/0306034}{{\tt arXiv:hep-ph/0306034
  [hep-ph]}}.
%%CITATION = HEP-PH/0306034;%%.

\bibitem{Cheng:2003ju}
H.-C. Cheng and I.~Low, ``{TeV symmetry and the little hierarchy problem},''
  \href{http://dx.doi.org/10.1088/1126-6708/2003/09/051}{{\em JHEP} {\bf 09}
  (2003)  051},
\href{http://arxiv.org/abs/hep-ph/0308199}{{\tt arXiv:hep-ph/0308199
  [hep-ph]}}.
%%CITATION = HEP-PH/0308199;%%.

\bibitem{Chen:2003fm}
M.-C. Chen and S.~Dawson, ``{One loop radiative corrections to the rho
  parameter in the littlest Higgs model},''
  \href{http://dx.doi.org/10.1103/PhysRevD.70.015003}{{\em Phys. Rev.} {\bf
  D70} (2004)  015003},
\href{http://arxiv.org/abs/hep-ph/0311032}{{\tt arXiv:hep-ph/0311032
  [hep-ph]}}.
%%CITATION = HEP-PH/0311032;%%.

\bibitem{Delgado:2015aha}
A.~Delgado, M.~Garcia-Pepin, B.~Ostdiek, and M.~Quiros, ``{Dark Matter from the
  Supersymmetric Custodial Triplet Model},''
  \href{http://dx.doi.org/10.1103/PhysRevD.92.015011}{{\em Phys. Rev.} {\bf
  D92} (2015) no.~1, 015011},
\href{http://arxiv.org/abs/1504.02486}{{\tt arXiv:1504.02486 [hep-ph]}}.
%%CITATION = ARXIV:1504.02486;%%.

\bibitem{Kribs:2018oad}
G.~D. Kribs, A.~Martin, and T.~Tong, ``{Effective Theories of Dark Mesons with
  Custodial Symmetry},'' \href{http://dx.doi.org/10.1007/JHEP08(2019)020}{{\em
  JHEP} {\bf 08} (2019)  020},
\href{http://arxiv.org/abs/1809.10183}{{\tt arXiv:1809.10183 [hep-ph]}}.
%%CITATION = ARXIV:1809.10183;%%.

\bibitem{Choi:2019zeb}
S.-M. Choi, H.~M. Lee, Y.~Mambrini, and M.~Pierre, ``{Vector SIMP dark matter
  with approximate custodial symmetry},''
  \href{http://dx.doi.org/10.1007/JHEP07(2019)049}{{\em JHEP} {\bf 07} (2019)
  049},
\href{http://arxiv.org/abs/1904.04109}{{\tt arXiv:1904.04109 [hep-ph]}}.
%%CITATION = ARXIV:1904.04109;%%.

\bibitem{Veltman:1977kh}
M.~J.~G. Veltman, ``{Limit on Mass Differences in the Weinberg Model},''
\href{http://dx.doi.org/10.1016/0550-3213(77)90342-X}{{\em Nucl. Phys.} {\bf
  B123} (1977)  89--99}.
%%CITATION = NUPHA,B123,89;%%.

\bibitem{PDG}
{\bf Particle Data Group} Collaboration, M.~T. et~al. {\em Phys. Rev.} {\bf
  D98} (2018)  030001.

\bibitem{Maksymyk:1993zm}
I.~Maksymyk, C.~Burgess, and D.~London, ``{Beyond S, T and U},''
  \href{http://dx.doi.org/10.1103/PhysRevD.50.529}{{\em Phys. Rev. D} {\bf 50}
  (1994)  529--535}, \href{http://arxiv.org/abs/hep-ph/9306267}{{\tt
  arXiv:hep-ph/9306267}}.

\bibitem{Burgess:1993mg}
C.~Burgess, S.~Godfrey, H.~Konig, D.~London, and I.~Maksymyk, ``{A Global fit
  to extended oblique parameters},''
  \href{http://dx.doi.org/10.1016/0370-2693(94)91322-6}{{\em Phys. Lett. B}
  {\bf 326} (1994)  276--281}, \href{http://arxiv.org/abs/hep-ph/9307337}{{\tt
  arXiv:hep-ph/9307337}}.

\bibitem{Kundu:1996ah}
A.~Kundu and P.~Roy, ``{A General treatment of oblique parameters},''
  \href{http://dx.doi.org/10.1142/S0217751X97001079}{{\em Int. J. Mod. Phys. A}
  {\bf 12} (1997)  1511--1530}, \href{http://arxiv.org/abs/hep-ph/9603323}{{\tt
  arXiv:hep-ph/9603323}}.

\bibitem{Barbieri:2004qk}
R.~Barbieri, A.~Pomarol, R.~Rattazzi, and A.~Strumia, ``{Electroweak symmetry
  breaking after LEP-1 and LEP-2},''
  \href{http://dx.doi.org/10.1016/j.nuclphysb.2004.10.014}{{\em Nucl. Phys.}
  {\bf B703} (2004)  127--146},
\href{http://arxiv.org/abs/hep-ph/0405040}{{\tt arXiv:hep-ph/0405040
  [hep-ph]}}.
%%CITATION = HEP-PH/0405040;%%.

\bibitem{Aad:2012tfa}
{\bf ATLAS Collaboration} Collaboration, G.~Aad {\em et al.}, ``{Observation of
  a new particle in the search for the Standard Model Higgs boson with the
  ATLAS detector at the LHC},''
  \href{http://dx.doi.org/10.1016/j.physletb.2012.08.020}{{\em Phys.Lett.} {\bf
  B716} (2012)  1--29},
\href{http://arxiv.org/abs/1207.7214}{{\tt arXiv:1207.7214 [hep-ex]}}.
%%CITATION = ARXIV:1207.7214;%%.

\bibitem{Chatrchyan:2012ufa}
{\bf CMS Collaboration} Collaboration, S.~Chatrchyan {\em et al.},
  ``{Observation of a new boson at a mass of 125 GeV with the CMS experiment at
  the LHC},'' \href{http://dx.doi.org/10.1016/j.physletb.2012.08.021}{{\em
  Phys.Lett.} {\bf B716} (2012)  30--61},
\href{http://arxiv.org/abs/1207.7235}{{\tt arXiv:1207.7235 [hep-ex]}}.
%%CITATION = ARXIV:1207.7235;%%.

\bibitem{Weinberg:1979sa}
S.~Weinberg, ``{Baryon and Lepton Nonconserving Processes},''
  \href{http://dx.doi.org/10.1103/PhysRevLett.43.1566}{{\em Phys. Rev. Lett.}
  {\bf 43} (1979)  1566--1570}.

\bibitem{Weinberg:1980bf}
S.~Weinberg, ``{Varieties of Baryon and Lepton Nonconservation},''
\href{http://dx.doi.org/10.1103/PhysRevD.22.1694}{{\em Phys. Rev.} {\bf D22}
  (1980)  1694}.
%%CITATION = PHRVA,D22,1694;%%.

\bibitem{Buchmuller:1985jz}
W.~Buchmuller and D.~Wyler, ``{Effective Lagrangian Analysis of New
  Interactions and Flavor Conservation},''
\href{http://dx.doi.org/10.1016/0550-3213(86)90262-2}{{\em Nucl. Phys.} {\bf
  B268} (1986)  621--653}.
%%CITATION = NUPHA,B268,621;%%.

\bibitem{Grzadkowski:2010es}
B.~Grzadkowski, M.~Iskrzynski, M.~Misiak, and J.~Rosiek, ``{Dimension-Six Terms
  in the Standard Model Lagrangian},''
  \href{http://dx.doi.org/10.1007/JHEP10(2010)085}{{\em JHEP} {\bf 10} (2010)
  085},
\href{http://arxiv.org/abs/1008.4884}{{\tt arXiv:1008.4884 [hep-ph]}}.
%%CITATION = ARXIV:1008.4884;%%.

\bibitem{Elias-Miro:2013eta}
J.~Elias-Miró, C.~Grojean, R.~S. Gupta, and D.~Marzocca, ``{Scaling and tuning
  of EW and Higgs observables},''
  \href{http://dx.doi.org/10.1007/JHEP05(2014)019}{{\em JHEP} {\bf 05} (2014)
  019},
\href{http://arxiv.org/abs/1312.2928}{{\tt arXiv:1312.2928 [hep-ph]}}.
%%CITATION = ARXIV:1312.2928;%%.

\bibitem{Jenkins:2009dy}
E.~E. Jenkins and A.~V. Manohar, ``{Algebraic Structure of Lepton and Quark
  Flavor Invariants and CP Violation},''
  \href{http://dx.doi.org/10.1088/1126-6708/2009/10/094}{{\em JHEP} {\bf 10}
  (2009)  094},
\href{http://arxiv.org/abs/0907.4763}{{\tt arXiv:0907.4763 [hep-ph]}}.
%%CITATION = ARXIV:0907.4763;%%.

\bibitem{Hanany:2010vu}
A.~Hanany, E.~E. Jenkins, A.~V. Manohar, and G.~Torri, ``{Hilbert Series for
  Flavor Invariants of the Standard Model},''
  \href{http://dx.doi.org/10.1007/JHEP03(2011)096}{{\em JHEP} {\bf 03} (2011)
  096},
\href{http://arxiv.org/abs/1010.3161}{{\tt arXiv:1010.3161 [hep-ph]}}.
%%CITATION = ARXIV:1010.3161;%%.

\bibitem{Lehman:2015via}
L.~Lehman and A.~Martin, ``{Hilbert Series for Constructing Lagrangians:
  expanding the phenomenologist's toolbox},''
  \href{http://dx.doi.org/10.1103/PhysRevD.91.105014}{{\em Phys. Rev.} {\bf
  D91} (2015)  105014},
\href{http://arxiv.org/abs/1503.07537}{{\tt arXiv:1503.07537 [hep-ph]}}.
%%CITATION = ARXIV:1503.07537;%%.

\bibitem{Henning:2015daa}
B.~Henning, X.~Lu, T.~Melia, and H.~Murayama, ``{Hilbert series and operator
  bases with derivatives in effective field theories},''
  \href{http://dx.doi.org/10.1007/s00220-015-2518-2}{{\em Commun. Math. Phys.}
  {\bf 347} (2016) no.~2, 363--388},
\href{http://arxiv.org/abs/1507.07240}{{\tt arXiv:1507.07240 [hep-th]}}.
%%CITATION = ARXIV:1507.07240;%%.

\bibitem{Henning:2017fpj}
B.~Henning, X.~Lu, T.~Melia, and H.~Murayama, ``{Operator bases, $S$-matrices,
  and their partition functions},''
  \href{http://dx.doi.org/10.1007/JHEP10(2017)199}{{\em JHEP} {\bf 10} (2017)
  199},
\href{http://arxiv.org/abs/1706.08520}{{\tt arXiv:1706.08520 [hep-th]}}.
%%CITATION = ARXIV:1706.08520;%%.

\bibitem{Lehman:2014jma}
L.~Lehman, ``{Extending the Standard Model Effective Field Theory with the
  Complete Set of Dimension-7 Operators},''
  \href{http://dx.doi.org/10.1103/PhysRevD.90.125023}{{\em Phys. Rev.} {\bf
  D90} (2014) no.~12, 125023},
\href{http://arxiv.org/abs/1410.4193}{{\tt arXiv:1410.4193 [hep-ph]}}.
%%CITATION = ARXIV:1410.4193;%%.

\bibitem{Lehman:2015coa}
L.~Lehman and A.~Martin, ``{Low-derivative operators of the Standard Model
  effective field theory via Hilbert series methods},''
  \href{http://dx.doi.org/10.1007/JHEP02(2016)081}{{\em JHEP} {\bf 02} (2016)
  081},
\href{http://arxiv.org/abs/1510.00372}{{\tt arXiv:1510.00372 [hep-ph]}}.
%%CITATION = ARXIV:1510.00372;%%.

\bibitem{Henning:2015alf}
B.~Henning, X.~Lu, T.~Melia, and H.~Murayama, ``{2, 84, 30, 993, 560, 15456,
  11962, 261485, ...: Higher dimension operators in the SM EFT},''
  \href{http://dx.doi.org/10.1007/JHEP09(2019)019,
  10.1007/JHEP08(2017)016}{{\em JHEP} {\bf 08} (2017)  016},
  \href{http://arxiv.org/abs/1512.03433}{{\tt arXiv:1512.03433 [hep-ph]}}.
[Erratum: JHEP09,019(2019)].
%%CITATION = ARXIV:1512.03433;%%.

\bibitem{Marinissen:2020jmb}
C.~B. Marinissen, R.~Rahn, and W.~J. Waalewijn, ``{..., 83106786, 114382724,
  1509048322, 2343463290, 27410087742, ... Efficient Hilbert Series for
  Effective Theories},'' \href{http://arxiv.org/abs/2004.09521}{{\tt
  arXiv:2004.09521 [hep-ph]}}.

\bibitem{Alonso:2013hga}
R.~Alonso, E.~E. Jenkins, A.~V. Manohar, and M.~Trott, ``{Renormalization Group
  Evolution of the Standard Model Dimension Six Operators III: Gauge Coupling
  Dependence and Phenomenology},''
  \href{http://dx.doi.org/10.1007/JHEP04(2014)159}{{\em JHEP} {\bf 04} (2014)
  159},
\href{http://arxiv.org/abs/1312.2014}{{\tt arXiv:1312.2014 [hep-ph]}}.
%%CITATION = ARXIV:1312.2014;%%.

\bibitem{Wells:2015uba}
J.~D. Wells and Z.~Zhang, ``{Effective theories of universal theories},''
  \href{http://dx.doi.org/10.1007/JHEP01(2016)123}{{\em JHEP} {\bf 01} (2016)
  123},
\href{http://arxiv.org/abs/1510.08462}{{\tt arXiv:1510.08462 [hep-ph]}}.
%%CITATION = ARXIV:1510.08462;%%.

\bibitem{Wells:2015cre}
J.~D. Wells and Z.~Zhang, ``{Renormalization group evolution of the universal
  theories EFT},'' \href{http://dx.doi.org/10.1007/JHEP06(2016)122}{{\em JHEP}
  {\bf 06} (2016)  122},
\href{http://arxiv.org/abs/1512.03056}{{\tt arXiv:1512.03056 [hep-ph]}}.
%%CITATION = ARXIV:1512.03056;%%.

\bibitem{Trott:2014dma}
M.~Trott, ``{On the consistent use of Constructed Observables},''
  \href{http://dx.doi.org/10.1007/JHEP02(2015)046}{{\em JHEP} {\bf 02} (2015)
  046},
\href{http://arxiv.org/abs/1409.7605}{{\tt arXiv:1409.7605 [hep-ph]}}.
%%CITATION = ARXIV:1409.7605;%%.

\bibitem{Trott:2017yhn}
M.~Trott, ``{EWPD in the SMEFT and the $\mathcal{O}(y_t^2,\lambda)$ one loop
  $Z$ decay width},'' in {\em {52nd Rencontres de Moriond on EW Interactions
  and Unified Theories}}, pp.~63--70.
\newblock 2017.
\newblock \href{http://arxiv.org/abs/1705.05652}{{\tt arXiv:1705.05652
  [hep-ph]}}.

\bibitem{Brivio:2017bnu}
I.~Brivio and M.~Trott, ``{Scheming in the SMEFT... and a reparameterization
  invariance!},'' \href{http://dx.doi.org/10.1007/JHEP07(2017)148}{{\em JHEP}
  {\bf 07} (2017)  148}, \href{http://arxiv.org/abs/1701.06424}{{\tt
  arXiv:1701.06424 [hep-ph]}}. [Addendum: JHEP 05, 136 (2018)].

\bibitem{Arzt:1994gp}
C.~Arzt, M.~B. Einhorn, and J.~Wudka, ``{Patterns of deviation from the
  standard model},'' \href{http://dx.doi.org/10.1016/0550-3213(94)00336-D}{{\em
  Nucl. Phys.} {\bf B433} (1995)  41--66},
\href{http://arxiv.org/abs/hep-ph/9405214}{{\tt arXiv:hep-ph/9405214
  [hep-ph]}}.
%%CITATION = HEP-PH/9405214;%%.

\bibitem{Einhorn:2013kja}
M.~B. Einhorn and J.~Wudka, ``{The Bases of Effective Field Theories},''
  \href{http://dx.doi.org/10.1016/j.nuclphysb.2013.08.023}{{\em Nucl. Phys.}
  {\bf B876} (2013)  556--574},
\href{http://arxiv.org/abs/1307.0478}{{\tt arXiv:1307.0478 [hep-ph]}}.
%%CITATION = ARXIV:1307.0478;%%.

\bibitem{Craig:2019wmo}
N.~Craig, M.~Jiang, Y.-Y. Li, and D.~Sutherland, ``{Loops and trees in generic
  EFTs},''
\href{http://arxiv.org/abs/2001.00017}{{\tt arXiv:2001.00017 [hep-ph]}}.
%%CITATION = ARXIV:2001.00017;%%.

\bibitem{Jenkins:2013zja}
E.~E. Jenkins, A.~V. Manohar, and M.~Trott, ``{Renormalization Group Evolution
  of the Standard Model Dimension Six Operators I: Formalism and lambda
  Dependence},'' \href{http://dx.doi.org/10.1007/JHEP10(2013)087}{{\em JHEP}
  {\bf 10} (2013)  087},
\href{http://arxiv.org/abs/1308.2627}{{\tt arXiv:1308.2627 [hep-ph]}}.
%%CITATION = ARXIV:1308.2627;%%.

\bibitem{Jenkins:2013wua}
E.~E. Jenkins, A.~V. Manohar, and M.~Trott, ``{Renormalization Group Evolution
  of the Standard Model Dimension Six Operators II: Yukawa Dependence},''
  \href{http://dx.doi.org/10.1007/JHEP01(2014)035}{{\em JHEP} {\bf 01} (2014)
  035},
\href{http://arxiv.org/abs/1310.4838}{{\tt arXiv:1310.4838 [hep-ph]}}.
%%CITATION = ARXIV:1310.4838;%%.

\bibitem{Liao:2016qyd}
Y.~Liao and X.-D. Ma, ``{Operators up to Dimension Seven in Standard Model
  Effective Field Theory Extended with Sterile Neutrinos},''
  \href{http://dx.doi.org/10.1103/PhysRevD.96.015012}{{\em Phys. Rev.} {\bf
  D96} (2017) no.~1, 015012},
\href{http://arxiv.org/abs/1612.04527}{{\tt arXiv:1612.04527 [hep-ph]}}.
%%CITATION = ARXIV:1612.04527;%%.

\bibitem{Elias-Miro:2013mua}
J.~Elias-Miro, J.~Espinosa, E.~Masso, and A.~Pomarol, ``{Higgs windows to new
  physics through d=6 operators: constraints and one-loop anomalous
  dimensions},'' \href{http://dx.doi.org/10.1007/JHEP11(2013)066}{{\em JHEP}
  {\bf 11} (2013)  066}, \href{http://arxiv.org/abs/1308.1879}{{\tt
  arXiv:1308.1879 [hep-ph]}}.

\bibitem{Brivio:2017vri}
I.~Brivio and M.~Trott, ``{The Standard Model as an Effective Field Theory},''
  \href{http://dx.doi.org/10.1016/j.physrep.2018.11.002}{{\em Phys. Rept.} {\bf
  793} (2019)  1--98},
\href{http://arxiv.org/abs/1706.08945}{{\tt arXiv:1706.08945 [hep-ph]}}.
%%CITATION = ARXIV:1706.08945;%%.

\bibitem{Baglio:2018bkm}
J.~Baglio, S.~Dawson, and I.~M. Lewis, ``{NLO effects in EFT fits to $W^+W^-$
  production at the LHC},''
  \href{http://dx.doi.org/10.1103/PhysRevD.99.035029}{{\em Phys. Rev. D} {\bf
  99} (2019) no.~3, 035029}, \href{http://arxiv.org/abs/1812.00214}{{\tt
  arXiv:1812.00214 [hep-ph]}}.

\bibitem{Skiba:2010xn}
W.~Skiba, \href{http://dx.doi.org/10.1142/9789814327183_0001}{``{Effective
  Field Theory and Precision Electroweak Measurements},''} in {\em {TASI
  2009}}, pp.~5--70.
\newblock 2011.
\newblock
\href{http://arxiv.org/abs/1006.2142}{{\tt arXiv:1006.2142 [hep-ph]}}.
\newblock
%%CITATION = ARXIV:1006.2142;%%.

\bibitem{Khandker:2012zu}
Z.~U. Khandker, D.~Li, and W.~Skiba, ``{Electroweak Corrections from Triplet
  Scalars},'' \href{http://dx.doi.org/10.1103/PhysRevD.86.015006}{{\em Phys.
  Rev. D} {\bf 86} (2012)  015006}, \href{http://arxiv.org/abs/1201.4383}{{\tt
  arXiv:1201.4383 [hep-ph]}}.

\bibitem{deBlas:2014mba}
J.~de~Blas, M.~Chala, M.~Perez-Victoria, and J.~Santiago, ``{Observable Effects
  of General New Scalar Particles},''
  \href{http://dx.doi.org/10.1007/JHEP04(2015)078}{{\em JHEP} {\bf 04} (2015)
  078}, \href{http://arxiv.org/abs/1412.8480}{{\tt arXiv:1412.8480 [hep-ph]}}.

\bibitem{Chiang:2015ura}
C.-W. Chiang and R.~Huo, ``{Standard Model Effective Field Theory: Integrating
  out a Generic Scalar},''
  \href{http://dx.doi.org/10.1007/JHEP09(2015)152}{{\em JHEP} {\bf 09} (2015)
  152}, \href{http://arxiv.org/abs/1505.06334}{{\tt arXiv:1505.06334
  [hep-ph]}}.

\bibitem{Ellis:2016enq}
S.~A.~R. Ellis, J.~Quevillon, T.~You, and Z.~Zhang, ``{Mixed heavy--light
  matching in the Universal One-Loop Effective Action},''
  \href{http://dx.doi.org/10.1016/j.physletb.2016.09.016}{{\em Phys. Lett. B}
  {\bf 762} (2016)  166--176}, \href{http://arxiv.org/abs/1604.02445}{{\tt
  arXiv:1604.02445 [hep-ph]}}.

\bibitem{Dawson:2017vgm}
S.~Dawson and C.~W. Murphy, ``{Standard Model EFT and Extended Scalar
  Sectors},'' \href{http://dx.doi.org/10.1103/PhysRevD.96.015041}{{\em Phys.
  Rev. D} {\bf 96} (2017) no.~1, 015041},
  \href{http://arxiv.org/abs/1704.07851}{{\tt arXiv:1704.07851 [hep-ph]}}.

\bibitem{Heeck:2014zfa}
J.~Heeck, ``{Unbroken B – L symmetry},''
  \href{http://dx.doi.org/10.1016/j.physletb.2014.10.067}{{\em Phys. Lett.}
  {\bf B739} (2014)  256--262},
\href{http://arxiv.org/abs/1408.6845}{{\tt arXiv:1408.6845 [hep-ph]}}.
%%CITATION = ARXIV:1408.6845;%%.

\bibitem{DAmbrosio:2002vsn}
G.~D'Ambrosio, G.~F. Giudice, G.~Isidori, and A.~Strumia, ``{Minimal flavor
  violation: An Effective field theory approach},''
  \href{http://dx.doi.org/10.1016/S0550-3213(02)00836-2}{{\em Nucl. Phys.} {\bf
  B645} (2002)  155--187},
\href{http://arxiv.org/abs/hep-ph/0207036}{{\tt arXiv:hep-ph/0207036
  [hep-ph]}}.
%%CITATION = HEP-PH/0207036;%%.

\bibitem{delAguila:2000rc}
F.~del Aguila, M.~Perez-Victoria, and J.~Santiago, ``{Observable contributions
  of new exotic quarks to quark mixing},''
  \href{http://dx.doi.org/10.1088/1126-6708/2000/09/011}{{\em JHEP} {\bf 09}
  (2000)  011}, \href{http://arxiv.org/abs/hep-ph/0007316}{{\tt
  arXiv:hep-ph/0007316}}.

\bibitem{delAguila:2008pw}
F.~del Aguila, J.~de~Blas, and M.~Perez-Victoria, ``{Effects of new leptons in
  Electroweak Precision Data},''
  \href{http://dx.doi.org/10.1103/PhysRevD.78.013010}{{\em Phys. Rev. D} {\bf
  78} (2008)  013010}, \href{http://arxiv.org/abs/0803.4008}{{\tt
  arXiv:0803.4008 [hep-ph]}}.

\bibitem{Crivellin:2020ebi}
A.~Crivellin, F.~Kirk, C.~A. Manzari, and M.~Montull, ``{Global Electroweak Fit
  and Vector-Like Leptons in Light of the Cabibbo Angle Anomaly},''
  \href{http://arxiv.org/abs/2008.01113}{{\tt arXiv:2008.01113 [hep-ph]}}.

\bibitem{Jiang:2018pbd}
M.~Jiang, N.~Craig, Y.-Y. Li, and D.~Sutherland, ``{Complete One-Loop Matching
  for a Singlet Scalar in the Standard Model EFT},''
  \href{http://dx.doi.org/10.1007/JHEP02(2019)031}{{\em JHEP} {\bf 02} (2019)
  031},
\href{http://arxiv.org/abs/1811.08878}{{\tt arXiv:1811.08878 [hep-ph]}}.
%%CITATION = ARXIV:1811.08878;%%.

\bibitem{Haisch:2020ahr}
U.~Haisch, M.~Ruhdorfer, E.~Salvioni, E.~Venturini, and A.~Weiler, ``{Singlet
  night in Feynman-ville: one-loop matching of a real scalar},''
  \href{http://dx.doi.org/10.1007/JHEP04(2020)164}{{\em JHEP} {\bf 04} (2020)
  164},
\href{http://arxiv.org/abs/2003.05936}{{\tt arXiv:2003.05936 [hep-ph]}}.
%%CITATION = ARXIV:2003.05936;%%.

\end{thebibliography}\endgroup

\end{document}